\documentclass[twocolumn,twocolappendix]{aastex631}

\pdfoutput=1

\usepackage{enumerate}
\usepackage{enumitem}
\usepackage{natbib,ifthen}
\usepackage{graphicx}
\usepackage{fancyhdr}
\usepackage{amssymb,amsmath}
\usepackage{multirow,bigdelim}
\usepackage{wasysym}  
\usepackage{amssymb}
\usepackage{chngcntr}

\newcommand*{\ditto}{---\texttt{"}---}

\newenvironment{packed_enum}{
\begin{enumerate}
  \setlength{\itemsep}{6.5pt}
  \setlength{\parskip}{0pt}
  \setlength{\parsep}{0pt}
}{\end{enumerate}}

\shorttitle{Higher-Order Millimeter Hydrogen RLs in the LMC}
\shortauthors{Sewi{\l}o et al.}

\begin{document}


\title{The Detection of Higher-Order Millimeter Hydrogen Recombination Lines in the Large Magellanic Cloud}


\correspondingauthor{Marta Sewi{\l}o}
\email{marta.m.sewilo@nasa.gov}

\author[0000-0003-2248-6032]{Marta Sewi{\l}o}
\affiliation{Exoplanets and Stellar Astrophysics Laboratory, NASA Goddard Space Flight Center, Greenbelt, MD 20771, USA}
\affiliation{Department of Astronomy, University of Maryland, College Park, MD 20742, USA}
\affiliation{Center for Research and Exploration in Space Science and Technology, NASA Goddard Space Flight Center, Greenbelt, MD 20771} 

\author[0000-0002-2062-1600]{Kazuki Tokuda}
\affiliation{Department of Earth and Planetary Sciences, Faculty of Sciences, Kyushu University, Nishi-ku, Fukuoka 819-0395, Japan}
\affiliation{National Astronomical Observatory of Japan, National Institutes of Natural Sciences, 2-21-1 Osawa, Mitaka, Tokyo 181-8588, Japan}

\author[0000-0003-4444-5602]{Stan E. Kurtz}
\affiliation{Instituto de Radioastronom\'{i}a y Astrof\'{i}sica, Universidad Nacional Aut\'{o}noma de M\'{e}xico, Apdo. Postal 3-72, 58090 Morelia, Michoac\'{a}n, Mexico}

\author[0000-0001-6752-5109]{Steven B. Charnley}
\affiliation{Astrochemistry Laboratory, NASA Goddard Space Flight Center, Greenbelt, MD 20771, USA}

\author[0000-0002-9277-8025]{Thomas M{\"o}ller}
\affiliation{I. Physikalisches Institut der Universit{\"a}t zu K{\"o}ln, Z{\"u}lpicher Str. 77, 50937, K{\"o}ln, Germany}

\author[0000-0002-1143-6710]{Jennifer Wiseman}
\affiliation{Exoplanets and Stellar Astrophysics Laboratory, NASA Goddard Space Flight Center, Greenbelt, MD 20771, USA}

\author[0000-0002-3925-9365]{C.-H. Rosie Chen}
\affiliation{Max-Planck-Institut f{\"u}r Radioastronomie, Auf dem H{\"u}gel 69 D-53121 Bonn, Germany}

\author[0000-0002-4663-6827]{Remy Indebetouw}
\affiliation{Department of Astronomy, University of Virginia, PO Box 400325, Charlottesville, VA 22904, USA}
\affiliation{National Radio Astronomy Observatory, 520 Edgemont Rd, Charlottesville, VA 22903, USA}

\author[0000-0002-3078-9482]{\'{A}lvaro S\'{a}nchez-Monge}
\affiliation{Institut de Ci\`encies de l'Espai (ICE, CSIC), Can Magrans s/n, E-08193, Bellaterra, Barcelona, Spain}
\affiliation{Institut d'Estudis Espacials de Catalunya (IEEC), Barcelona, Spain}

\author[0000-0002-6907-0926]{Kei E. I. Tanaka}
\affiliation{Tokyo Institute of Technology, 2 Chome-12-1 Ookayama, Meguro City, Tokyo 152-8550, Japan}

\author[0000-0003-2141-5689]{Peter Schilke}
\affiliation{I. Physikalisches Institut der Universit{\"a}t zu K{\"o}ln, Z{\"u}lpicher Str. 77, 50937, K{\"o}ln, Germany}

\author[0000-0001-7826-3837]{Toshikazu Onishi}
\affiliation{Department of Physics, Graduate School of Science, Osaka Metropolitan University, 1-1 Gakuen-cho, Naka-ku, Sakai, Osaka 599-8531, Japan}

\author[0000-0002-8217-7509]{Naoto Harada}
\affiliation{Department of Earth and Planetary Sciences, Faculty of Sciences, Kyushu University, Nishi-ku, Fukuoka 819-0395, Japan}


\begin{abstract}
We report the first extragalactic detection of the higher-order millimeter hydrogen recombination lines ($\Delta n>2$). The $\gamma$-, $\epsilon$-, and $\eta$-transitions  have been detected toward the millimeter continuum source N\,105--1\,A in the star-forming region N\,105 in the Large Magellanic Cloud (LMC) with the Atacama Large Millimeter/submillimeter Array (ALMA). We use the H40$\alpha$ line, the brightest of the detected recombination lines (H40$\alpha$, H36$\beta$, H50$\beta$, H41$\gamma$, H57$\gamma$, H49$\epsilon$, H53$\eta$, and H54$\eta$), and/or the 3 mm free-free continuum emission to determine the physical parameters of N\,105--1\,A (the electron temperature, emission measure, electron density, and size) and study ionized gas kinematics. We compare the physical properties of N\,105--1\,A to a large sample of Galactic compact and ultracompact (UC) H\,{\sc ii} regions and conclude that N\,105--1\,A is similar to the most luminous ($L>10^5$ $L_{\odot}$) UC H\,{\sc ii} regions in the Galaxy. N\,105--1\,A is ionized by an O5.5 V star, it is deeply embedded in its natal molecular clump, and likely associated with a (proto)cluster. We incorporate high-resolution molecular line data including CS, SO, SO$_2$, and CH$_3$OH ($\sim$0.12 pc),  and HCO$^{+}$ and CO ($\sim$0.087 pc) to explore the molecular environment of N\,105--1\,A.  Based on the CO data, we find evidence for a cloud-cloud collision that likely triggered star formation in the region.  We find no clear outflow signatures, but the presence of filaments and streamers indicates on-going accretion onto the clump hosting the UC H\,{\sc ii} region.  Sulfur chemistry in N\,105--1\,A is consistent with the accretion shock model predictions. 
\end{abstract}

\section{Introduction}
\label{s:intro}

The hydrogen radio and millimeter/submillimeter recombination lines (RRL and mm-RLs, respectively) are excellent tools to measure the temperature and column density of ionized gas in star-forming regions, and provide information on the ionized gas kinematics (e.g., \citealt{dupree1970}; \citealt{brown1978}; \citealt{gordon2002}; \citealt{tanaka2016}). They allow us to peer into the dense and dusty regions of the molecular clouds  inaccessible through observations at other wavelengths. 

Recombination lines are formed by a radiative recombination in which a free electron recombines with an ion. In this process, the electron ends up in any of the bound electronic states and the atom emits a photon that carries away the energy lost by the electron. Following the recombination, the electron cascades down the energy levels from the level it recombined to (with the principal quantum number $n$$>$1) to the ground state ($n$=1) by a sequence of spontaneous emissions, producing a series of emission lines (the recombination lines) at wavelengths that depend on the difference in energy between the levels. For higher $n$, this difference in energy is smaller, and thus the atom emits at longer wavelengths.  If the electron recombines to a highly-excited state (large $n$), the process will produce spectral lines from radio to ultraviolet (UV) wavelengths.  The recombination to the ground state of the hydrogen atom emits the Lyman $\alpha$ (Ly$\alpha$) spectral line in the UV, the strongest emission line observed toward astrophysical objects. Atoms other than hydrogen produce recombination lines as well (such as helium, carbon, and oxygen); however, they have much lower abundances compared to hydrogen and thus the recombination lines are weaker (e.g., \citealt{gordon2002}).  Recombination lines can also be emitted by ions; the detection of the He\,{\sc ii}, C\,{\sc ii}, and O\,{\sc ii} recombination lines has been reported in the literature (e.g., \citealt{chaisson1976}; \citealt{liu2023}).

The RRL/mm-RL transitions are identified by the name of the element, the principal quantum number of the final level, and the change in the principal quantum number ($\Delta n$) indicated with successive letters in the Greek alphabet. The recombination lines of H from level $n+\Delta n$ to $n$ are denoted (H$n\alpha$, H$n\beta$, H$n\gamma$, H$n\delta$, H$n\epsilon$, H$n\zeta$, H$n\eta$...) for $\Delta n$ = (1, 2, 3, 4, 5, 6, 7...). The $\alpha$-transitions ($\Delta n$=1) have the largest Einstein coefficient for spontaneous emission ($A$) and thus are the most likely transitions, producing the brightest recombination lines that can be detected at large distances.  The hydrogen RL $\alpha$- and $\beta$-transitions were first detected outside the Galaxy in the 1970s. The higher-order ($\Delta n$$>$2) mm/radio transitions had not been detected until our observations reported in this paper.  

The first extragalactic detection of radio recombination lines was reported by \citet{mezger1970}.  The H109$\alpha$, He109$\alpha$, and H137$\beta$ transitions at 6 cm (5 GHz) were detected toward the star-forming region 30 Doradus in the Large Magellanic Cloud (LMC) with the Parkes 64 m radio telescope (half-power beam width, HPBW$\approx$4$'$).  The H109$\alpha$ line was later detected toward seven other regions in the LMC in addition to 30 Dor (out of 14 surveyed) by \citet{mcgee1974}.  In the higher resolution observations of 30 Dor with the Australia Telescope Compact Array (ATCA; HPBW$\sim$15$^{''}$), \citet{peck1997} detected the H90$\alpha$, H113$\beta$, and He90$\alpha$ ($\sim$8.9 GHz),  H92$\alpha$ (8.3 GHz), and again the H109$\alpha$ (5 GHz) recombination lines.  30 Dor has been the LMC star-forming region best-studied in RRLs. The H30$\alpha$ and H40$\alpha$ lines are now routinely detected toward the LMC star-forming regions with the Atacama Large Millimeter/submillimeter Array (ALMA), in programs targeting cold molecular gas, such as CO (e.g., \citealt{indebetouw2013}: 30 Doradus; \citealt{saigo2017}: N\,159--East; \citealt{nayak2019}: N\,79). The only detection of a carbon RL in the LMC (C30$\alpha$ in N\,79) is uncertain since it may in fact be a helium RL (He30$\alpha$). 

\begin{figure*}
\centering
\includegraphics[width=0.485\textwidth]{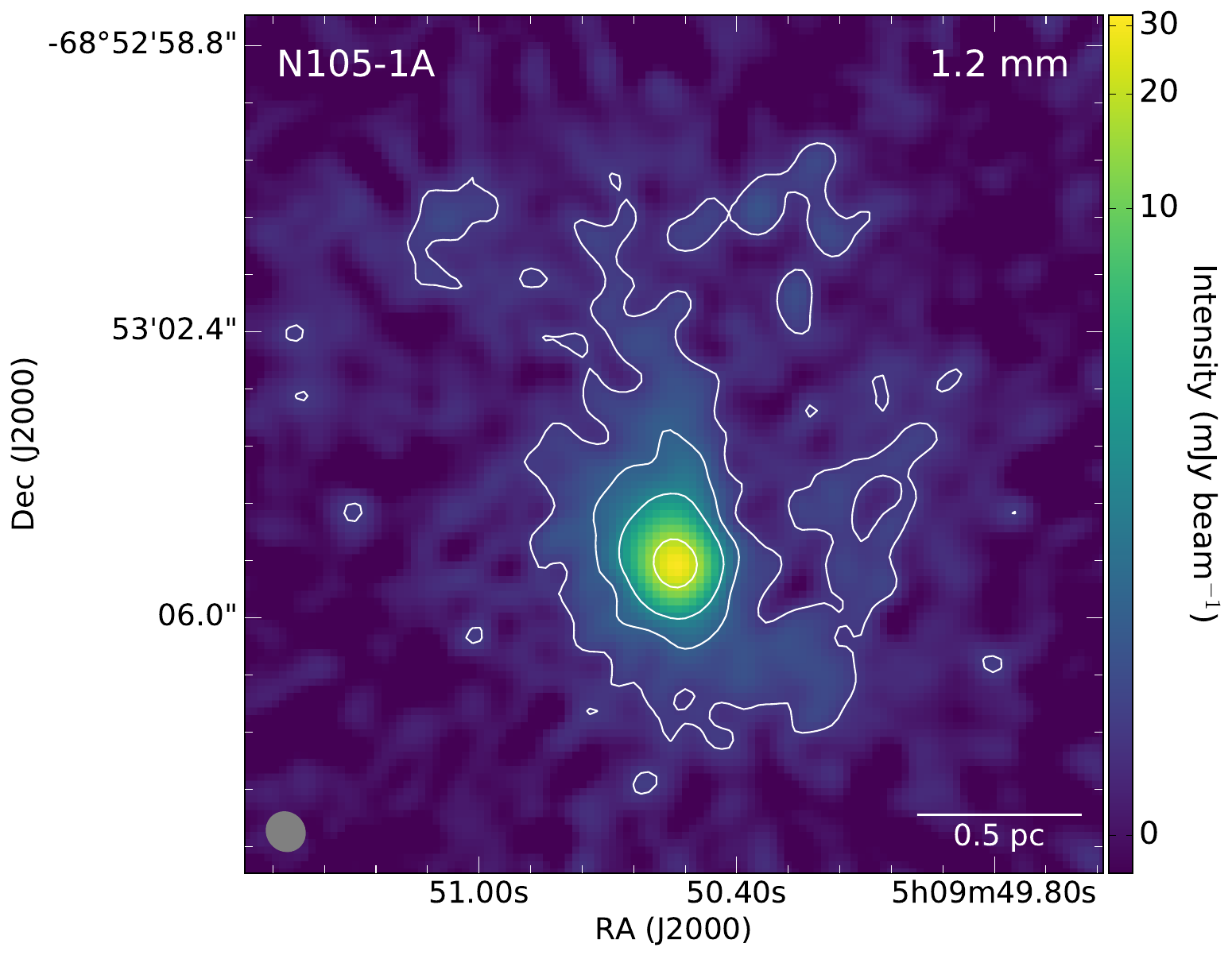}
\includegraphics[width=0.485\textwidth]{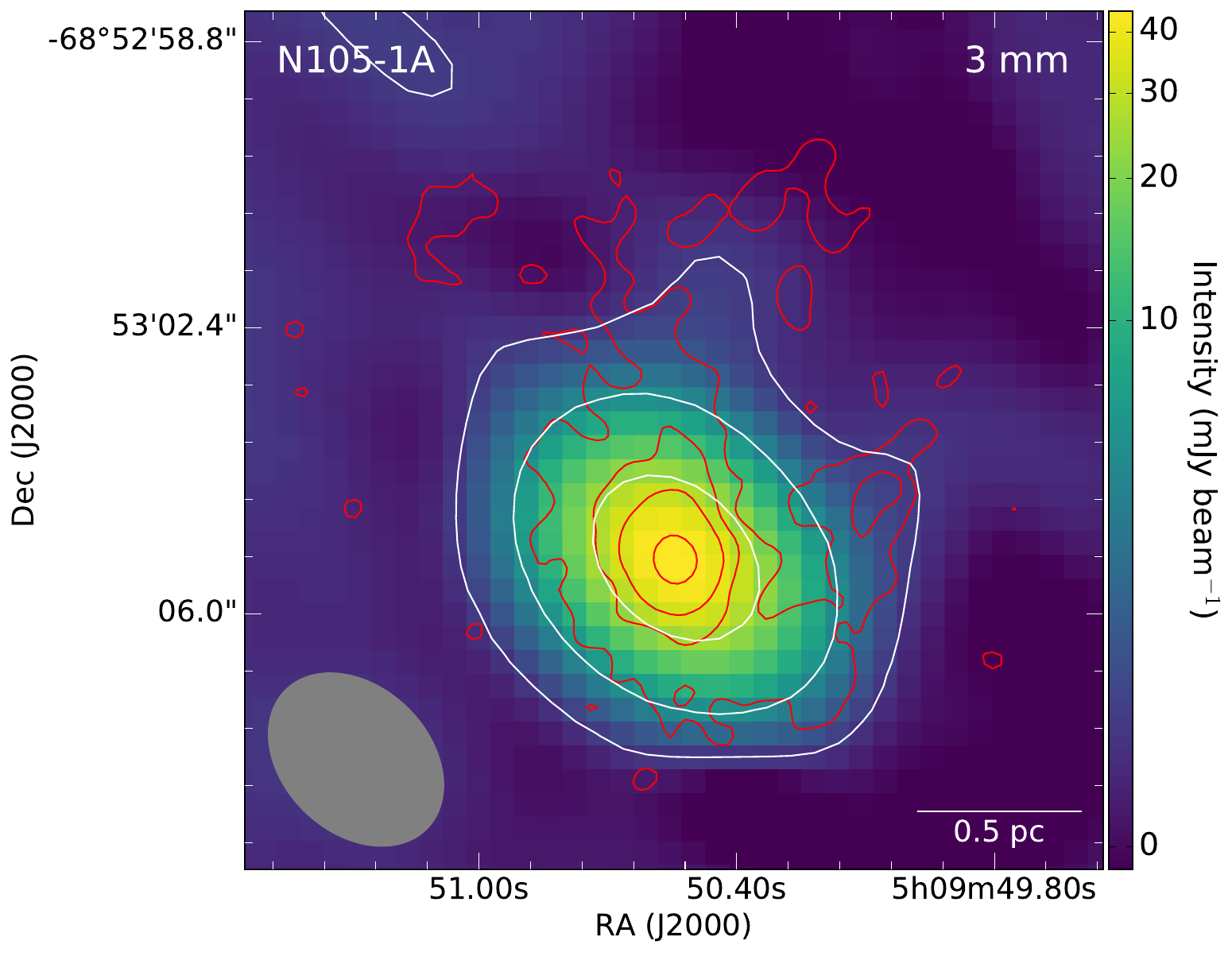}
\caption{The 1.2 mm (left) and 3 mm (right) continuum images of N\,105--1\,A.  In the left/right panel, white contours correspond to (3, 10, 30, 250)/(3, 20, 100) times $6.9\times10^{-5}$/$2.5\times10^{-4}$ Jy beam$^{-1}$, the 1.2/3 mm continuum image rms noise ($\sigma$). Red contours in the right panel are the same as white contours in the left panel.  The size of the ALMA synthesized beam is indicated in the bottom left corner in each image:  $0\rlap.{''}51\times0\rlap.{''}47$ for 1.2 mm and $2\rlap.{''}49\times1\rlap.{''}83$ for the 3 mm continuum image, respectively. \label{f:cont}} 
\end{figure*}

The first extragalactic detections of  RRLs beyond the Magellanic Clouds were made toward M\,82 (H166$\alpha$, \citealt{shaver1977}; H102$\alpha$,  \citealt{bell1977}; H92$\alpha$, \citealt{chaisson1977}) and NGC\,253 (H102$\alpha$, \citealt{seaquist1977}).  After these first observations of extragalactic RRLs, no new detections were reported until the H53$\alpha$ mm-RL was detected toward NGC\,2146 over 10 years later (\citealt{puxley1991}).  RLs in the millimeter and submillimeter range have proven to be excellent tools in studying the star formation activity in dusty regions in the centers of nearby galaxies (e.g., \citealt{scoville2013}). ALMA observations have provided new detections of the H$n\alpha$ mm-RLs toward nearby galaxies (e.g., \citealt{bendo2015}; \citealt{bendo2016}; \citealt{bendo2017}; \citealt{michiyama2020}) and the first detection of the fainter H$n\beta$  and He$n\alpha$ lines toward NGC\,253 (\citealt{meier2015}; \citealt{martin2021}). At cosmological distances, \citet{emig2019} claimed the first detection of the RRL after stacking 13 $\alpha$-transitions with principal quantum numbers $n=266-301$, detected in the spectrum of the radio quasar 3C\,190 ($z=1.1946$) with LOw Frequency ARray (LOFAR; \citealt{vanhaarlem2013}).  

In this paper, we report the detection of the H40$\alpha$, H36$\beta$, H50$\beta$, H41$\gamma$, H57$\gamma$, H49$\epsilon$, H53$\eta$, and H54$\eta$ mm-RLs and a tentative detection of the H55$\theta$ line with ALMA toward the 1.2 mm continuum source N\,105--1\,A in the N\,105 star-forming region  in the LMC (LHA\,120--N\,105, \citealt{henize1956}; DEM\,L86,  \citealt{davies1976}).  The LMC is the Milky Way's satellite galaxy, the nearest star-forming galaxy, located at the distance of $49.59\pm0.09$ (statistical) $\pm0.54$ (systematic) kpc (\citealt{pietrzynski2019}). The (sub)mm-RL $\gamma$-, $\epsilon$-, $\eta$-, and $\theta$-transitions are detected outside the Galaxy for the first time. 

The environment of the LMC is distinct from that found in our Galaxy - it is characterized by a low metallicity (lower abundances of gaseous atoms heavier than He; $Z_{\rm LMC}$ $\sim$ 0.3--0.5 $Z_{\odot}$, e.g., \citealt{russell1992}; \citealt{westerlund1997}; \citealt{rolleston2002}), lower dust-to-gas ratio (e.g., \citealt{dufour1975,dufour1984}; \citealt{koornneef1984}; \citealt{duval2014}), higher intensity of the interstellar UV radiation (e.g., \citealt{browning2003}; \citealt{welty2006}), and lower cosmic-ray density (e.g., \citealt{abdo2010}; \citealt{knodlseder2013}).  Metal (such as C, O, and N) abundances are expected to set the electron temperature of H\,{\sc ii} regions since the collisionally excited lines from metals are the primary cooling mechanism in the ionized gas (e.g., \citealt{shaver1983}; \citealt{balser2011} and references therein). Thus, low abundances of metals in the LMC are expected to produce high temperatures.  All the properties of the LMC's environment can affect the chemistry of the sources emitting mm-RLs/RRLs. The relatively small distance to the LMC enables studies of individual stars and protostars in this unique environment that is similar to galaxies at the peak of star formation in the Universe ($z$$\sim$1.5--2; e.g., \citealt{pei1999}; \citealt{mehlert2002}; \citealt{madau2014}).

The paper is organized as follows. In Section~\ref{s:data}, we describe our ALMA observations and the archival data used in our analysis. We summarize previous studies on N\,105--1\,A in Section~\ref{s:1A}. The data analysis results and discussion are presented in Section~\ref{s:results} and Section~\ref{s:discussion}, respectively. In Section~\ref{s:summary}, we provide the summary and conclusions of our study.


\section{ALMA Observations} 
\label{s:data}

The N\,105--1 field hosting the source with the detection of mm-RLs (N\,105--1\,A) was observed with ALMA 12 m Array in Band 6 as part of the Cycle 7 project 2019.1.01720.S (PI M. Sewi{\l}o). The data were calibrated with version 5.6.1-8  of the ALMA pipeline in CASA (Common Astronomy Software Applications; \citealt{casa2022}). The observations were executed twice on October 21, 2019 with 43 antennas and baselines from 15 m to 783 m.  The (bandpass, flux, phase) calibrators were (J0519$-$4546, J0519$-$4546, J0440$-$6952) and (J0538$-$4405, J0538$-$4405, J0511$-$6806) for the first and second run, respectively. N\,105--1 was observed again on October 23, 2019 with 43 antennas, baselines from 15 m to 782 m, and the same calibrators.  The total on-source integration time was $\sim$13.1 minutes including both executions.

The spectral setup included four 1875 MHz spectral windows centered on frequencies of 242.4 GHz, 244.8 GHz, 257.9 GHz, and 259.7 GHz. 
Continuum was subtracted in the uv domain from the line spectral window and imaged.  The CASA task \texttt{tclean} was used for imaging using the Hogbom deconvolver, standard gridder, Briggs weighting with a robust parameter of 0.5, and auto-multithresh masking.  The rms sensitivity of 0.05 mJy per 0$\rlap.{''}51\times0\rlap.{''}47$ beam (4.4 mK) was achieved in the continuum. Sensitivity  of 
1.97 mJy (0.15 K) per 0$\rlap.{''}54\times0\rlap.{''}50$ beam was achieved in the 242.4 GHz spectral cube;
1.88 mJy (0.15 K) per 0$\rlap.{''}53\times0\rlap.{''}49$ beam in the 244.8 GHz cube; 
2.05 mJy (0.16 K) per 0$\rlap.{''}51\times0\rlap.{''}47$ beam in the 257.9 GHz cube;
2.28 mJy (0.18 K) per 0$\rlap.{''}51\times0\rlap.{''}47$ beam in the 259.7 GHz cube;
all four with 488.3 kHz (0.6 km s$^{-1}$) channels.  The images were corrected for primary beam attenuation.  The Band 6 molecular line data for the ALMA N\,105--1 field were analyzed and discussed in \citet{sewilo2022n105}.

In our analysis, we also utilize the ALMA Band 3 12m- and 7m-Array observations of two young stellar objects (YSOs) located in N\,105--1 (one coinciding with 1\,A; see Fig.~\ref{f:grid}) from our Cycle 5 project 2017.1.00093.S (PI T. Onishi).  We used the data from observations of both YSOs for imaging since the fields of view overlap considerably.  In this paper, we analyze spectral windows centered on the H40$\alpha$ line (a rest frequency of 99.02295~GHz; 12m data only to use with our Band 6 12m data) and $^{13}$CO (1--0) (110.20135  GHz, an upper state energy, $E_{\rm U}=5.3$ K; combined 12m and 7m data).

The 12m observations toward N\,105--1 were executed on April 2 and April 3, 2018 with 41 and 43 antennas and baselines between 15 m and 543 m and 15 m and 500 m, respectively. The (bandpass, flux, phase) calibrators were (J0635$-$7516, J0635$-$7516, J0529$-$7245) for both runs.  The total on-source integration time was $\sim$4 minutes.  The 1875 MHz spectral window centered on the H40$\alpha$  line was divided into 1920 channels of $\sim$976.6 kHz ($\sim$2.96 km s$^{-1}$) each.  The 59 MHz spectral window centered on the $^{13}$CO (1--0) line was divided into 1920 channels of $\sim$30.7 kHz ($\sim$0.084 km s$^{-1}$) each. The data were calibrated with CASA  5.1.1-5.  

Continuum was subtracted in the {\sc uv} domain from the H40$\alpha$ line spectral window. Sensitivity of  4.4 mJy per $2\rlap.{''}56\times1\rlap.{''}95$ beam (0.11 K) was achieved in the H40$\alpha$ data cube.  The H40$\alpha$ spectral cube has a cell size of  $0\rlap.{''}3\times0\rlap.{''}3 \times 2.96$ km s$^{-1}$. Sensitivity of 0.25 mJy per $2\rlap.{''}49\times1\rlap.{''}83$ beam (6.8 mK) was achieved in the continuum (99.023 GHz or 3.03 mm) constructed from the line-free channels. The images were corrected for primary beam attenuation.

\begin{figure*}
\centering
\includegraphics[width=\textwidth]{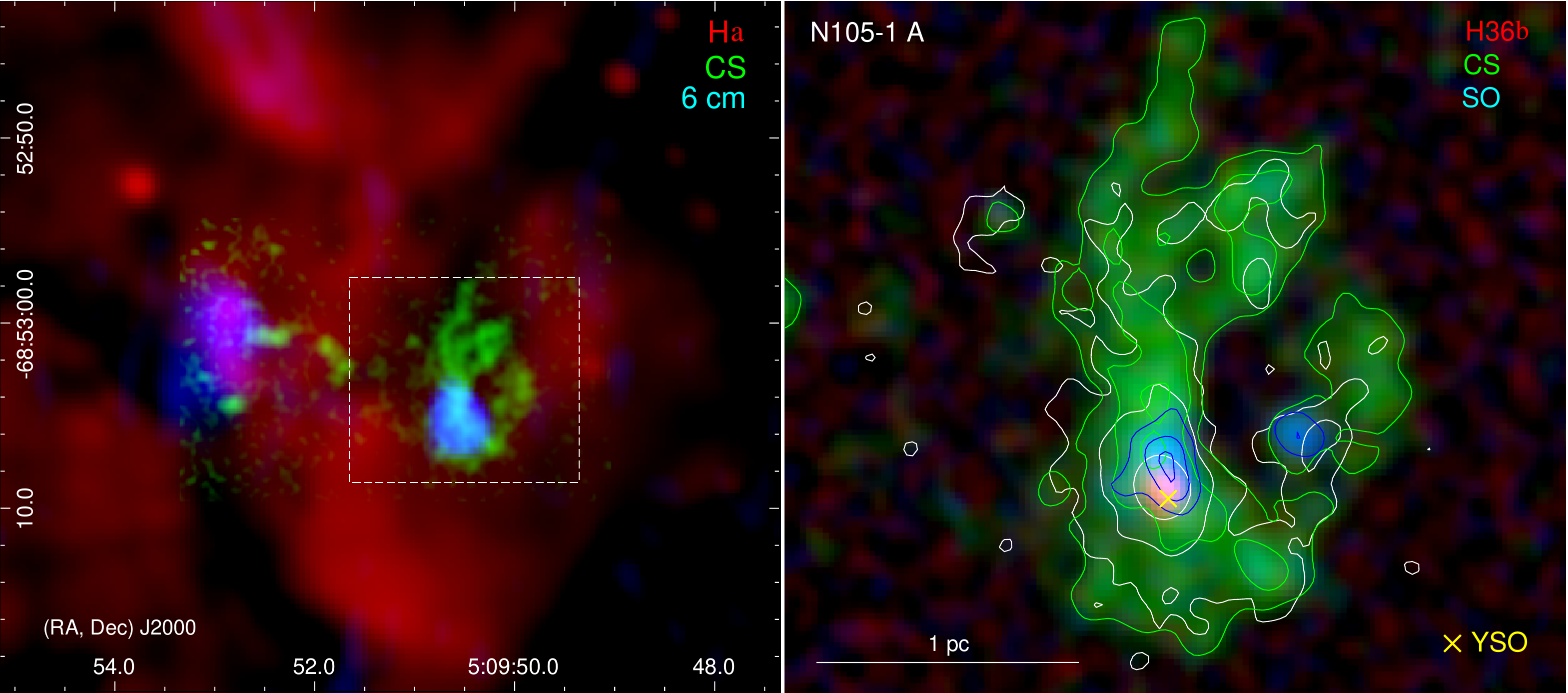}
\caption{Left: Three-color mosaic of the northern part of the N\,105 star-forming region combining the H$\alpha$ image from the MCELS2 survey with the MOSAIC II camera on the Blanco 4m Telescope at CTIO (red; a pixel size of $0\rlap.{''}27\times0\rlap.{''}27$, PI You-Hua Chu), the CS (5--4) image of the ALMA field N\,105--1 from \citet[green; an angular resolution $\sim$0$\rlap.{''}$5]{sewilo2022n105}, and the ATCA 4.8 GHz / 6 cm image (blue; $\sim$2$''$, \citealt{indebetouw2004}). The white box indicates the area shown in the image in the right panel. Right:  Three-color mosaic of N\,105--1\,A combining the H36$\beta$ recombination line (red, this paper) and the CS (5--4) (green; the same as in the left panel) and SO $^{3}\Sigma$ 6$_6$--5$_5$ (blue) images from \citet{sewilo2022n105}.  All the images have the same angular resolution of $\sim$0$\rlap.{''}$5 ($\sim$0.12 pc).  Green/blue contours correspond to the CS/SO emission with contour levels of  (20, 50, 90)\% of the CS/SO integrated intensity peak of 230.8/262.8 mJy beam$^{-1}$ km s$^{-1}$. White contours correspond to the 1.2 mm continuum emission with contour levels of (3, 10, 100) times the image rms noise level of $6.9\times10^{-5}$ Jy beam$^{-1}$. \label{f:region}} 
\end{figure*}

The 7m observations were executed eight times between December 29, 2017 and January 10, 2018 with 10 (1 execution) and 11 (7)  antennas, with baselines between 8.9 m and 48.9 m. The total on-source integration time was $\sim$19.7 minutes for each target YSO. Both bandpass and flux calibrators were J0522$-$3627 (7 executions) and J0006$-$0623 (1).  J0450$-$8101 (4 executions) and J0529$-$7245 (4) were used as phase calibrators.  The 62 MHz spectral window centered on the $^{13}$CO (1--0) line was divided into 2048 channels of $\sim$30.3 kHz ($\sim$0.082 km s$^{-1}$) each.  We combined the 12m and 7m data to image the $^{13}$CO (1--0) line. The $^{13}$CO (1--0) spectral cube has a cell size of  $0\rlap.{''}3 \times 0\rlap.{''}3 \times 0.2$ km s$^{-1}$ and it is corrected for primary beam attenuation.  The sensitivity of 16.9 mJy per $2\rlap.{''}29\times1\rlap.{''}67$ beam  (0.44 K) was achieved in the $^{13}$CO (1--0) data cube.

The 1.2 mm and 3 mm continuum images are shown in Fig.~\ref{f:cont}. The Band 3 spectrum from the spectral window covering H recombination lines and all the Band 6 spectra (first presented and analyzed in \citealt{sewilo2022n105}) are shown in Appendix A.  The mm-RL spectra for 1\,A (both Band 6 and Band 3) were extracted for the analysis as the mean within the contour corresponding to 50\% of the 1.2 mm continuum emission peak.

\subsection{ALMA Archival Data}
\label{s:arch}

We retrieved the Band 7 $^{12}$CO (3--2) and HCO$^{+}$ (4--3)  data from the Cycle 7 project 2019.1.01770.S from the ALMA archive to explore, respectively, the diffuse and dense gas surrounding N\,105--1\,A. Target Lh06 corresponding to our ALMA field N\,105--1, was observed as a part of the ``MAGellanic Outflow and chemistry Survey'' (MAGOS; PI K. Tanaka). A detailed description of the observations can be found in Tanaka et al. (2023, in prep.). The CO and HCO$^{+}$ data were calibrated and imaged using CASA 6.1.2.7. We used the CASA task \texttt{tclean} for imaging with the multi-scale deconvolver and the Briggs weighting with a robust parameter of 0.5. 

The CO (3--2) line with the rest frequency of 345,795.99 MHz and $E_{\rm U}$ of 11.5 K was included in the 937.5 MHz-wide spectral window centered at 345.3 GHz and divided into 1920 channels of 488.28 kHz each (or 0.42 km s$^{-1}$). The resulting CO (3--2) spectral cube has a $0\rlap.{''}41\times0\rlap.{''}33$ beam and sensitivity of 4.5 mJy beam$^{-1}$ (or 0.35 K) at a channel width of 1 km s$^{-1}$.  

The HCO$^{+}$ (4--3) line with the rest frequency of 356,734.22 MHz  and $E_{\rm U}$ of 42.8 K was included in the 1.875 GHz-wide spectral window centered at 357.1 GHz and divided into 3840 channels of 488.32 kHz each (or 0.41 km s$^{-1}$).  The resulting HCO$^{+}$ (4--3) spectral cube has a $0\rlap.{''}40\times0\rlap.{''}32$ beam and sensitivity of 7.0 mJy beam$^{-1}$ (or 0.52 K) at a channel width of 0.5 km s$^{-1}$. 

The 345.798 GHz ($\sim$870 $\mu$m) continuum image was constructed from line-free channels in the Band 7 data;  the sensitivity of 0.32 mJy per $0\rlap.{''}47\times0\rlap.{''}$39 beam was achieved in the continuum. 

\section{Previous Studies on N\,105--1\,A}
\label{s:1A}

The N\,105 star-forming region is located at the western edge of the LMC bar (e.g., \citealt{ambrocio1998}) and is associated with the OB association LH\,31 (e.g., \citealt{lucke1970}; 18 OB stars and two Wolf-Rayet stars)  and the sparse cluster NGC\,1858 (e.g., \citealt{bica1996};  age estimates in a range 8--17 Myr; \citealt{vallenari1994}; \citealt{alcaino1986}). The ALMA observations discussed in this paper are coincident with the brightest part of the optical nebula (N\,105A) and the peak of the molecular cloud in N\,105 (traced by $^{12}$CO 1--0, e.g., \citealt{wong2011}), dense gas emission peaks (traced by HCO$^{+}$ and HCN 1--0; \citealt{seale2012}), and the thermal radio continuum source MC\,23 or B0510--6857 (e.g., \citealt{mcgee1972}, \citealt{ellingsen1994}, \citealt{filipovic1998}).  The source with the detection of higher-order mm-RLs, N\,105--1\,A,  is the brightest 8.6 GHz (3 cm) and 4.8 GHz (6 cm) source in N\,105 (B0510$-$6857\,W) detected by \citet{indebetouw2004} with the Australia Telescope Compact Array (ATCA; a resolution of $\sim$1$\rlap.{''}$5 and $\sim$2$''$ at 3 cm and 6 cm, respectively) and it was classified as an ultracompact (UC) H\,{\sc ii} region. 

The infrared source coincident with N\,105--1\,A / B0510$-$6857\,W was reported in the literature as a candidate protostar by \citet{epchtein1984} based on five-band near-infrared (near-IR) photometry. The YSO classification was later supported by \citet[their source N\,105A IRS1]{oliveira2006} with high spatial resolution near-IR spectroscopic observations.  The 3--4 $\mu$m spectrum from the Infrared Spectrometer And Array Camera (ISAAC) on the ESO-VLT displays a very red continuum, very strong Br$\alpha$ and Pf$\gamma$ recombination line emission, and a non-detection of the Pf$\delta$ line which was postulated to be the result of a high dust column density in front of the source ($A_{\rm V}$$\sim$40 mag).  Further evidence supporting the interpretation of the source as an embedded protostar includes broad Br$\alpha$ line wings likely indicating the presence of an outflow, and the fact that the source is very bright in $L'$-band and it is extremely red ($K_{\rm S}$-$L'$ = 3.9 mag).  Strong hydrogen RLs indicate that the massive YSO has started ionizing its immediate surroundings. The recent VLT/KMOS near-IR spectroscopic observations reported in \citet{sewilo2022n105} also revealed strong hydrogen RLs - a full Brackett series emission in the {\it H}$+${\it K} bands.   

N\,105--1\,A was also classified as a YSO based on the {\it Spitzer} Space Telescope mid-IR photometric (\citealt{whitney2008}, source \#318 or SSTISAGE1C\,J050950.53$-$685305.4;  \citealt{gruendl2009}, source 050950.53$-$685305.5; \citealt{carlson2012}) and spectroscopic ({\it Spitzer}'s The InfraRed Spectrograph, IRS 5--38 $\mu$m, \citealt{seale2012}; \citealt{jones2017}) observations. The {\it Spitzer}/IRS spectrum exhibits relatively strong fine-structure lines such as [S\,{\sc IV}] 10.5 $\mu$m, [Ne\,{\sc II}] 12.8 $\mu$m, [Ne\,{\sc III}]  15.5 $\mu$m,  [S\,{\sc III}] 18.7 $\mu$m  and  33.5 $\mu$m,  and  [S\,{\sc III}] 34.8 $\mu$m, and weak emission lines from the Polycyclic Aromatic Hydrocarbons (PAHs).  The spectral energy distribution (SED, from near- to mid-IR) of N\,105--1\,A was well-fit with the \citet{robitaille2006} YSO radiative transfer models by \citet{carlson2012} with the best-fit stellar mass and luminosity of $31.3\pm2.6$ M$_{\odot}$ and ($1.4\pm0.2)\times10^5$ L$_{\odot}$, respectively.

N\,105--1\,A was included in the molecular spectral line analysis of the 1.2 mm continuum sources in the ALMA field N\,105--1 performed by \citet{sewilo2022n105}.  The CH$_3$OH, SO$_2$, SO, CS, H$^{13}$CO$^{+}$, H$^{13}$CN, and (tentatively) H$_2$CS lines are detected toward 1\,A (see Appendix~\ref{s:appA} for more details).  The detection of multiple CH$_3$OH and SO$_2$ lines allowed for an independent rotational temperature determination for each of these species.  

The distributions of the H$\alpha$, CS, SO, H36$\beta$, 1.2 mm, and 6 cm emission are compared in Fig.~\ref{f:region}.  The high-resolution H$\alpha$ image shows that 1\,A is associated with the extended low-intensity H$\alpha$ emission around the mm/radio continuum peak, while the mm continuum emission extending to the north from the main peak overlaps with what appears to be the H$\alpha$-dark region. The CS emission has a filamentary structure and extends even further to the north ($\sim$8$''$ or 1.9 pc), filling in the H$\alpha$-dark region. The H36$\beta$ mm-RL  emission (and the emission from all other mm-RLs except H57$\gamma$; see below) coincides with the mm/radio continuum peak, while the CS and SO peaks are offset to the north.   No molecular line emission for species detected by \citet{sewilo2022n105} coincides with the mm/radio continuum peak. 

N\,105--1\,A is associated with cold CH$_3$OH ($12\pm1$ K) and hot SO$_2$ ($96\pm20$ K). The presence of the hot SO$_2$ component led  \citet{sewilo2022n105} to suggest that 1\,A may host a hot molecular core.  Hot cores are compact ($D\lesssim0.1$ pc), hot ($T_{\rm kin}\gtrsim100$ K), and dense ($n_{\rm H}\gtrsim10^{6-7}$ cm$^{-3}$) regions around forming massive stars (e.g., \citealt{garay1999}; \citealt{kurtz2000}; \citealt{cesaroni2005}; \citealt{palau2011}). They produce rich molecular line spectra with many lines from complex organic molecules (COMs; containing 6 or more atoms including C, \citealt{herbst2009}). COMs are the products of interstellar grain-surface chemistry  (released to the gas phase by thermal evaporation or shock sputtering), or post-desorption gas chemistry (e.g., \citealt{herbst2009}; \citealt{oberg2016}; \citealt{oberg2021}). There are only a handful of known bona fide hot cores outside the Galaxy, all are located in the LMC (\citealt{shimonishi2016b}; \citealt{sewilo2018}; \citealt{sewilo2019}; \citealt{shimonishi2020}; \citealt{sewilo2022n105}).  

\citet{sewilo2022n105} argue that cold CH$_3$OH, but hot SO$_2$ detected toward the 1.2 mm continuum source in 1\,A may indicate that the SO$_2$ emission originates from the area offset from the continuum source and/or CH$_3$OH  is sub-thermally excited.  \citet{sewilo2022n105} conclude that it is more likely that the hot core in 1\,A is coincident with the SO$_2$ peak offset from the continuum peak by $\sim$0$\rlap.{''}$6 where CH$_3$OH is 10 K warmer and that it is externally illuminated since no IR source is detected in the existing observations; it was suggested that the hot core might be associated with outflow shocks.

\begin{figure}[b!]
\centering
\includegraphics[width=0.48\textwidth]{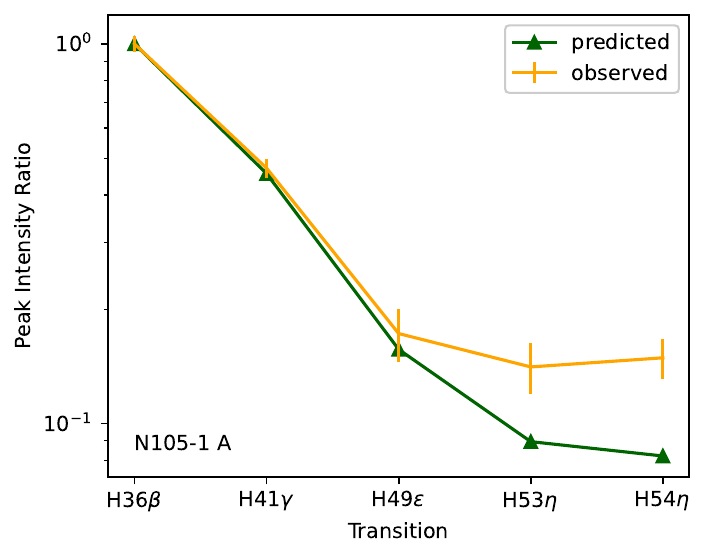} 
\caption{A comparison between the predicted and observed peak intensity ratios of hydrogen mm-RLs  (the peak intensity of a given mm-RL with respect to the peak intensity of the H36$\beta$ transition) detected in Band 6 toward N\,105--1\,A (see Section~\ref{s:HRL} for details). The uncertainties of the observed peak intensity ratios shown in the plot were calculated by error propagation using the 1$\sigma$ uncertainties of the mm-RL peak intensities provided in Table~\ref{t:fitRLs}. \label{f:intensities}}
\end{figure}


\section{Results and Analysis}
\label{s:results}

\subsection{Hydrogen Recombination Line Emission Toward N\,105--1\,A}
\label{s:HRL}

We detected five hydrogen mm-RLs toward N\,105--1\,A in our Band 6 observations: H36$\beta$ (260.03278~GHz), H41$\gamma$ (257.63549~GHz), H49$\epsilon$ (241.86116~GHz), H53$\eta$ (257.19399~GHz), and H54$\eta$ (243.94239~GHz). An additional transition, H55$\theta$ (258.52592~GHz), is tentatively detected. These are all hydrogen RLs with $\Delta n\leq8$ (up to $\theta$-transitions) falling within the frequency range of our observations (e.g., \citealt{gordon2002}). Three more mm-RLs (H40$\alpha$, H50$\beta$, and H57$\gamma$ at 99.02295 GHz, 99.22521 GHz, and 98.67189 GHz, respectively) were detected toward N\,105--1\,A in our unrelated, lower-resolution ALMA Band 3 project targeting high-mass YSOs in the LMC (see Section~\ref{s:data}). 

\begin{figure*}[ht!]
\centering
\includegraphics[width=0.485\textwidth]{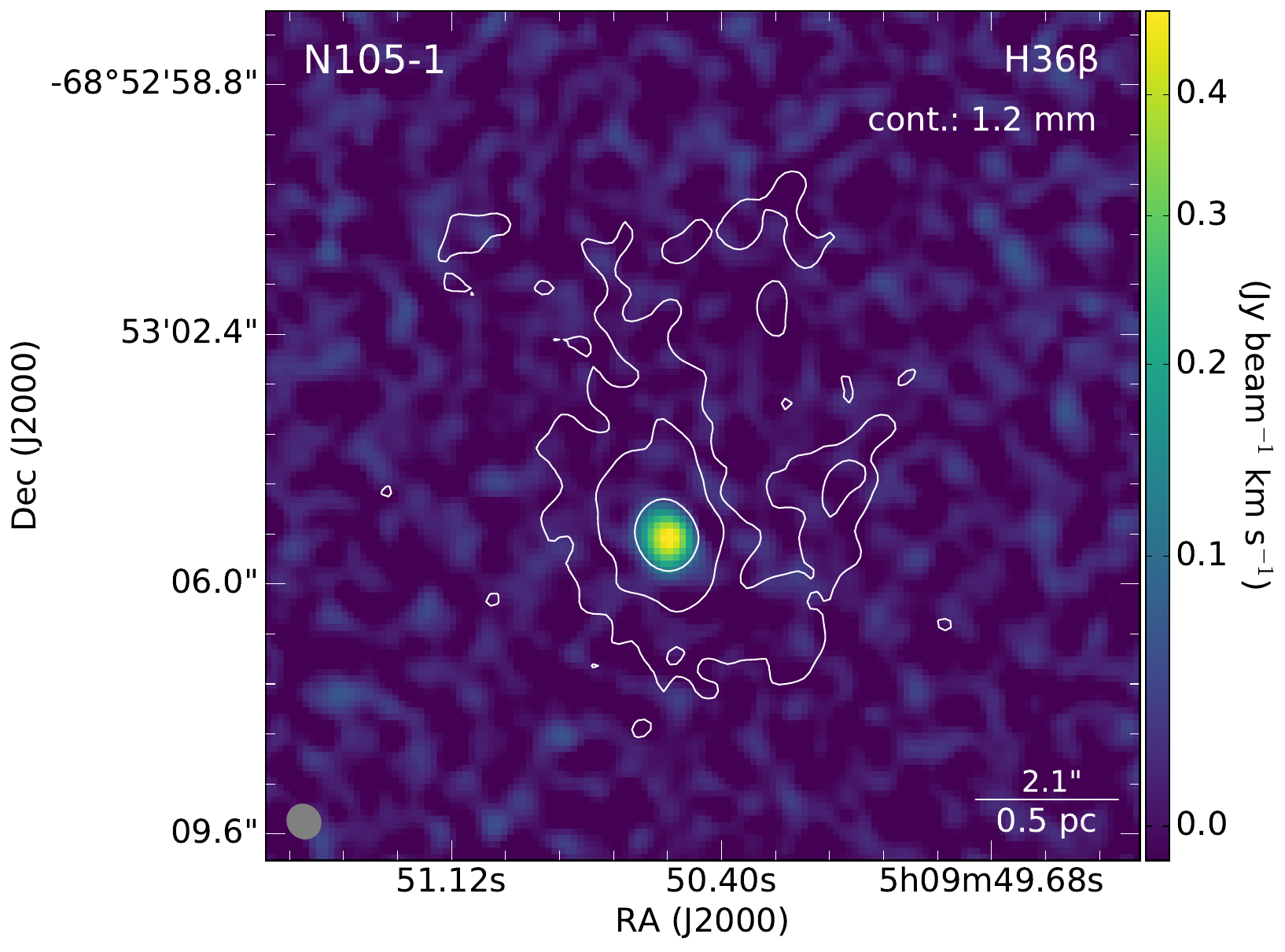}
\hfill
\includegraphics[width=0.485\textwidth]{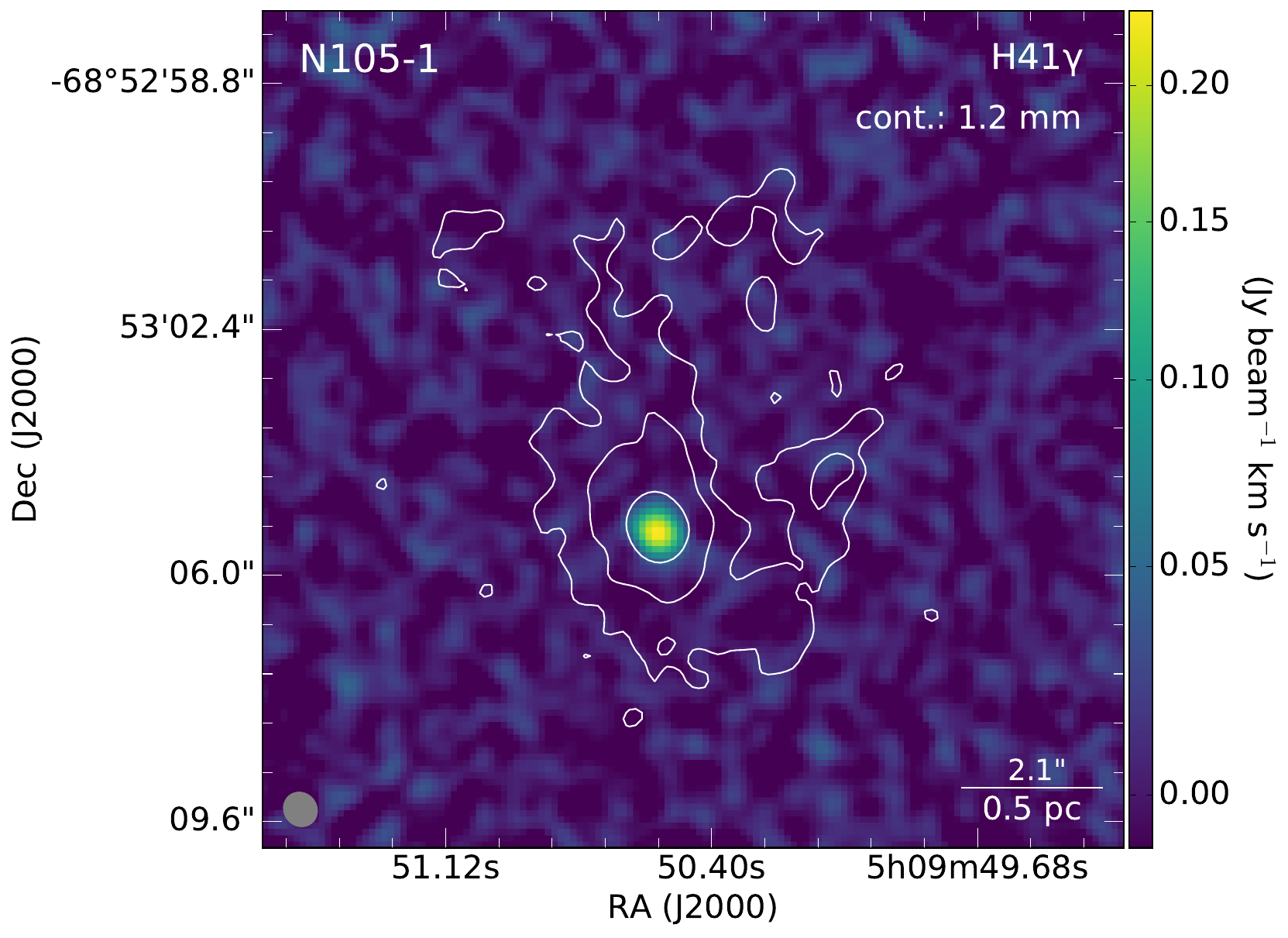}
\hfill
\includegraphics[width=0.325\textwidth]{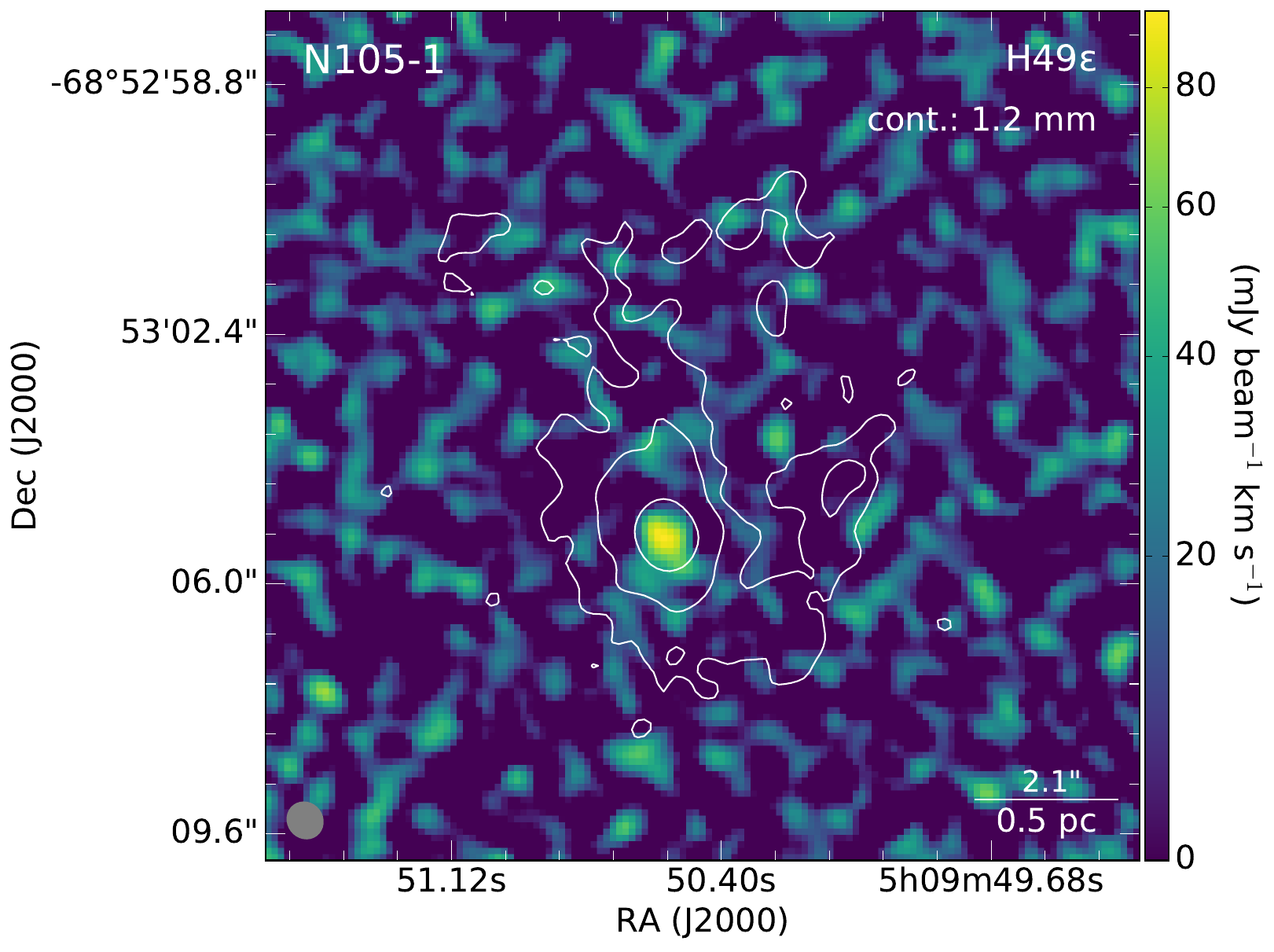}
\hfill
\includegraphics[width=0.325\textwidth]{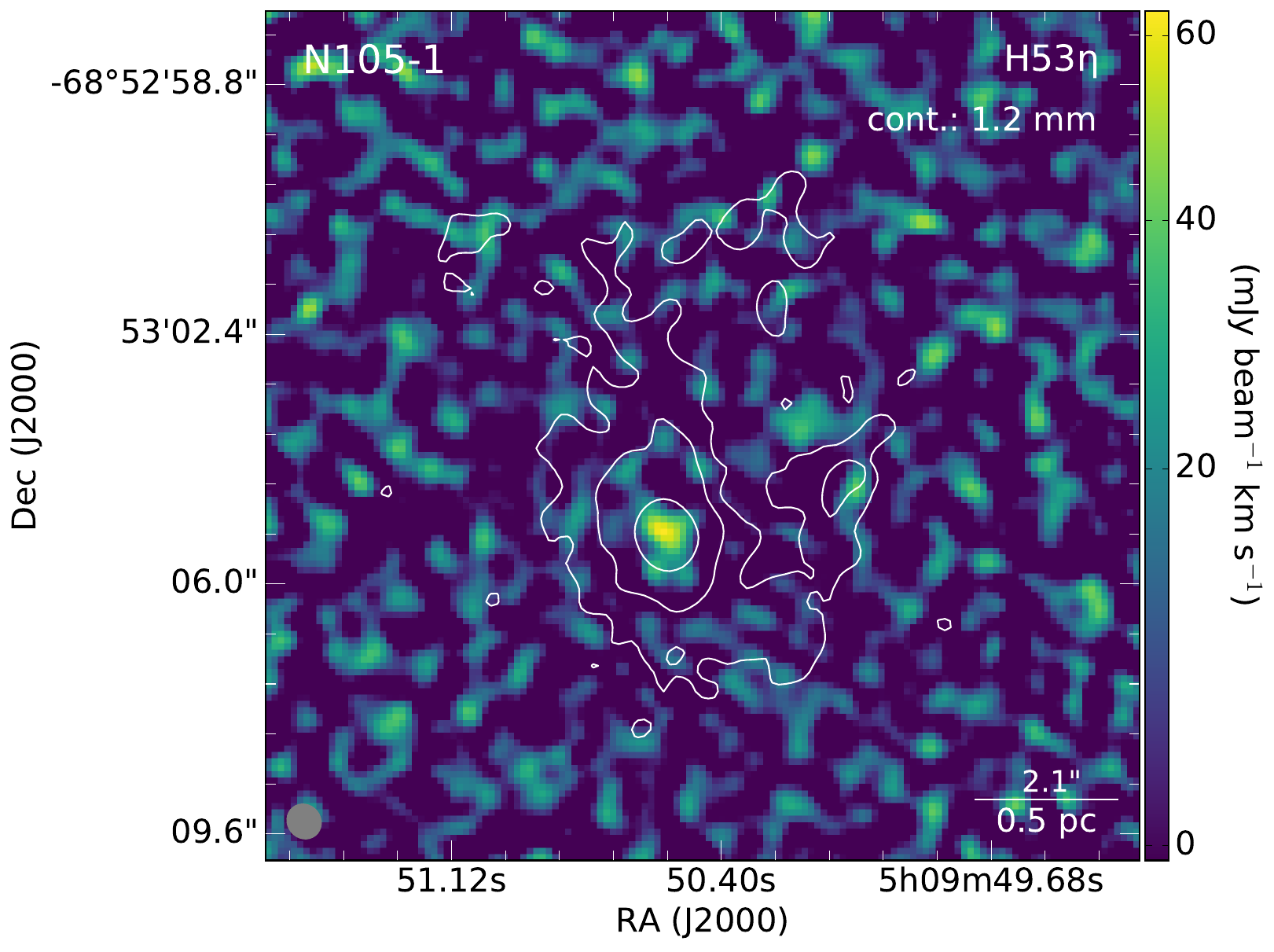}
\hfill
\includegraphics[width=0.325\textwidth]{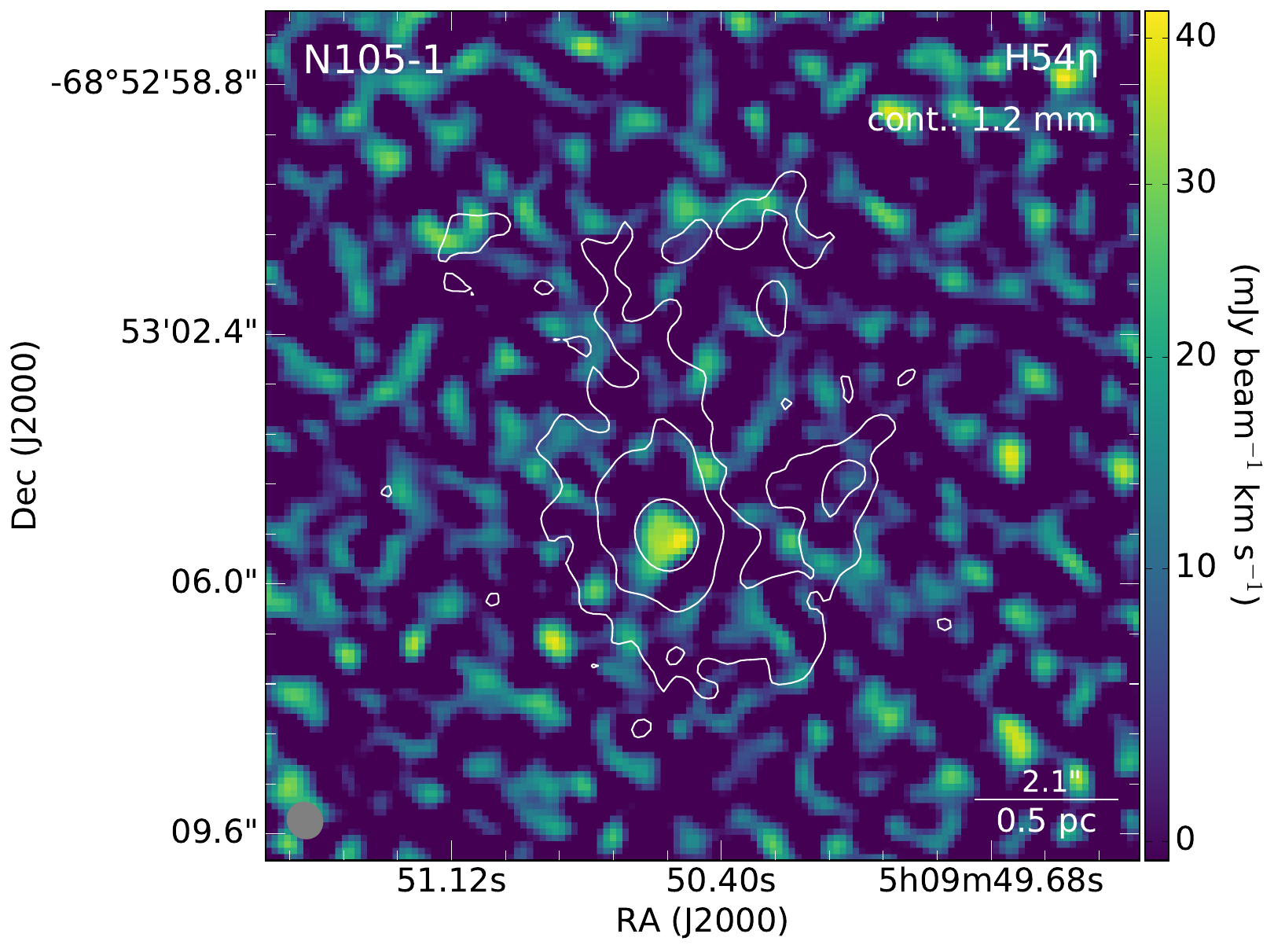}
\caption{From left to right, top to bottom:  Band 6 H36$\beta$, H41$\gamma$, H49$\epsilon$, H53$\eta$, and H53$\eta$ recombination line integrated intensity images of N\,105--1.  The white contours in each image correspond to the 1.2 mm continuum emission with contour levels of (3, 10, 100)$\sigma$.  \label{f:RRLB6}}
\end{figure*}

\begin{figure*}[ht!]
\centering
\includegraphics[width=0.325\textwidth]{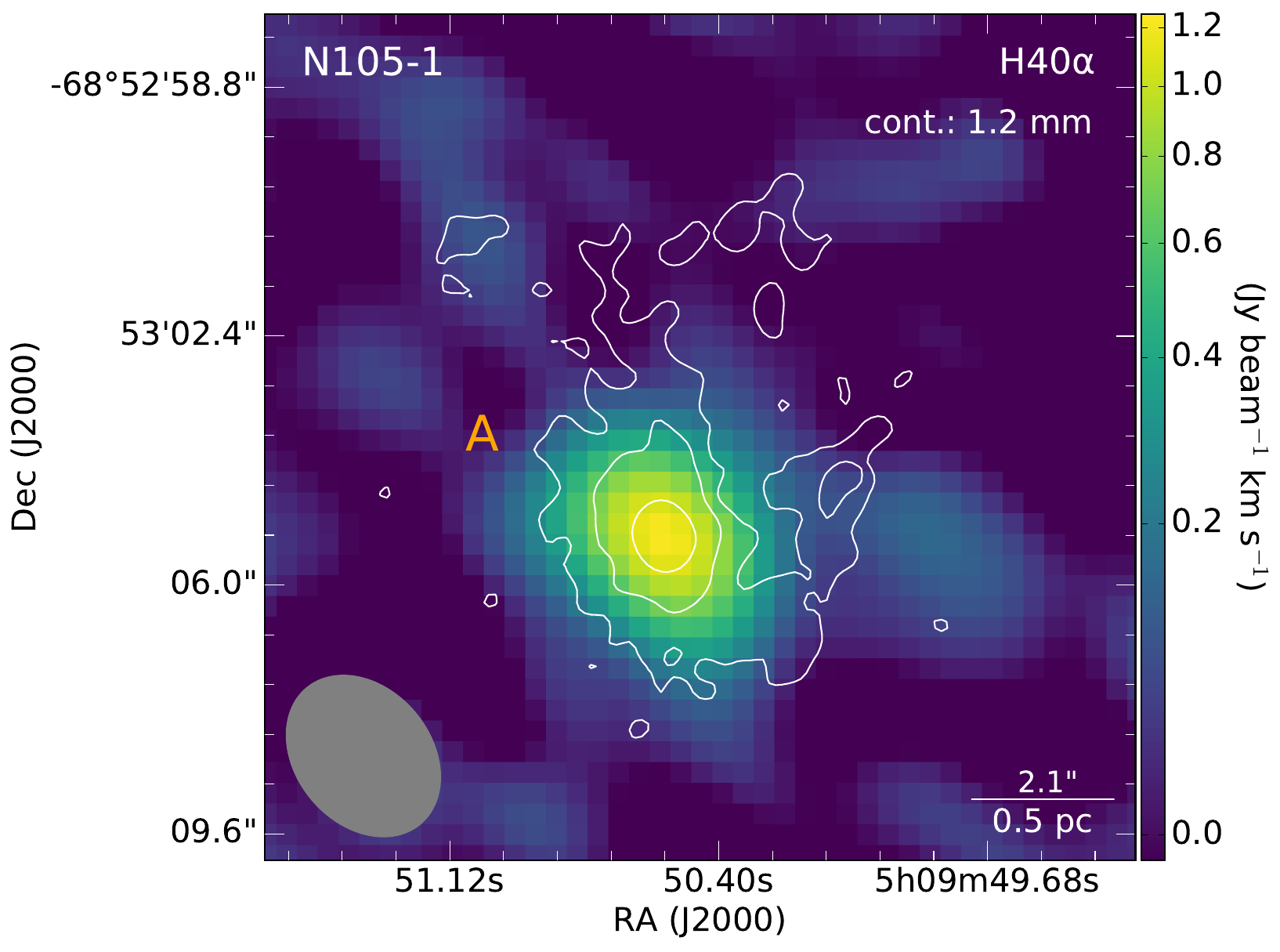}
\hfill
\includegraphics[width=0.325\textwidth]{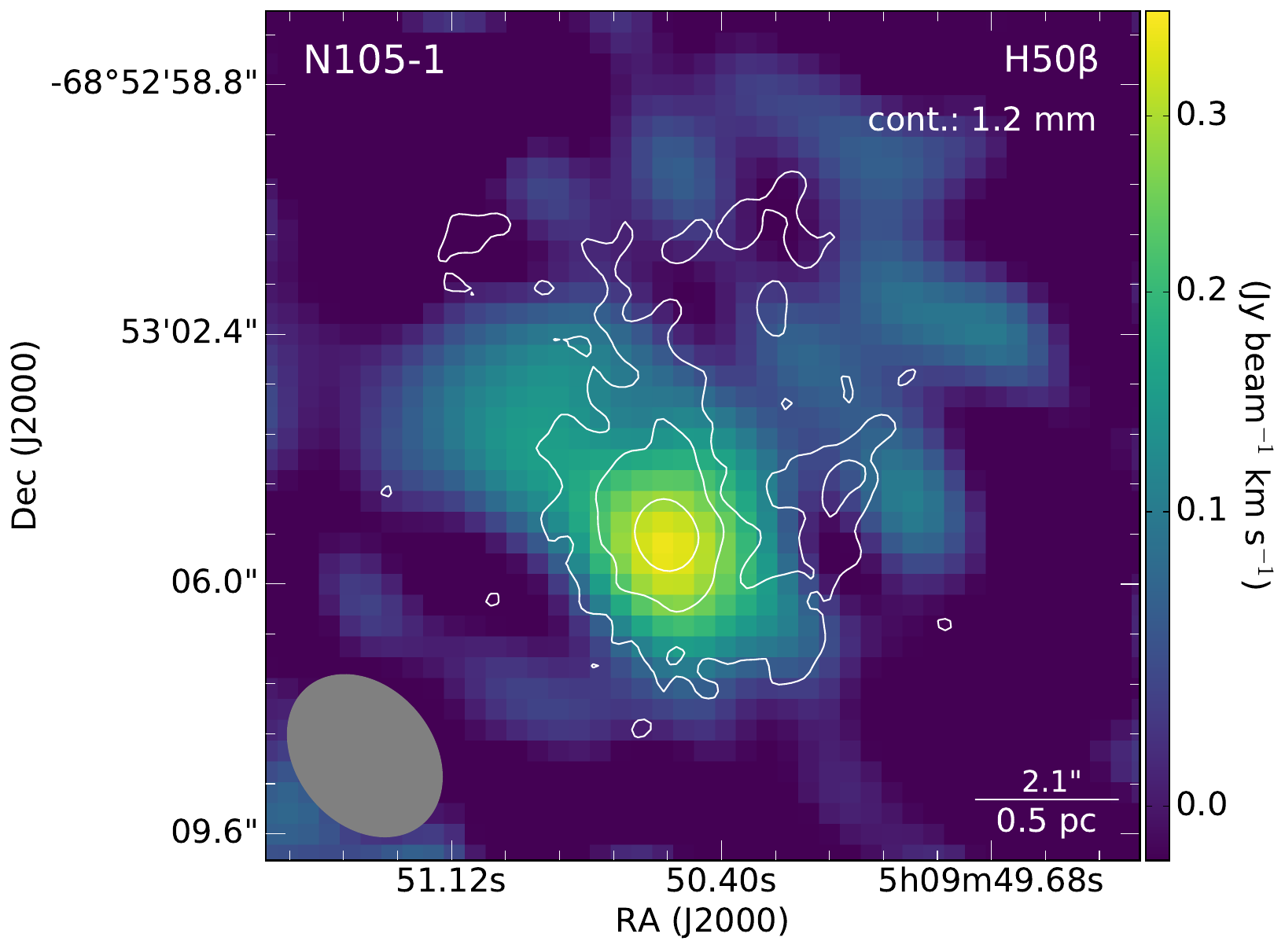}
\hfill
\includegraphics[width=0.325\textwidth]{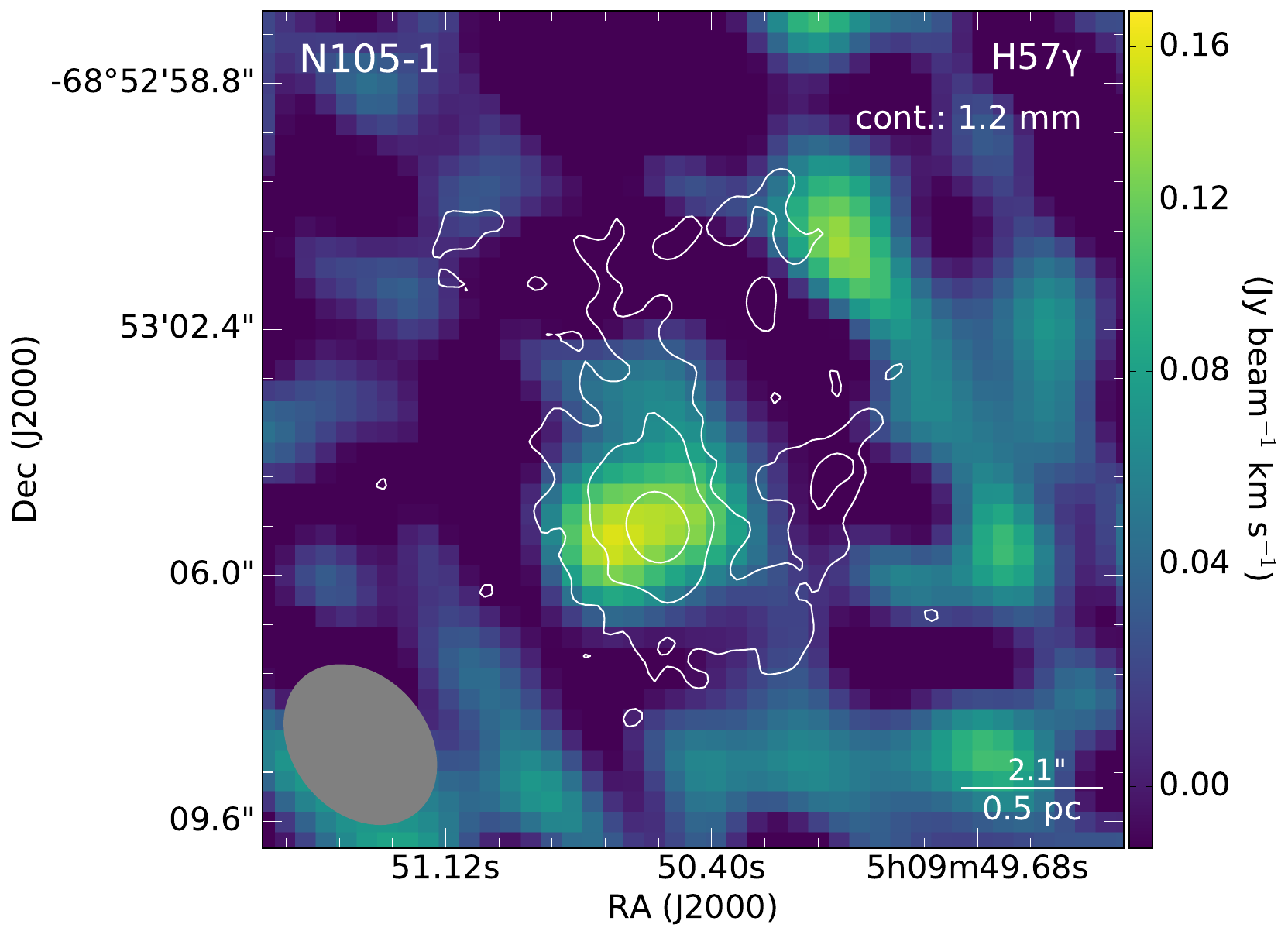}
\caption{From left to right:  Band 3 H40$\alpha$, H50$\beta$, and H57$\gamma$ recombination line integrated intensity images of N\,105--1\,A.  The white contours are the same as in Fig.~\ref{f:RRLB6}.    \label{f:RRLB3}}
\end{figure*}

\begin{figure*}[ht]
\centering
\includegraphics[width=0.48\textwidth]{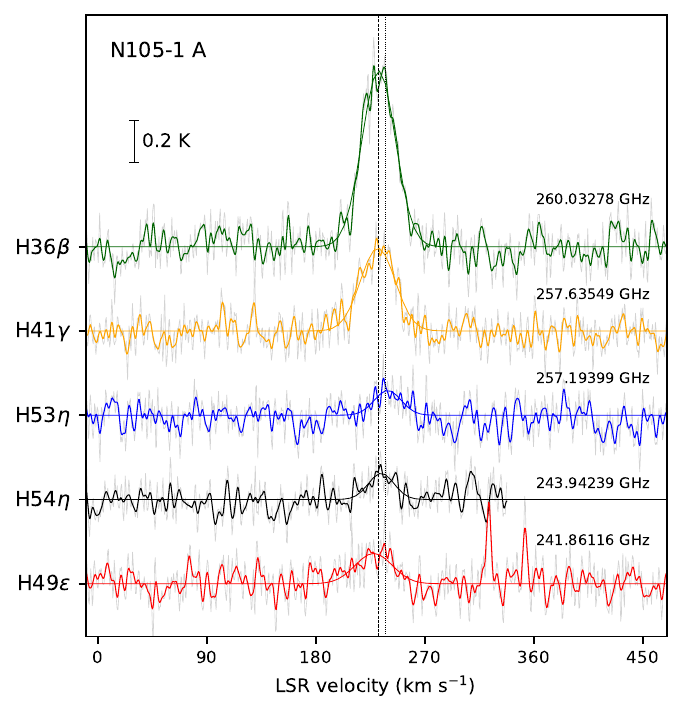}
\includegraphics[width=0.48\textwidth]{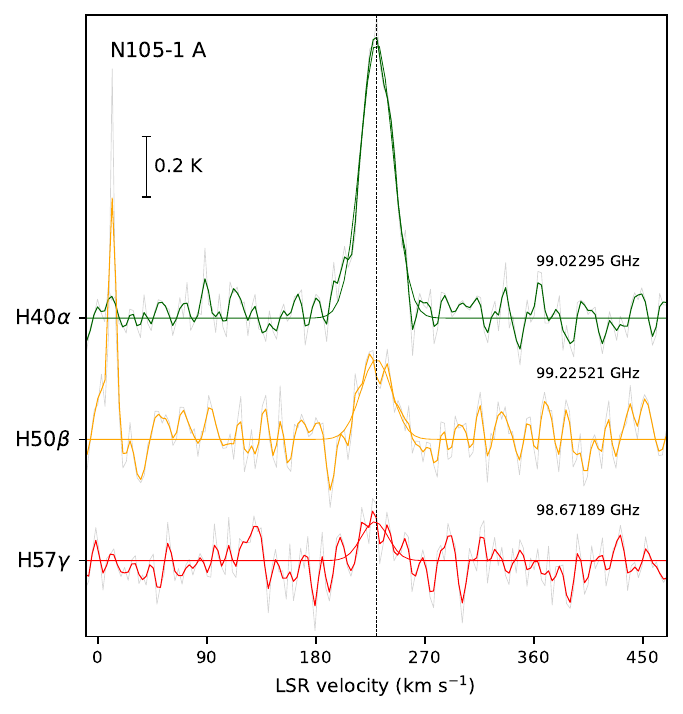} 
\caption{Left: Hydrogen recombination lines observed toward N\,105--1\,A in Band 6 (from top to bottom in order of decreasing frequency):  H36$\beta$, H41$\gamma$, H53$\eta$, H54$\eta$, and H49$\epsilon$.  Right: Hydrogen recombination lines observed toward N\,105--1\,A in Band 3 (from top to bottom):  H40$\alpha$, H50$\beta$, and H57$\gamma$.  The observed spectra are shown in light gray; the spectra shown in color were smoothed using the Hanning window with the window size of 11 (Band 6) and 5 (Band 3; using PyAstronomy.pyasl.smooth).  The dashed and dotted lines in the left panel indicate the central velocity of the H36$\beta$ (231.9~km~s$^{-1}$) and H$^{13}$CO$^{+}$ (237.2~km~s$^{-1}$) lines, respectively, while the dashed  line in the right panel indicates the central velocity of the H40$\alpha$  line (230.1~km~s$^{-1}$). The fitted Gauss functions are overlaid. The results of the fitting are reported in Table~\ref{t:fitRLs}.\label{f:RLs}}
\end{figure*}

We verified the identification of the faintest reliably-detected mm-RLs, H53$\eta$ and H54$\eta$, by comparing their observed peak intensities relative to the peak intensity of the H36$\beta$ line (the brightest RL in our Band 6 observations), to those predicted by theory, assuming local thermodynamic equilibrium (LTE) and optically thin conditions (the assumptions supported by our subsequent analysis; see below and Sections~\ref{s:xclass}--\ref{s:phys}). We used theoretical intensities from Table~B2 in \citet{gordon2002}, which reproduces entries in Table~1 of \citet{towle1996}.  The relative mm-RL intensities are plotted in Fig.~\ref{f:intensities}.  There is a good agreement for H41$\gamma$ and H49$\epsilon$ lines.  For the H53$\eta$ and H54$\eta$ lines, the observed relative intensities deviate from those predicted by the theory, but this deviation is small enough to be explained by the uncertainties in their analysis ($\sim$6$\sigma$-detections). The spatial distribution of the H53$\eta$ and H54$\eta$ line emission is in good agreement with other mm-RLs (see Figs.~\ref{f:RRLB6} and \ref{f:RRLB3}) and the lines are too broad for molecular lines (Fig.~\ref{f:RLs}).   

To our knowledge, only the $\alpha$- and $\beta$-transitions have been observed outside the Galaxy to date, making the $\gamma$-, $\epsilon$-, and $\eta$-transitions observed toward N\,105--1\,A the first extragalactic detections. The $\eta$-transitions are the first to be detected at millimeter wavelengths.  

The H36$\beta$, H41$\gamma$, H49$\epsilon$, H53$\eta$, and H54$\eta$ integrated intensity images are shown in Fig.~\ref{f:RRLB6} ($\sim$0$\rlap.{''}$5 resolution), and the H40$\alpha$, H50$\beta$, and H57$\gamma$ integrated intensity images are shown in Fig.~\ref{f:RRLB3} ($\sim$2$\rlap.{''}$2). The mm-RL emission peaks for all observed transitions except H57$\gamma$ (detected in the 3 mm band) are coincident with the millimeter and radio continuum peaks, but all are offset from the CS, SO, and SO$_2$ molecular line emission peaks (by $\sim$1--1.4 times the beam size at 1.2 mm; see Fig.~\ref{f:region} and Figs.~\ref{f:RRLB6}--\ref{f:RRLB3}).  The  H57$\gamma$ emission peak appears to be offset to the east from the continuum peak by $\sim$$0\rlap.{''}6$; since this offset corresponds to a relatively small fraction of the 3 mm beam ($\sim$30\%) and the signal-to-noise ratio for this line is low  (lower than for any other mm-RL detected toward 1\,A), we do not attempt to assign any physical meaning to it. 

We fitted Gauss functions to all mm-RLs using the CASSIS interactive spectrum analyzer \citep{vastel2015}. Table~\ref{t:fitRLs} shows the Gaussian fitting results:  the peak velocity ($v_{LSR}$),  linewidth ($\Delta v_{\rm FWHM}$, the full-width at half maximum), peak intensity ($T_b$), and the integrated intensity ($\int T_b  dv$) for each transition.  No correction for beam dilution was applied. The observed mm-RLs  with the fitted Gauss profiles are shown in Fig.~\ref{f:RLs}.

A comparison of the observed intensity ratios of H$n\alpha$ and higher-order lines observed at approximately the same frequency to theoretical predictions provides an empirical test for departures from LTE.  For a pair of lines with similar frequencies the size of the beam will be similar and thus  complications due to the inhomogeneity in the nebula will be significantly reduced or eliminated.  The H40$\alpha$ and H50$\beta$ lines detected toward 1\,A are an ideal line pair to investigate the conditions under which the observed recombination lines are formed.   The theoretical value of the H50$\beta$/H40$\alpha$ line ratio is 0.275 (\citealt{menzel1969}; \citealt{dupree1970}) under the assumption that the impact broadening can be neglected and both lines are optically thin and formed in LTE.  The observed H50$\beta$/H40$\alpha$ line ratio toward 1\,A of 0.29 $\pm$ 0.06 is in good agreement with the theoretical value, indicating that the region is likely to be in LTE.

\begin{deluxetable*}{cccccc}
\centering
\tablecaption{Gauss Fitting Results for Hydrogen Recombination Lines  Detected Toward N\,105--1\,A\tablenotemark{\footnotesize a,b} \label{t:fitRLs}}
\tablewidth{0pt}
\tablehead{
\multicolumn{1}{c}{Transition} &
\colhead{Frequency} &
\colhead{$v_{LSR}$} & 
\colhead{$\Delta v$\tablenotemark{\footnotesize c}} & 
\colhead{$T_b$} & 
\colhead{$\int T_b  dv$} \\
\colhead{} &
\colhead{(GHz)} &
\colhead{(km s$^{-1}$)} &
\colhead{(km s$^{-1}$)} & 
\colhead{(K)} &
\colhead{(K km s$^{-1}$)} 
}
\startdata
\multicolumn{6}{c}{Band 6} \\
\hline
H36$\beta$ & 260.03278 & 231.9(0.6)  & 32.3(1.3)  & 0.84(0.03) & 29.0(1.5)\\
H41$\gamma$ & 257.63549 & 231.1(0.8) & 33.5(1.8)  & 0.39(0.02) & 14.2(1.0)\\
H49$\epsilon$ & 241.86116 & 227.7(2.7) & 35.7(6.4) & 0.14(0.02) & 5.5(1.3)\\
H53$\eta$ & 257.19399 & 239.3(1.9) & 25.7(4.5) & 0.12(0.02) & 3.3(0.7)\\
H54$\eta$ & 243.94239 & 233.9(1.4) & 24.4(3.4) & 0.12(0.02)&  3.3(0.6) \\
\hline
\multicolumn{6}{c}{Band 3} \\
\hline
H40$\alpha$ &  99.02295 & 230.1(0.6) & 31.1(1.5) & 0.91(0.04) & 30.6(1.9)  \\
H50$\beta$   &  99.22521 & 229.9(2.9) & 29.3(6.9) & 0.26(0.05) & 8.4(2.6)  \\
H57$\gamma$   &  98.67189 & 228.6(5.7) & 25.5(13.5) & 0.13(0.06) &  3.6(2.4)\\
\enddata
\tablenotetext{a}{Both the Band 3 and Band 6 mm-RL spectra for 1\,A were extracted for the analysis as the mean within the contour corresponding to 50\% of the 1.2 mm continuum emission peak, using data cubes with the native angular resolution. No correction for beam dilution was applied to $T_{\rm b}$ and $\int T_b  dv$.}
\tablenotetext{b}{Band 6: The fitting was performed on the spectra after applying three iterations of Hanning smoothing, resulting in a channel width of $\Delta v_{\rm instr}$ = (4.51, 4.54, 4.83, 4.54, 4.78) km~s$^{-1}$ for (H36$\beta$, H41$\gamma$, H49$\epsilon$, H53$\eta$, H54$\eta$) lines.  Band 3:  The fitting was performed on the spectra after applying one iteration of Hanning smoothing, resulting in a channel width of $\Delta v_{\rm instr}$ = 5.91 km~s$^{-1}$ for the H40$\alpha$, H50$\beta$, H57$\gamma$ lines.}
\tablenotetext{c}{The full width at half-maximum (FWHM) corrected for instrumental broadening: $\Delta v = \sqrt{\Delta v_{\rm obs}^2 - \Delta v_{\rm instr}^2}$, where $\Delta v_{\rm obs} $ is the observed linewidth and  $\Delta v_{\rm instr}$ is the channel width after Hanning smoothing. }
\end{deluxetable*}

\subsection{\textsc{XCLASS} Modeling of Recombination Lines}
\label{s:xclass}

We model the observed mm-RLs using the eXtended CASA Line Analysis Software Suite (XCLASS\footnote{\url{https://xclass.astro.uni-koeln.de/}}; \citealt{moller2017}) with additional extensions (T. M\"{o}ller, priv. comm.). In XCLASS, a contribution of each mm-RL to a model spectrum is described by a certain number of components, with each component described by the source size ($\theta_{\rm source}$), electron temperature ($T_{\rm e}$),  emission measure (EM),  line width ($\Delta v$),  velocity offset ($v_{\rm offset}$), and the position along the line of sight.  The physics behind the model is summarized in Appendix~\ref{app:xclass}.

All the model parameters are fitted to the observational data using the Levenberg-Marquardt algorithm \citep{marquardt1963algorithm} provided by the optimization package \textsc{MAGIX}\footnote{\url{https://magix.astro.uni-koeln.de/}} (\citealt{moller2013}). The errors of the mm-RL parameters are estimated using the \texttt{emcee}\footnote{\url{https://emcee.readthedocs.io/en/stable/}} package \citep{foreman2013}, which implements the affine-invariant ensemble sampler of \citet{goodman2010} to perform a Markov chain Monte Carlo (MCMC) algorithm, approximating the posterior distribution of the model parameters by random sampling in a probabilistic space. Details of the error estimation are described in \citet{moller2021}. To get a more reliable error estimation, the errors for the emission measures and the electron densities are calculated on a logarithmic scale, i.e., these parameters are converted to their log10 values before applying the MCMC algorithm and converted back to linear scale after finishing the error estimation procedure.

In our XCLASS model, we include all mm-RL transitions up to $\theta$-transitions ($\Delta n=8$) covered by our ALMA observations in both Band 3 and Band 6: H40$\alpha$, H36$\beta$, H50$\beta$, H41$\gamma$, H57$\gamma$, H49$\epsilon$, H67$\epsilon$, H53$\eta$, H54$\eta$, H74$\eta$, H55$\theta$, and H77$\theta$; spectra of N\,105--1\,A centered on the rest frequencies of these mm-RLs are shown in Fig.~\ref{f:xclassLTE}.  We also include several molecular species detected toward N\,105--1\,A (CH$_3$OH, H$^{13}$CO$^{+}$, H$^{13}$CN, CS, H$_2$CS, SO$_2$, and SO) to identify any contributions from the molecular lines to the mm-RL profiles; the LTE-parameters for these molecules are taken from \citet{sewilo2022n105}. Recombination lines of atoms other than H (such as He, C, N, O, and S) are not reliably detected toward N\,105--1\,A.  

We perform the XCLASS analysis for three sets of lines: all 12 mm-RL transitions from Band 3 and Band 6, Band 3 transitions only, and Band 6 transitions only.  For each set, we examine four different scenarios. In the first scenario, we assume LTE conditions and describe all mm-RLs by a single component, i.e., we use a single set of $T_{\rm e}$ and EM to model all mm-RL transitions. In the second scenario, we again assume that all mm-RLs are in LTE, but now we use an additional (second) component. In the remaining two scenarios, we model all mm-RL transitions in non-LTE, using one or two components. 

We are unable to obtain a good model fit (i.e., reliably determine $T_{\rm e}$ and EM) for any combination of the line sets and scenarios. For line sets with single band mm-RL transitions, this is likely because many lines have a very low signal-to-noise ratio. For the line set combining all Band 3 and Band 6 mm-RL transitions, the difference in the angular resolution between the Band 3 and Band 6 observations (and the resulting beam dilution in Band 3) prevents us from obtaining a satisfactory fit to the data.  We did not find any significant differences between the LTE and non-LTE model fits, indicating that LTE is a reasonable assumption. 

In Fig.~\ref{f:xclassLTE} in Appendix~\ref{app:xclass}, we present an example XCLASS synthetic spectrum fitted to Band 6 mm-RL transitions, representing the LTE model with one component. The corner plot of the MCMC error estimate for the same model is shown in Fig.~\ref{f:xclasscornerLTE}; it indicates that it is not possible to constrain $T_{\rm e}$ and EM based on this model fit.  Figure~\ref{f:xclassLTE} still provides us with some useful information. Firstly, it demonstrates that we detected (or marginally detected in the case of the H55$\theta$ line) all hydrogen mm-RLs with $\Delta n\leq8$ covered by our Band 6 observations. Secondly, Fig.~\ref{f:xclassLTE} shows that the observed mm-RL intensities in Band 6 are consistent with the LTE model. Lastly, extending the model to lower frequencies reveals that the intensities of mm-RLs detected in Band 3 are significantly lower than predicted by the model, indicating that beam dilution effects are considerable in the Band 3 observations having about five times larger beam than in Band 6.   

\subsection{Physical Properties of the Ionized Gas in N\,105--1\,A Based on the H40$\alpha$ and 3 mm Continuum Data}
\label{s:phys}

We determine the physical properties of the ionized gas in N\,105--1\,A using the standard analytical methods incorporating a continuum free-free and/or an H$n\alpha$ recombination line emission.  The H40$\alpha$ line is the only $\alpha$-transition currently available for 1\,A.  

\subsubsection{The Electron Temperature} 
\label{s:temp}

Assuming the recombination lines are emitted under pure LTE conditions and pressure broadening of the lines by electron impacts is negligible, the observed recombination line to continuum ratio can be utilized to directly estimate the LTE electron temperature ($T_{\rm e}^{*}$) of the ionized gas (e.g., \citealt{dupree1970} and references therein; \citealt{roelfsema1992}; \citealt{gordon2002}) from the following formula for H$n\alpha$ lines (\citealt{wilson2009}):

\begin{eqnarray}
\label{e:Te}
\frac{I_{\rm L}}{I_{\rm C}} \left(\frac{\Delta v}{{\rm km\,s^{-1}}}\right) \approx  \frac{6985}{a(T_{\rm e}, \nu)} \left(\frac{\nu_{\rm L}}{{\rm GHz}}\right)^{1.1}  \\ 
\cdot \left(\frac{T_{\rm e}^{*}}{{\rm K}}\right)^{-1.15} \left(\frac{1}{1 + y^{+}}\right)   \nonumber 
\end{eqnarray}
where 
$I_{\rm L}$ is the line intensity, 
$I_{\rm C}$ is the free-free continuum intensity,
$\Delta v$ is the FWHM line width, 
$\nu_{\rm L}$ is the H$n\alpha$ line rest frequency, 
$y^{\rm +} $ is a singly ionized helium-to-hydrogen abundance ratio ($N_{\rm He^{+}}/N_{\rm H^{+}}$), $a(T_e, \nu)$ is a dimensionless factor of the order of 1 slowly varying with $T_e$ and $\nu$, tabulated in \citet{mezger1967}. At the frequency of $\sim$100 GHz, $a$$\sim$0.9 for $T_e$ from 8,000 to 12,000 K, typical for H\,{\sc ii} regions. $I_{\rm L}/I_{\rm C}$ ratio allows $I_{\rm L}$ and $I_{\rm C}$ to be given in any convenient units. 

We utilize the H40$\alpha$ line ($\nu_{\rm L}$ = 99.02295 GHz) to estimate $T_{\rm e}^{*}$ in N\,105--1\,A. The $y^{\rm +}$ parameter cannot be determined for 1\,A in this work since no He recombination line has been detected toward this source in our observations.  A typical value of $y^{\rm +}$ in the Galaxy is $\sim$8--10\%, although significant variations have been found (e.g., \citealt{churchwell1974}; \citealt{lockman1975}; \citealt{balser2001}). Previous studies suggest that $y^{\rm +}$ in the LMC is similar to that measured in the Galaxy (e.g., \citealt{dufour1975}; \citealt{rosa1987}; \citealt{peck1997}; \citealt{tsamis2003}). \citet{dufour1975} determined $y^{\rm +}$ of 0.1 specifically for N\,105 using optical spectroscopy. 

We obtain $I_{\rm L} \,  \Delta v$ from the H40$\alpha$ integrated intensity, $\int I_{\rm L} dv$; for Gaussian line profiles, $\int I_{\rm L} dv \approx 1.064 \, I_{\rm L} \,  \Delta v$ (e.g., \citealt{brown1978}). We determine $\int I_{\rm L} dv$ of $1110\pm10$ mJy km s$^{-1}$ from the H40$\alpha$ integrated intensity (moment 0) map by measuring the emission within the 3$\sigma$ contour. We  measured the 99 GHz continuum flux density of $43.2\pm0.4$ mJy within the same area from the 99 GHz continuum map. The 3 mm emission is expected to be mostly free-free, but some contribution from the dust thermal emission is likely. We utilized the archival 870 $\mu$m continuum image to estimate the dust emission at 3 mm under the assumption that all the emission at 870 $\mu$m originates from dust. We measured the 870 $\mu$m continuum flux density (Band 7, B7) on the image smoothed to the beam of the 3 mm image and scaled it to the expected 3 mm continuum dust-only flux density (B3) using the formula:  $S_{\rm B3} = S_{\rm B7}\,(\nu_{\rm B3}/\nu_{\rm B7})^{2+\beta}$, where $\beta$ is a dust emissivity index.  We adopt  $\beta$ of 1.7 for N\,105 (\citealt{gordon2014}; see also \citealt{sewilo2022n105}). The comparison of the predicted dust-only and total flux density measured from the same region at 3 mm indicates that only 3.6\% of the 3 mm emission is the dust thermal emission.  To calculate $T_{\rm e}^{*}$, we removed the contribution from the dust emission from the 3 mm (99 GHz) flux density.

From  Eq. 1, we determine $T_{\rm e}^{*}$ of $10\,940\pm1\,350$ K toward 1\,A which is consistent with $T_{\rm e}$ measurements toward Galactic H\,{\sc ii} regions (see Section~\ref{s:Te}); the uncertainties were calculated by error propagation after incorporating the 10\% flux calibration error into the flux density uncertainties. Since the observed H50$\beta$/H40$\alpha$ line ratio is close to its theoretical LTE value in 1\,A,  $T_{\rm e}^{*}$ is a good representation of the true electron temperature ($T_{\rm e}$).

\subsubsection{Emission Measure and Electron Density}
\label{s:EM}

We calculate the emission measure (EM) toward N\,105--1\,A using the following formula (e.g., \citealt{wilson2009}): 
\begin{eqnarray} \label{e:EML}
{\rm EM}
\,({\rm pc\, cm^{-6}}) = 1.74\times10^{-3} \left(\frac{T_{\rm e}}{{\rm K}}\right)^{3/2}  \left(\frac{\nu_{\rm L}}{{\rm GHz}}\right) \\
\nonumber 
\cdot \, \eta^{-1} \left(\frac{T_{\rm L} }{{\rm K}}\right)  \left(\frac{\Delta v}{{\rm km}\, {\rm s}^{-1}}\right),
\end{eqnarray}
where $T_{\rm L}$ is the H$n\alpha$ line intensity in K, $\eta$ is the beam filling factor, and other parameters are the same as in Eq.~\ref{e:Te}.  We use $T_{\rm e}=10\,940$ K (Section~\ref{s:temp}). We calculate $T_{\rm L} \, \Delta v$ from the integrated intensity of the H40$\alpha$ line ($\int I_{\rm L} dv$ converted to K km s$^{-1}$) as described above. The beam filling factor $\eta = \theta_s^2 / (\theta_s^2 + \theta_{\rm beam}^2)$, where $\theta_s^2$ and $\theta_{\rm beam}^2$ are the FWHM size of the source and the synthesized beam, respectively.  For 1\,A, we obtain $\theta_s$ of $0\rlap.{''}55\pm0\rlap.{''}10$  based on  the 2D Gauss fitting to the 3 mm image with the CASA task \texttt{imfit}; it is the geometric mean of the major and minor source FWHMs deconvolved from the beam.  We obtain EM of $(8.9\pm1.7)\times10^7$ pc cm$^{-6}$ toward N\,105--1\,A. 

The emission measure can also be estimated based on the free-free continuum emission (EM$_{\rm cont}$); it is expressed by the  formula (e.g., \citealt{mezger1967}; \citealt{roelfsema1992}): 
\begin{eqnarray} \label{e:EMC}
{\rm EM_{cont}}\,({\rm pc\, cm^{-6}}) = \frac{\tau_{\rm C}}{8.235\times10^{-2} \, a(T_{\rm e}, \nu) \, T_{\rm e}^{-1.35} \, \nu^{-2.1}},
\end{eqnarray}
where $\tau_{\rm C}$ is the continuum optical depth,  $\nu$ is the continuum frequency in GHz, and $a(T_{\rm e}, \nu)$ is the same as in Eq.~\ref{e:Te}. We calculate the peak $\tau_{\rm C}$ and EM$_{\rm cont}$ based on the 3 mm continuum data. 

For the 99.023 GHz / 3 mm free-free continuum peak intensity of $43.64\pm2.10$ mJy per $2\rlap.{''}49\times1\rlap.{''}83$ beam (96.4\% of the observed continuum peak of 45.27 mJy beam$^{-1}$), we obtain the brightness temperature ($T_{\rm b}$) of $1.19\pm0.06$ K, assuming the Rayleigh-Jeans approximation. The uncertainties include the continuum image rms and a 10\% flux density calibration error. We estimate the peak continuum optical depth ($\tau_{\rm C}$) from the formula $T_{\rm b}=T_{\rm e} (1-e^{-\tau_{\rm C}})$ after correcting $T_{\rm b}$ for the beam dilution effect: $\tau_{\rm C} = -{\rm ln}(1-\eta^{-1}\,T_{\rm b}/T_{\rm e}) = (1.7\pm0.2)\times10^{-3}$.  From Eq.~\ref{e:EMC}, we obtain the peak EM$_{\rm cont}$ of $(1.0\pm0.2)\times10^8$ pc cm$^{-6}$ toward N\,105--1\,A, in a very good agreement with the EM derived from the mm-RL data.

To estimate the rms electron density, $n_{\rm e}$, we use the expression $n_{\rm e} = \sqrt{{\rm EM}/\Delta s}$, where $\Delta s$ (in pc) is the size of the source along the line of sight.  Assuming that the extent of the source along the line of sight is the same as in the plane of the sky, we adopt the source size obtained from the 2D Gauss fitting ($\theta_s$) as $\Delta s$ ($0.13\pm0.02$ pc). The resulting $n_{\rm e}$ is $(2.6\pm0.3)\times10^4$ cm$^{-3}$ for N\,105--1\,A.

The values of (FWHM size, EM, $n_{\rm e}$) of (0.13 pc, $\sim$$9\times10^7$ pc cm$^{-6}$, $\sim$$3\times10^4$ cm$^{-3}$) derived for N\,105--1\,A indicate that this LMC source is likely an UC H\,{\sc ii} region.  UC H\,{\sc ii} regions are one of the earliest phases of massive star formation, defined observationally as regions with sizes $\lesssim$0.1 pc, $n_{\rm e}\gtrsim10^4$ cm$^{-3}$, and ${\rm EM}\gtrsim10^7$ pc cm$^{-6}$ (e.g., \citealt{wood1989}; \citealt{kurtz2000}; \citealt{hoare2007}). The UC H\,{\sc ii} region phase follows the hypercompact (HC) H\,{\sc ii} region phase (regions ten times smaller and 100 times denser than UC H\,{\sc ii}s,  characterized by positive spectral indices and often showing broad RLs; e.g., \citealt{kurtz2005}; \citealt{hoare2007}) and precedes the compact H\,{\sc ii} region phase. The formation of the H\,{\sc ii} region signals the arrival of a massive protostar onto the main sequence and thus is the earliest manifestation of an OB star. The massive stars reach the main sequence while still deeply embedded in their natal molecular cloud and with ongoing accretion (e.g., \citealt{churchwell2002}; \citealt{zinnecker2007}; \citealt{tanaka2016}). The immediate surroundings of the ionizing star(s) of HC/UC H\,{\sc ii} regions are therefore expected to be very dynamic due to possible infall, outflows, stellar winds, accretion disk rotation, turbulence, and shocks.

\subsubsection{Ionizing Flux of the Exciting Star}
\label{s:lyman}


For a spherical, homogeneous, optically thin, and dust-free model, the number of Lyman continuum photons ($N_{\rm Ly}$) required to ionize the source can be expressed by the formula (e.g., \citealt{goudis1977} and references therein): 
\begin{eqnarray} \label{e:lyman}
N_{\rm Ly} (s^{-1}) = 4.76\times10^{48} \left(\frac{S_{\rm \nu}^{\rm ff}}{{\rm Jy}}\right)   \left(\frac{D}{{\rm kpc}}\right)^2  \\  
\nonumber 
\cdot  \left(\frac{\nu}{{\rm GHz}}\right)^{0.1} \left(\frac{T_{\rm e}}{{\rm K}}\right)^{-0.45},
\end{eqnarray}
where $S_{\rm \nu}^{\rm ff}$ is the free-free continuum integrated flux density at a frequency $\nu$,  and $D$ is a distance.  For $S_{\rm \nu}^{\rm ff}$, we adopt the 3 mm integrated flux density obtained using the 2D Gauss fitting with the CASA task \verb|imfit|: $48.5\pm0.8$ mJy (in an excellent agreement with the flux density integrated over the region enclosed by the 3$\sigma$ contour, see Table~\ref{t:photometry}), after removing the contribution from the thermal dust emission (3.6\%).  We adopt $T_{\rm e}$ determined in Section~\ref{s:temp} and the LMC distance of 49.6 kpc \citep{pietrzynski2019}

For 1\,A, Eq. 2 results in $N_{\rm Ly}$ of $1.32\times10^{49}$ photons per second (log\,$N_{\rm Ly}=49.12$) which corresponds to an  O5.5\,V star (log\,$N_{\rm Ly}$=49.11) according to  models of Galactic O stars by \citet{martins2005}. The model luminosity and spectral mass for an O5.5\,V star are $2.6\times10^5$ $L_{\odot}$ and 34.4 $M_{\odot}$, respectively. 

In an earlier work by \citet{smith2002}, the authors modeled ionizing flux densities for the metallicity range from 0.05 to 2 $Z_{\odot}$. In their models, the ionizing flux of log\,$N_{\rm Ly}$=49.12 corresponds to the O6\,V spectral type and luminosity of $\sim$$3\times10^5$ $L_{\odot}$ for the metallicity within the range observed in the LMC ($Z=0.4\,Z_{\odot}$).  \citet{smith2002} predict the same ionizing fluxes for the solar metallicity for the relevant spectral type range for dwarfs.


\subsection{SED Fitting: Physical Properties of the YSO Associated with N\,105--1\,A}
\label{s:yso} 

\begin{figure}[ht!]
\centering
\includegraphics[width=0.48\textwidth]{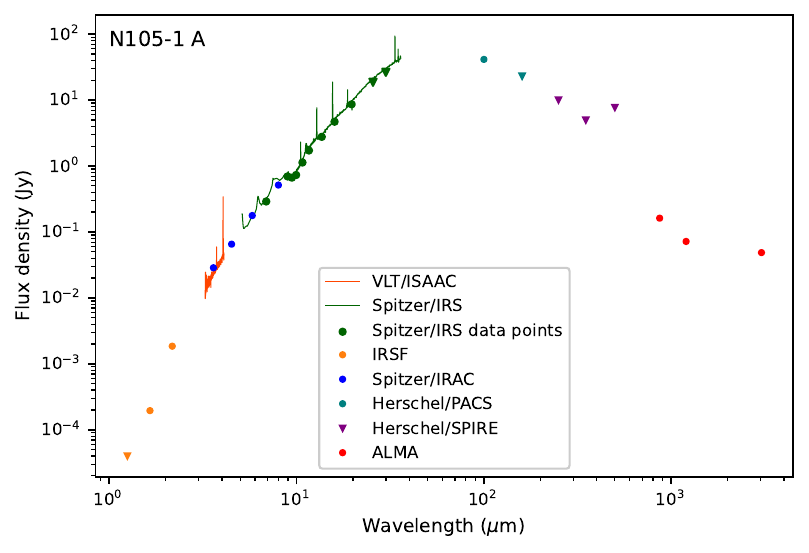} 
\caption{The spectral energy distribution (SED) for N\,105--1\,A, covering a wavelength range from 1.25 $\mu$m to 2.7 mm. Filled circles and triangles are valid flux values and flux upper limits, respectively. The flux error bars are smaller than the data points. The photometry and references are provided in Table~\ref{t:photometry}. \label{f:sed}}
\end{figure}

The IR source coinciding with N\,105--1\,A was identified as a YSO candidate in several previous studies (see Section~\ref{s:1A}). 
We fit the spectral energy distribution (SED) of YSO 050950.53$-$685305.5 with a set of radiative transfer model  SEDs for YSOs developed by \citet{robitaille2017} using the \citet{robitaille2007} SED fitting tool. The SED fitting was previously performed for the YSO associated with 1\,A \citep{carlson2012}, but without the long-wavelength photometry ($>$20 $\mu$m) that is crucial to constrain the evolutionary stage of the YSOs, and using the older version of the YSO models (\citealt{robitaille2006}).

To construct the multiwavelength SED of 1\,A, we compiled the photometric data covering a wavelength range from $\sim$1.3 $\mu$m to 3 mm. The SED is shown in Fig.~\ref{f:sed}, while flux densities with references and technical details are provided in Appendix~\ref{s:appPhot}. Figure~\ref{f:grid} shows the images of N\,105--1, from optical (H$\alpha$) to radio (6 cm) wavelengths.

We do not include the {\it Herschel} photometry except the 100 $\mu$m flux density in the fitting because at the longer {\it Herschel} wavelengths (at the lower spatial resolution), the emission from 1\,A is unresolved from the neighboring source to the east (see Fig.~\ref{f:grid}).  In addition, we set the two longest-wavelength data points out of 11 extracted from the {\it Spitzer}/IRS spectrum (see Appendix~\ref{s:appPhot}) as upper limits for the fitting due to the reduced {\it Spitzer}'s spatial resolution at these wavelengths ($>$20 $\mu$m; see also Fig.~\ref{f:grid}).  We  also use the 870 $\mu$m and 1.2 mm (dust-only) flux densities as upper limits because they include the emission from both the compact source and the extended component (they were measured within the corresponding 3$\sigma$ contour, see Table~\ref{t:photometry}).  The dust-only 3 mm flux density (3.6\% of the total 3 mm flux density; see Section~\ref{s:temp}) is used with the 20\% uncertainty to account for the flux calibration errors and the uncertainty in estimating the dust contribution to the free-free emission at this wavelength. 

\begin{figure}[t!]
\centering
\includegraphics[width=0.35\textwidth,angle=-90.]{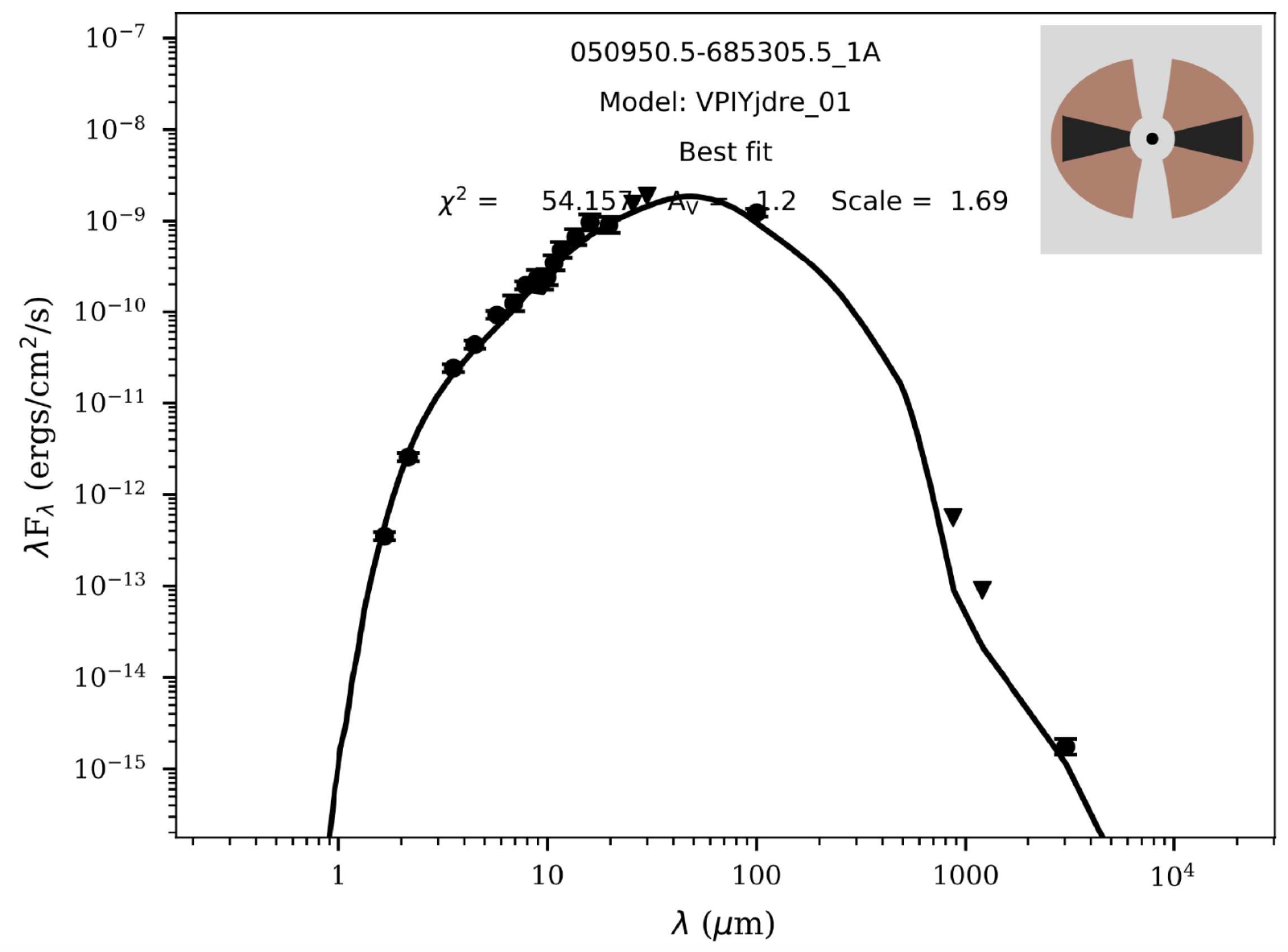} 
\caption{The SED of 1\,A showing the data points used for the fitting and with the best-fit YSO SED model from the \citet{robitaille2017} grid overplotted. Filled circles and triangles are valid flux values and flux upper limits, respectively. The values of a reduced $\chi^2$, interstellar visual extinction ($A_{\rm V}$), and a distance scale for the best-fit model are indicated in the plots. The icon in the upper right corner of the plot represents the basic layout of the best-fit YSO model \citep{robitaille2017}.  \label{f:bestfit}}
\end{figure}

The \citet{robitaille2017} YSO SED models consist of a combination of several components: a star, a disk, an infalling envelope, bipolar cavities, and an ambient medium. They were computed as 18 sets of models with increasing complexity, described by two up to 12 variables.  To find the best-fitting YSO model, \citet{robitaille2017} recommends to identify the best model set first using a Bayesian approach and adopt the model with the lowest $\chi^2$ from the most likely set (see also \citealt{sewilo2019}).  The most likely model set contains the highest fraction of  models that provide a ``good'' fit defined as models with $\chi^2-\chi^2_{\rm best} < F\,n_{\rm data}$ where $\chi^2_{\rm best}$ is $\chi^2$ of the best-fit model across all the model sets and all models, $n_{\rm data}$ is the number of data points used for the fitting, and $F$ is is a threshold parameter that is determined empirically.  In case of 1\,A, $F=6$ is a reasonable choice; however, the well-fit model statistics is very small. Ultimately, the best-fit model is the model with the lowest $\chi^2$ across all the model sets and all models. The SED with the best-fit model overlaid is shown in Fig.~\ref{f:bestfit}.  The best-fit model comes from the model set \verb|spubsmi| and includes a central source, passive disk, Ulrich (rotationally flattened) envelope, bipolar cavities, an ambient interstellar medium (see an inset cartoon in Fig.~\ref{f:bestfit}, from \citealt{robitaille2017}). The presence of both the envelope and the disk in the best-fit model indicates that 050950.53$-$685305.5 is likely a Class I YSO.

We determine the luminosity of the central source of $1.3\times10^5$ $L_{\odot}$ from the stellar radius and effective temperature returned by the fitter for the best-fit YSO model using the Stefan-Boltzmann equation.   To obtain mass, we compare the position of the best-fit model in the Hertzsprung-Russell (H-R) diagram with the PARSEC evolutionary tracks that include the pre-main sequence (PMS) stage; the evolutionary tracks were calculated for the initial stellar masses from 0.1 to 350 $M_{\odot}$, and a range of metallicities \citep{bressan2012,chen2015}.  We adopt the mass of the closest PARSEC track in the H-R diagram as the mass of YSO 050950.53$-$685305.5. The age of the PARSEC photospheric model is interpolated from the closest point on the track to the model on the H-R diagram to constrain the age of the source.  The resulting mass and age of the YSO are 28 $M_{\odot}$ and $1.1\times10^5$ years, respectively. The luminosity and mass of 1\,A  obtained in our analysis are in good agreement with the SED fitting results of   \citet[see Section~\ref{s:1A}]{carlson2012}. 

The luminosity based on the SED fitting is two times lower and the mass is $\sim$20\% lower than the values calculated from the mm-RLs and the 3 mm continuum.  Despite these lower values, the SED results are generally consistent with the mm-RL and 3 mm values.  We consider the recombination line values for luminosity and mass to be more reliable, as they are based on direct observations of the ionizing photons.

\begin{figure*}[ht!]
\centering
\includegraphics[width=0.325\textwidth]{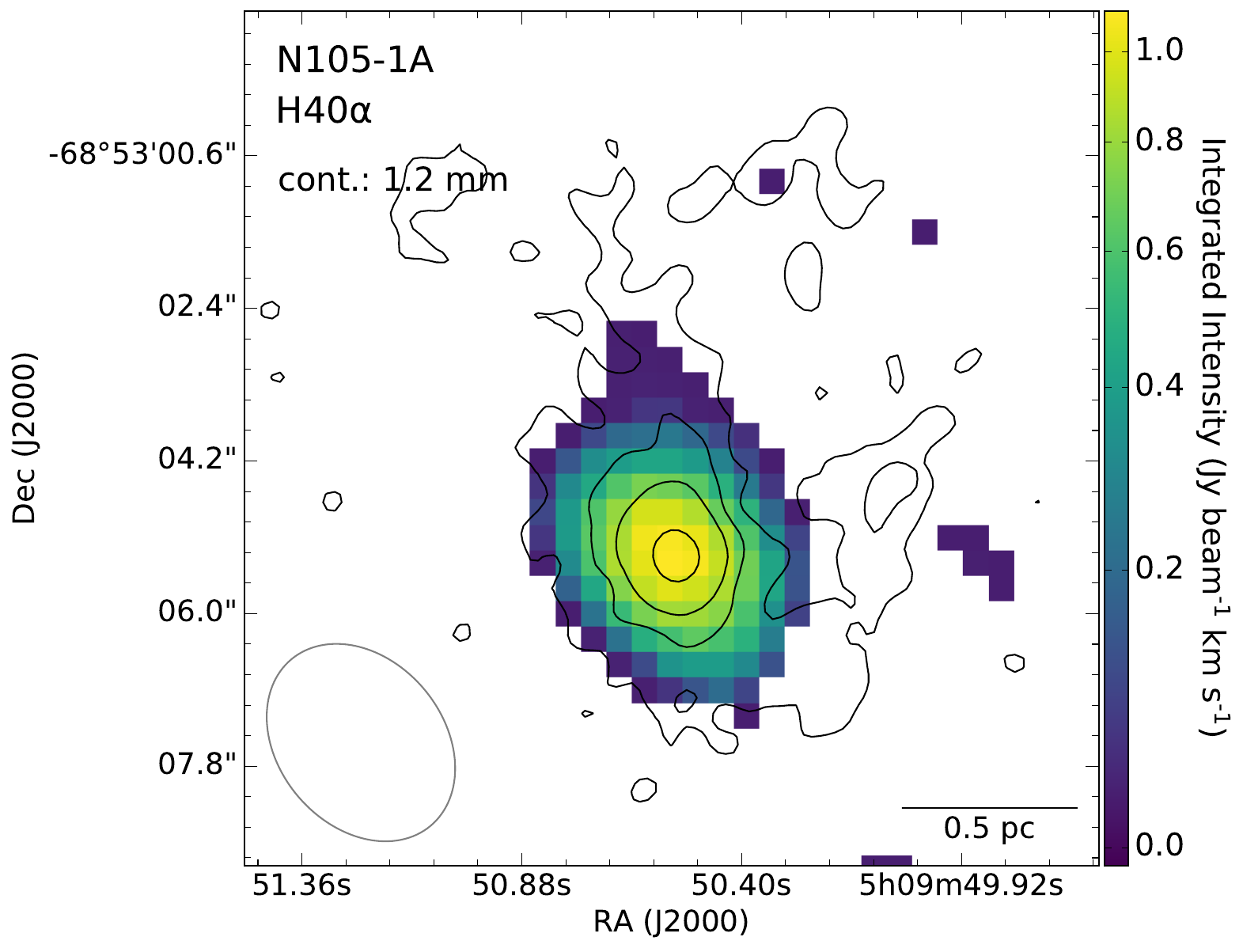}
\includegraphics[width=0.325\textwidth]{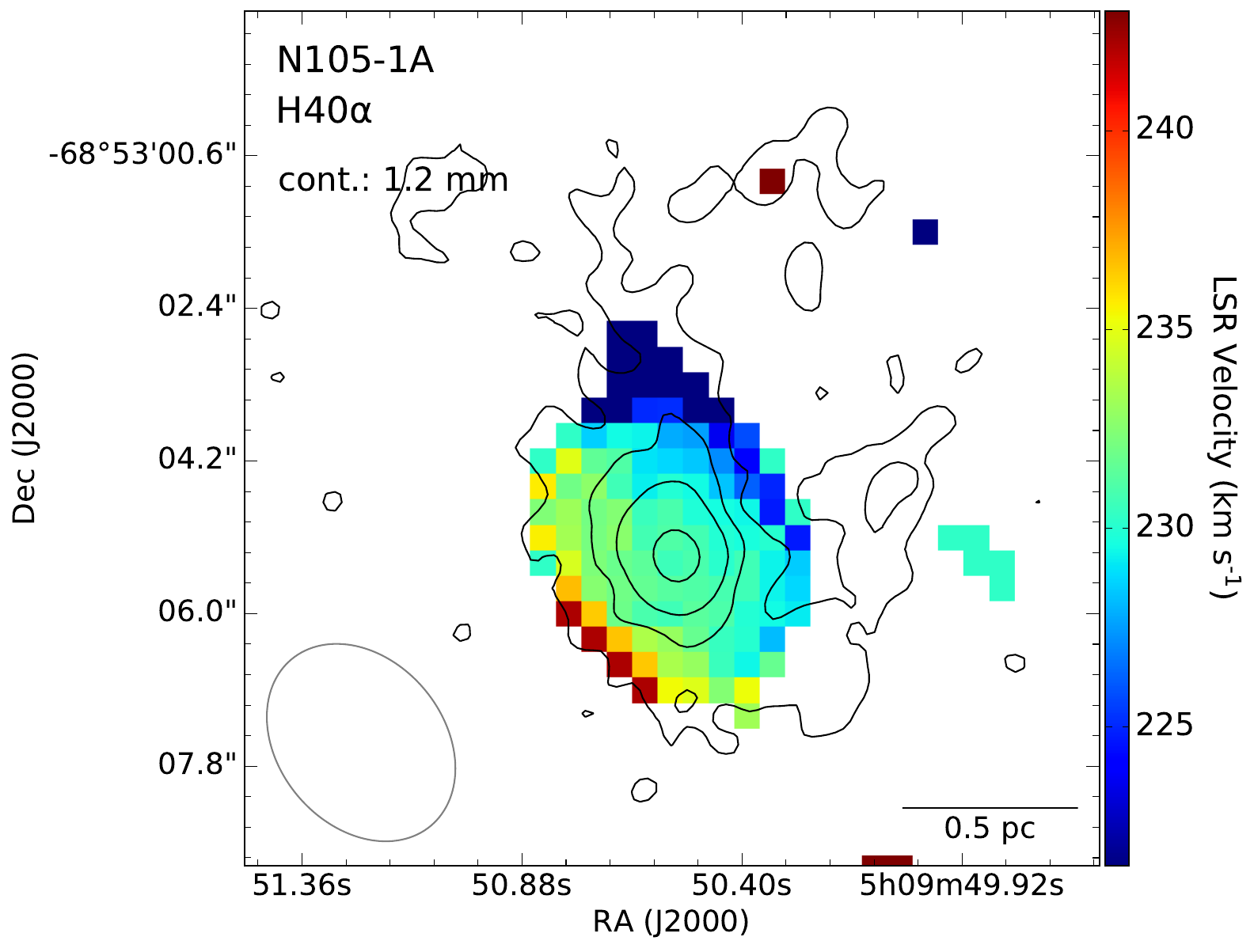}
\includegraphics[width=0.325\textwidth]{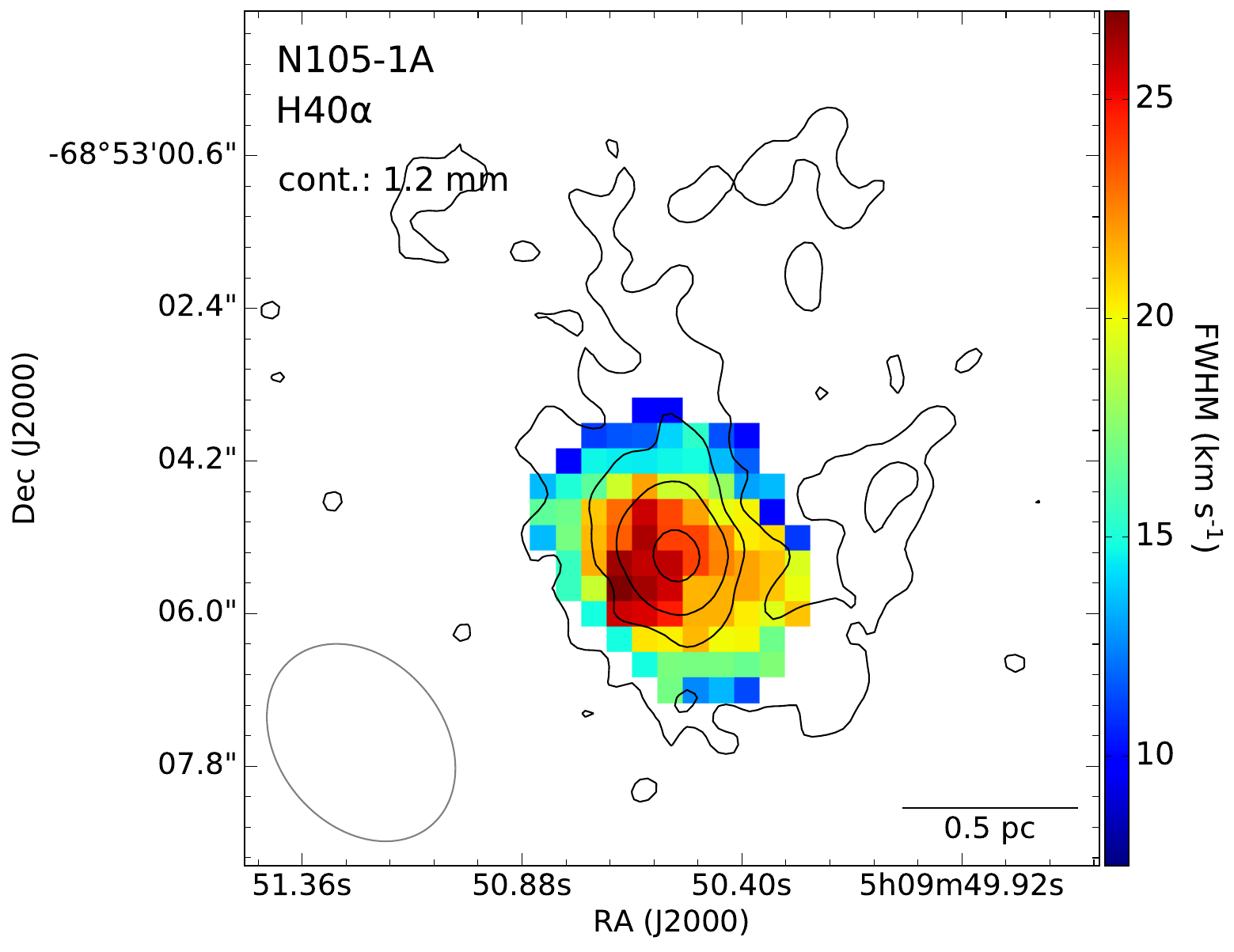}\\
\includegraphics[width=0.325\textwidth]{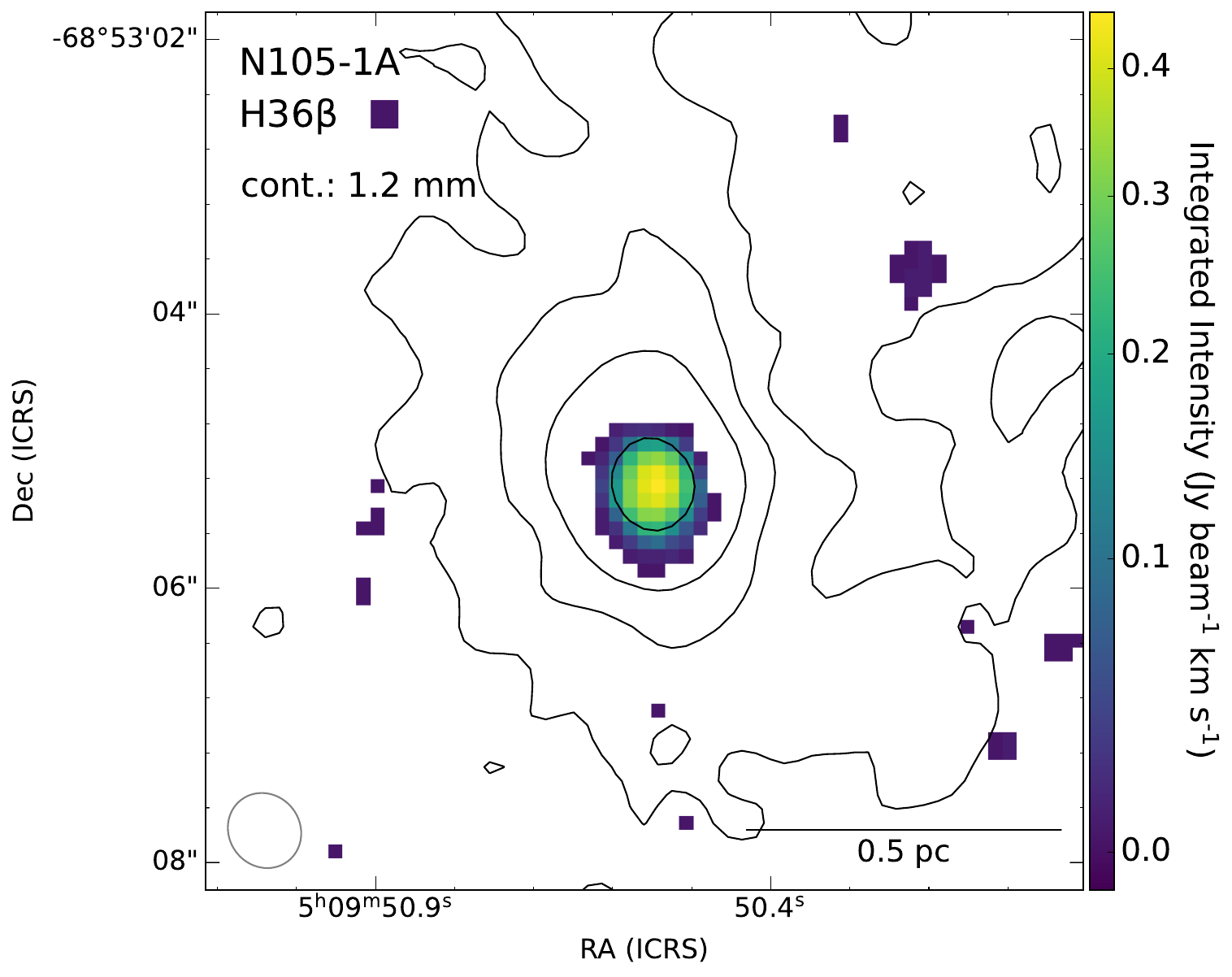}
\includegraphics[width=0.325\textwidth]{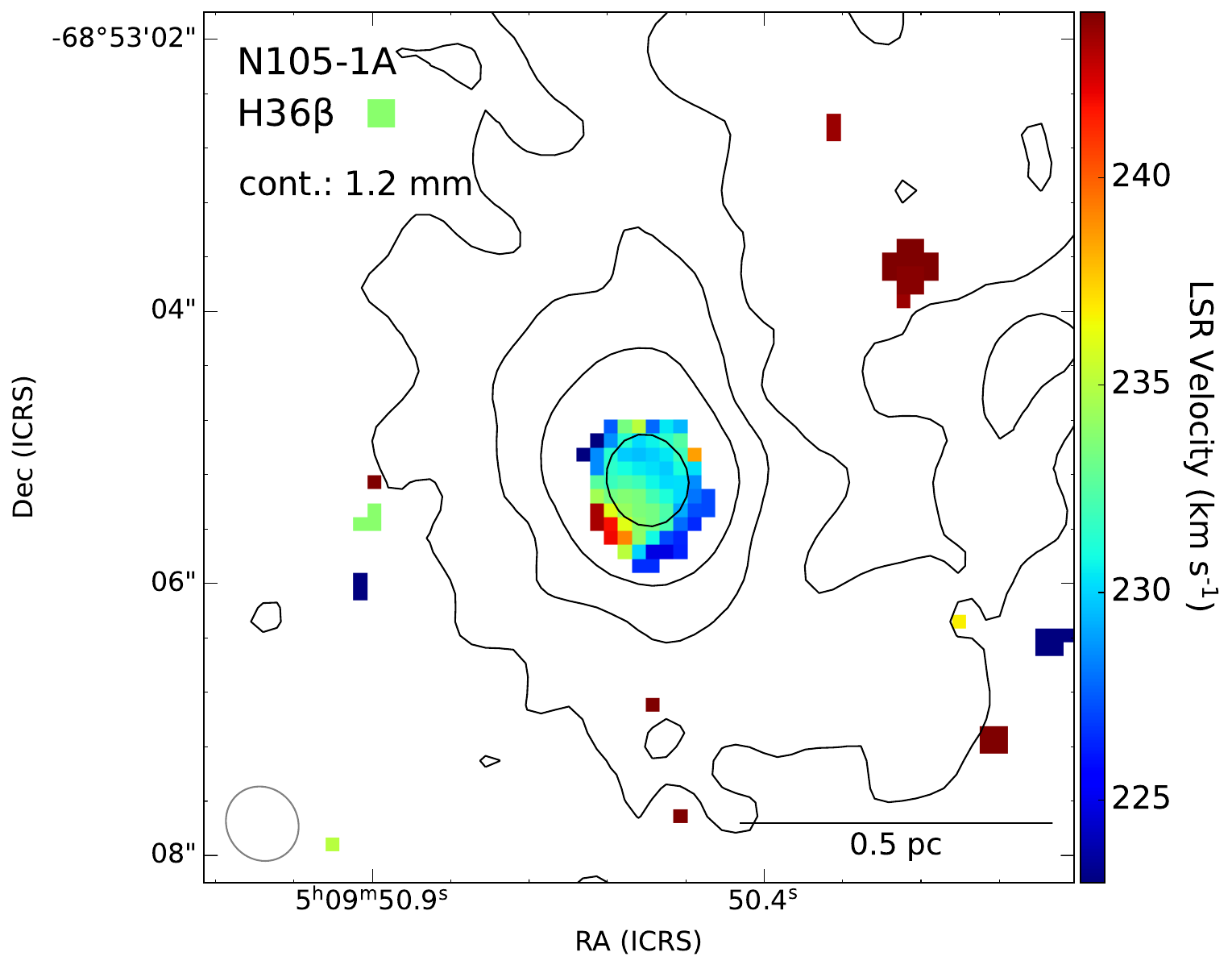}
\includegraphics[width=0.325\textwidth]{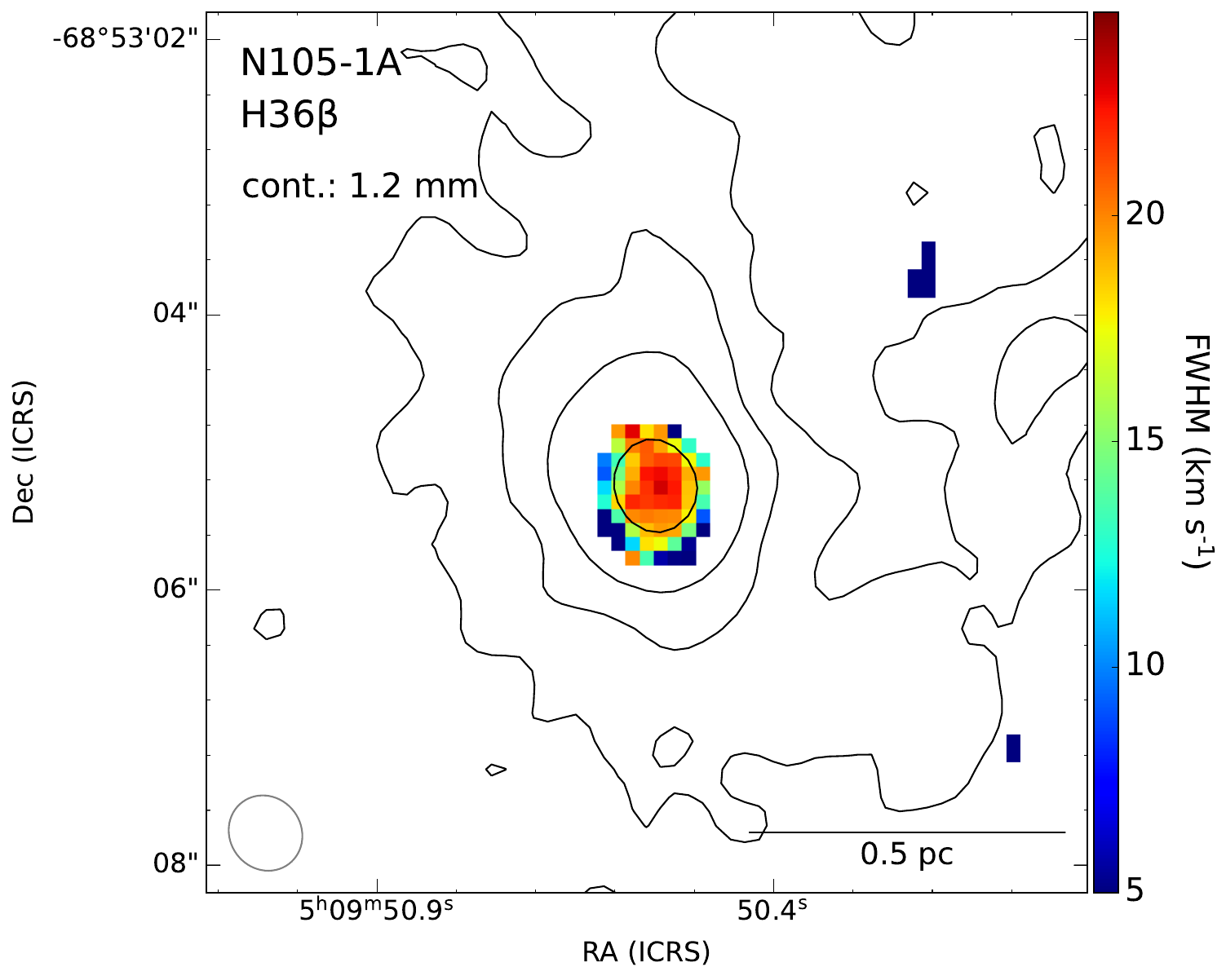}
\caption{The H40$\alpha$ (upper panel) and H36$\beta$ (lower panel) integrated line intensity (moment 0, left),  intensity weighted velocity (moment 1, center), and observed line width (right) images with the 1.2 mm continuum contours overlaid for reference, with contour levels the same as in Fig.~\ref{f:cont}.  To make moment maps, we only used the emission above 3$\sigma$. The velocity gradient of 17 km~s$^{-1}$ over 3$\rlap.{''}$5 or 20 km~s$^{-1}$ pc$^{-1}$ is detected in H40$\alpha$ in a roughly SE--NW direction. The size of the ALMA synthesized beam ($2\rlap.{''}56\times1\rlap.{''}95$ for H40$\alpha$ and $0\rlap.{''}51\times0\rlap.{''}47$ for H36$\beta$) is shown in the lower left corner in each image. \label{f:ionmom}}
\end{figure*}

\subsection{Ionized Gas Kinematics}
\label{s:iongaskin} 

There are three groups of the line broadening mechanisms that can affect the recombination line widths: the natural, Doppler, and pressure (collisional) broadening (e.g., \citealt{gordon2002}). The effects of the natural and collisional broadening are negligible for mm/submm RLs (e.g., \citealt{brocklehurst1972}; \citealt{roelfsema1992}; \citealt{gordon2002}), thus we only consider the Doppler thermal and non-thermal contributions to the line widths of mm-RLs detected toward N\,105--1\,A. 

The thermal line width can be determined from the formula: 
\begin{eqnarray}
\Delta v_{\rm th} = \sqrt{8\,{\rm ln}2\, \frac{k_{\rm B} T_{\rm e}}{\mu m_{\rm H}}} 
\end{eqnarray}
where $k_B$ is the Boltzmann constant, $T_{\rm e}$ is the gas electron temperature,  $\mu$ is the molecular weight in atomic units (for atomic hydrogen, $\mu_{\rm H}$ = 1.008 amu), and $m_{\rm H}$ is the mass of the hydrogen atom. For the temperature of the hydrogen gas of $10\,940\pm1\,350$ K, $\Delta v_{\rm th}$ is $22.3\pm1.4$ km s$^{-1}$.

The non-thermal (turbulence and large-scale motions) line width can be estimated using the equation:
\begin{eqnarray}
\Delta v_{\rm nth} = \sqrt{\Delta v^2 - \Delta v_{\rm th}^2}
\end{eqnarray}
where $\Delta v$ is the measured line width corrected for instrumental broadening (see Table~\ref{t:fitRLs}).  The average line width for the three brightest recombination lines is 32.3 km s$^{-1}$, resulting in $\Delta v_{\rm nth}$ of $\sim$23.4 km s$^{-1}$, similar to $\Delta v_{\rm th}$. 

The relatively large non-thermal component toward 1\,A likely indicates the presence of bulk motions in the region.  To demonstrate this, we  constructed the velocity and line width (FWHM) images for the two brightest recombination lines detected toward 1\,A: H40$\alpha$ (Band 3) and H36$\beta$ (Band 6; $\sim$5$\times$ higher angular resolution than the Band 3 observations); all the images are shown in Fig.~\ref{f:ionmom}.  The H40$\alpha$ velocity map reveals the velocity gradient of 17 km~s$^{-1}$ over 3$\rlap.{''}$5 (or 20 km~s$^{-1}$ pc$^{-1}$) along the line going through the peak of the 1.2 mm continuum emission in a roughly SE--NW direction. It is striking though that the velocity changes the most at the outskirts of the source while it remains roughly the same across its central part. The H40$\alpha$ lines are the broadest in the area southeast from the continuum peak, possibly hinting on some additional line broadening caused by external bulk motions.  The results on the kinematic structure of the ionized gas toward 1\,A based on the current H40$\alpha$ data have to be treated with caution since the signal-to-noise ratio of the H40$\alpha$ line is low in the outermost regions. The H36$\beta$ emission traces much smaller scales close to the central star;  the overall trend in velocity seems to be similar to that seen in H40$\alpha$.  The H36$\beta$ lines are the broadest at the continuum peak. 

The velocity gradients observed in the ionized gas may trace infall, outflow, rotation, or their combination, or highlight overlapping velocity components. The observed velocity patterns are influenced by the viewing angle, often making the interpretation difficult. Based on the near-IR spectroscopic data, \citet{oliveira2006} found evidence for the outflow in 1\,A (their source N\,105A IRS1), therefore the velocity gradient detected in mm-RLs in the same direction as observed in the near-IR data may trace the outflow.  However, we did not find any clear outflow signatures in the $^{12}$CO (3--2) data (see Section~\ref{s:shocksphot}).  The H40$\alpha$ and H36$\beta$ velocity gradients are not homogenous, therefore the rotation is not the most likely interpretation, but cannot be ruled out. 

\section{Discussion}
\label{s:discussion}

In this section, we compare the physical properties obtained for the LMC source N\,105--1\,A to those of Galactic H\,{\sc ii} regions to search for any differences that could be explained by different environments in these galaxies. We also investigate the distribution and kinematics of the ionized gas in N\,105--1\,A in the context of its molecular environment to further explore the nature of the source.

\subsection{Electron Temperature of N\,105--1\,A vs. Galactic H\,{\sc ii} Regions}
\label{s:Te}

The recombination line observations of Galactic star-forming regions revealed the $T_{\rm e}$ Galactocentric radial gradient which was interpreted as the metallicity gradient  (e.g., \citealt{churchwell1975}; \citealt{churchwell1978}; \citealt{shaver1983}).  The electron temperature increases while the metallicity decreases with increasing Galactocentric distance ($R_{\rm GC}$, e.g., \citealt{shaver1983}; \citealt{balser2011}; \citealt{fernandez2017}; \citealt{maciel2019}). The lower abundance of metal coolants (such as C and O) affects the balance between the heating and cooling within H\,{\sc ii} regions, resulting in warmer temperatures.  \citet{shaver1983} performed the RL observations (to obtain reliable $T_{\rm e}$) and optical spectra (to derive metal abundances using $T_{\rm e}$ from RLs) of a large sample of Galactic H\,{\sc ii} regions to investigate $T_{\rm e}$ and metallicity variations with $R_{\rm GC}$; they found that  $T_{\rm e}  =  (433\pm40)\,R_{\rm GC} + (3150\pm110)$, where $T_{\rm e}$ is in K and $R_{\rm GC}$ in kpc. 

Based on the \citet{shaver1983}'s results, we find that Galactic H\,{\sc ii} regions with $T_{\rm e}$ of 10\,940 K, the value we determined for N\,105--1\,A, are expected to be at $R_{\rm GC}$ of $\sim$18 kpc and have the oxygen abundance (${\rm 12+[O/H]}=-0.067\,R_{\rm GC}+9.38$) of $\sim$8.2 (or metallicity $Z$ of $\sim$0.32 Z$_{\odot}$). This is consistent with the oxygen abundance ($[12+{\rm log(O/H)}]_{\rm LMC}=8.4$) and metallicity ($Z_{\rm LMC}$ = 0.3--0.5 $Z_{\odot}$) measured in the LMC (e.g.,  \citealt{russell1992}).  The oxygen abundance variations determined in other studies provide similar results, e.g., $R_{\rm GC}$ at which the oxygen abundance is the same as in the LMC is $\sim$16 kpc based on abundance studies involving hundreds of Cepheid variables (\citealt{maciel2019}; \citealt{luck2011}).

\subsection{N\,105--1\,A vs. Galactic H\,{\sc ii} Regions with the Detection of mm-RLs}
\label{s:galcomp}

We compare the properties of N\,105--1\,A to those of a large sample of Galactic compact and UC H\,{\sc ii} regions, and several HC H\,{\sc ii} region candidates, with the detection of mm-RLs from \citet{kim2017}.  \citet{kim2017} detected mm-RLs with $n$ from 39 to 65, and $\Delta n$ = 1, 2, 3, and 4 ($\alpha$-, $\beta$-, $\gamma$-, and $\delta$-transitions) toward 178 sources using the IRAM 30\,m (HPBW$\sim$29$''$) and Mopra 22\,m (HPBW$\sim$36$''$) telescopes. Based on the known distances to 120 sources, the physical scales probed by the \citet{kim2017}'s single-dish observations range from 0.1 pc to $\sim$1 pc. These physical scales are similar to those we trace with ALMA in the LMC: 0.51 pc (3 mm / Band 3) and 0.12 pc (1.2 mm / Band 6), therefore the \citet{kim2017} catalog of compact/UC H\,{\sc ii} regions is an ideal dataset to utilize to compare the properties of the LMC and Galactic objects with mm-RL detections. We have started building a sample of compact/UC H\,{\sc ii} regions in the LMC based on the archival ALMA data and such comparisons will be more conclusive in the future. Here, we  focus on a single object with the first detection of the higher order hydrogen mm-RLs outside the Galaxy (up to $\Delta n$=7) to understand its nature.

\citet{kim2017} targeted 976 compact dust clumps selected from the APEX Telescope Large Area Survey of the Galaxy (ATLASGAL; \citealt{schuller2009}). About 10,000 dense clumps have been identified based on the ATLASGAL 870 $\mu$m continuum data, covering the pre-stellar, protostellar, and H\,{\sc ii} region massive star formation stages (\citealt{contreras2013}, \citealt{urquhart2014}; \citealt{csengeri2014}). The \citet{kim2017}'s IRAM/Mopra sample includes roughly the same number of the mid-IR bright and mid-IR quiet clumps that are among the brightest 870 $\mu$m clumps in their categories.  \citet{kim2017} detected mm-RLs toward 18\% of the targeted ATLASGAL massive clumps. Out of 178 clumps with the detection of an $\alpha$-transition, (65, 23, 22) sources were also detected in (H$n\beta$, H$n\gamma$, H$n\delta$). 

\begin{figure}[t!]
\centering
\includegraphics[width=0.485\textwidth]{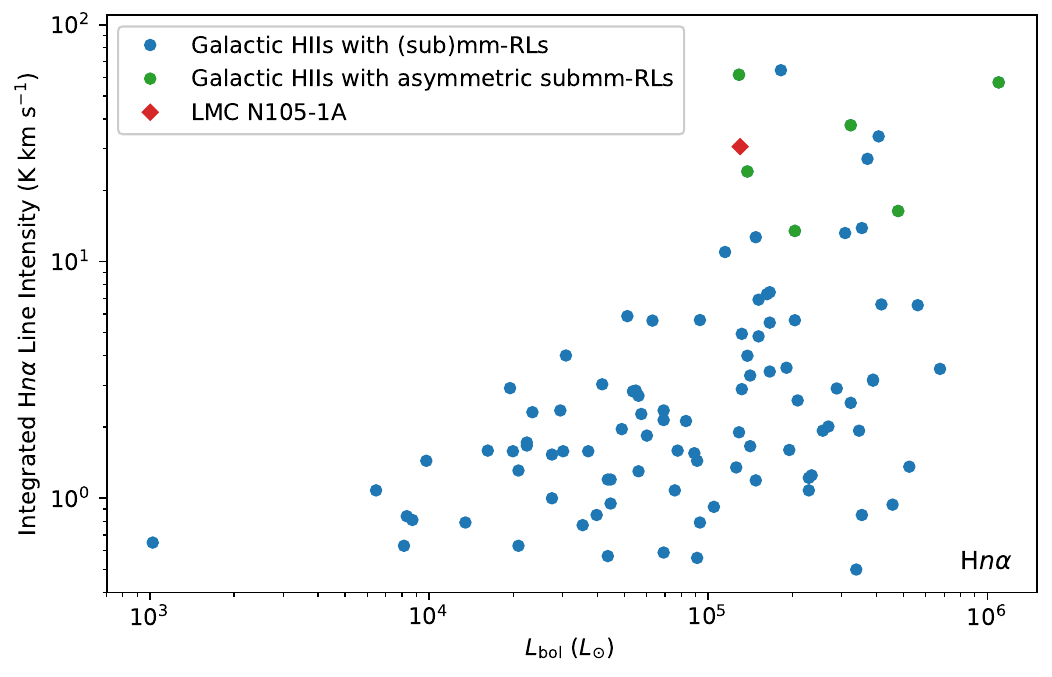}
\includegraphics[width=0.485\textwidth]{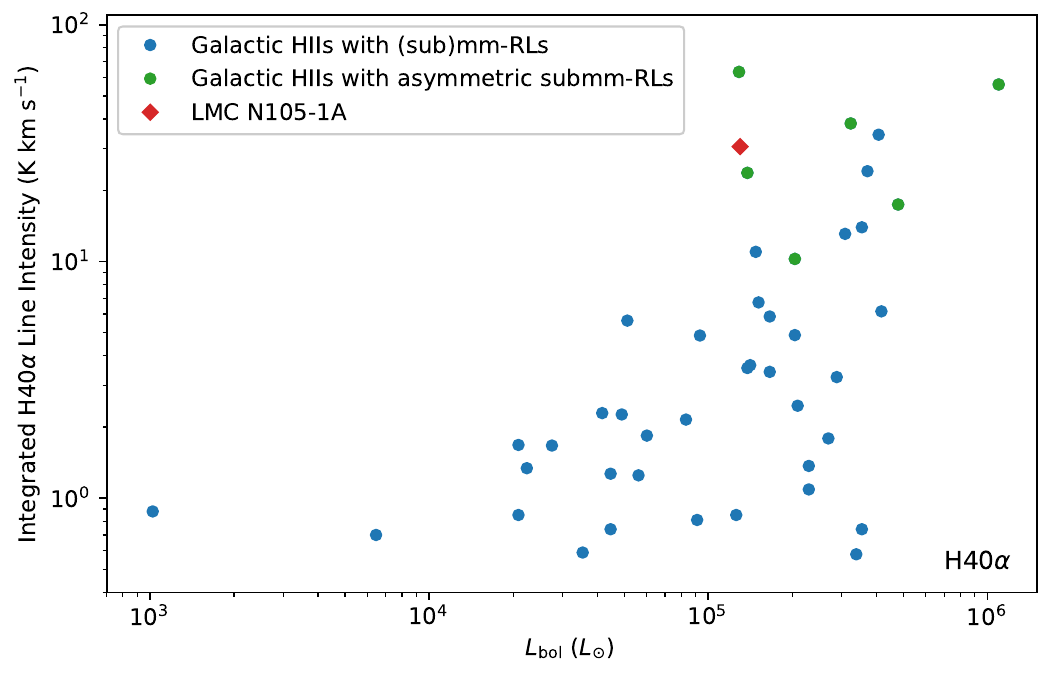}
\caption{Integrated H$n \alpha$ (stacked $\alpha$-transitions; upper panel) and H40$\alpha$ (lower panel) line intensities as a function of bolometric luminosity ($L_{\rm bol}$) for the Galactic H\,{\sc ii} regions with the detection of (sub)mm-RLs \citep{kim2017} with the position of N\,105--1\,A indicated with the red diamond. The integrated H40$\alpha$ line intensity (not corrected for beam dilution) is plotted for 1\,A in both plots. Galactic sources with asymmetric submm-RLs discussed in \citet{kim2018} are also overlaid as indicated in the legend.  The integrated H40$\alpha$ line intensity observed toward 1\,A is similar to that observed toward the most luminous Galactic sources from the \citet{kim2017,kim2018} sample. \label{f:IntHna}}
\end{figure}
About 75\% of the clumps with the detection of mm-RLs from \citet{kim2017} are associated with both the radio continuum and mid-IR emission (in the 22 $\mu$m band of the Wide-field Infrared Survey Explorer, WISE, survey), and were previously identified in the literature as compact or UC H\,{\sc ii} regions. Similarly to these Galactic sources, N\,105--1\,A is detected at radio and mid-IR wavelengths (including WISE at 22 $\mu$m, although due to a relatively low angular resolution, the emission is contaminated by the emission from source 1\,B).  N\,105--1\,A's 6 cm radio luminosity ($S_{\rm 6\,cm}\times D_{\rm LMC, kpc}^2$) of $\sim$$6.4\times10^4$ mJy kpc$^2$ lies well within the distribution of the 6 cm radio luminosities of the \citet{kim2017}'s H\,{\sc ii} regions with mm-RLs, with the median of $2.4\times10^5$ mJy kpc$^2$ (see Fig. 10 in \citealt{kim2017}).

In the top panel of Fig.~\ref{f:IntHna}, we show the integrated H$n \alpha$ line intensities of Galactic H\,{\sc ii} regions with the detection of mm-RLs from \citet[stacked H39$\alpha$, H40$\alpha$, H41$\alpha$, and H42$\alpha$ transitions; not corrected for beam dilution]{kim2017} as a function of bolometric luminosity ($L_{\rm bol}$), with the position of N\,105--1\,A indicated. Since the H40$\alpha$ transition is the only $\alpha$-transition observed toward 1\,A  to date, we also show the plot only including Galactic H\,{\sc ii} regions with the detection of the H40$\alpha$ transition.  The plots demonstrate that 1\,A lies within the parameter space covered by the Galactic sources. The integrated H40$\alpha$ line intensity of $\sim$31~K~km~s$^{-1}$ observed toward 1\,A (Table~\ref{t:fitRLs}) is among the highest values reported by \citet{kim2017} for Galactic H\,{\sc ii} regions. In their sample, the integrated H40$\alpha$ line intensity ranges from 0.5 to 63.24~K~km~s$^{-1}$, with only four sources (out of 60) above 30~K~km~s$^{-1}$:
AGAL010.624$-$00.384 (34.34~K~km~s$^{-1}$), 
AGAL034.258$+$00.154 (38.31),
AGAL043.166$+$00.011 (55.96), 
AGAL012.804$-$00.199 (63.24).
In case of the stacked $\alpha$-transitions (H$n \alpha$), the integrated H$n \alpha$ line intensity ranges from 0.37 to 64.28~K~km~s$^{-1}$ with five sources (out of 178) above 30~K~km~s$^{-1}$, including AGAL333.604$-$00.212 (64.28~K~km~s$^{-1}$) that was not detected in H40$\alpha$. The sources with the highest values of the integrated H40$\alpha$ and/or H$n \alpha$ line intensities are  among the most luminous in the \citet{kim2017} sample ($L_{\rm bol}>10^5$ $L_{\odot}$). 

For example, clump AGAL043.166$+$00.011 is the central part of one of the most active massive star-forming region in the Galaxy ($M = 4.3\times10^4$ M$_\odot$, \citealt{giannetti2017}), W\,49, and it is a massive protocluster candidate (\citealt{urquhart2018}). AGAL034.258$+$00.154 and AGAL010.624$-$00.384 are well-studied massive star-forming regions (G34.26+0.15 and G10.62-0.38) hosting UC H\, {\sc ii} regions and hot cores; both regions are unresolved in \citet{kim2017}'s single-dish observations.

N\,105--1\,A lies within the physical parameter space (FWHM size, EM, and $n_{\rm e}$; see Section~\ref{s:phys}) covered by the \citet{kim2017} sample of Galactic H\,{\sc ii} regions with mm-RLs (see their Fig. 16) and consistent with being at the UC\,{\sc ii} region stage of the massive star evolution. 

N\,105--1\,A corresponds to source IRAS\,05101$-$6855A in the main IRAS catalog \citep{iras2} and IRAS\,F05101-6856 in the IRAS Faint Source Catalog (with a better positional match and the same photometry within the uncertainties; \citealt{irasfaint}). The photometry of both IRAS\,05101$-$6855A and IRAS\,F05101-6856 fulfills the criteria for UC H\,{\sc ii} regions proposed by \citet{wood1989}.

\begin{figure*}[ht!]
\centering
\includegraphics[width=0.45\textwidth]{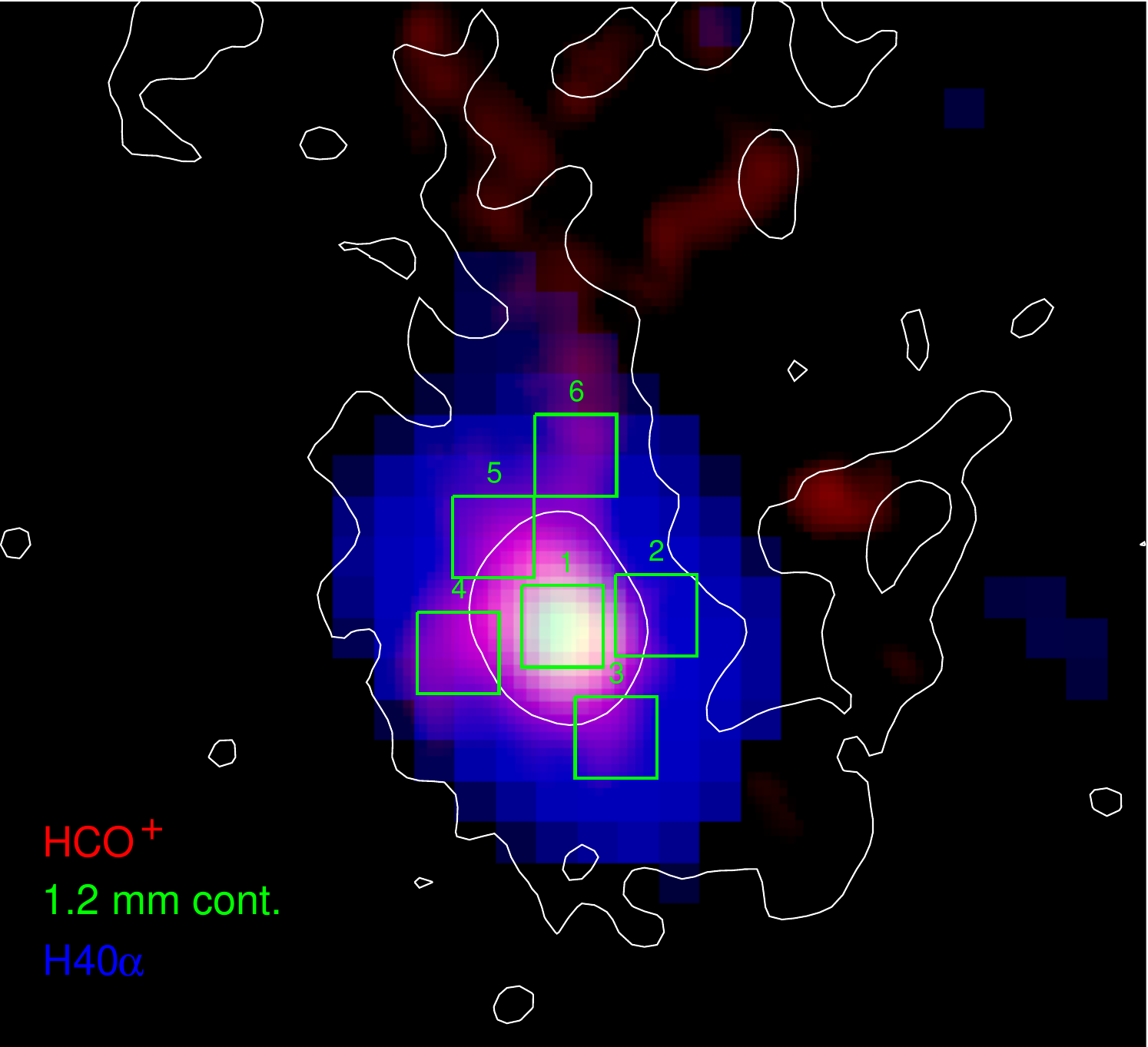}\\
\includegraphics[width=0.45\textwidth]{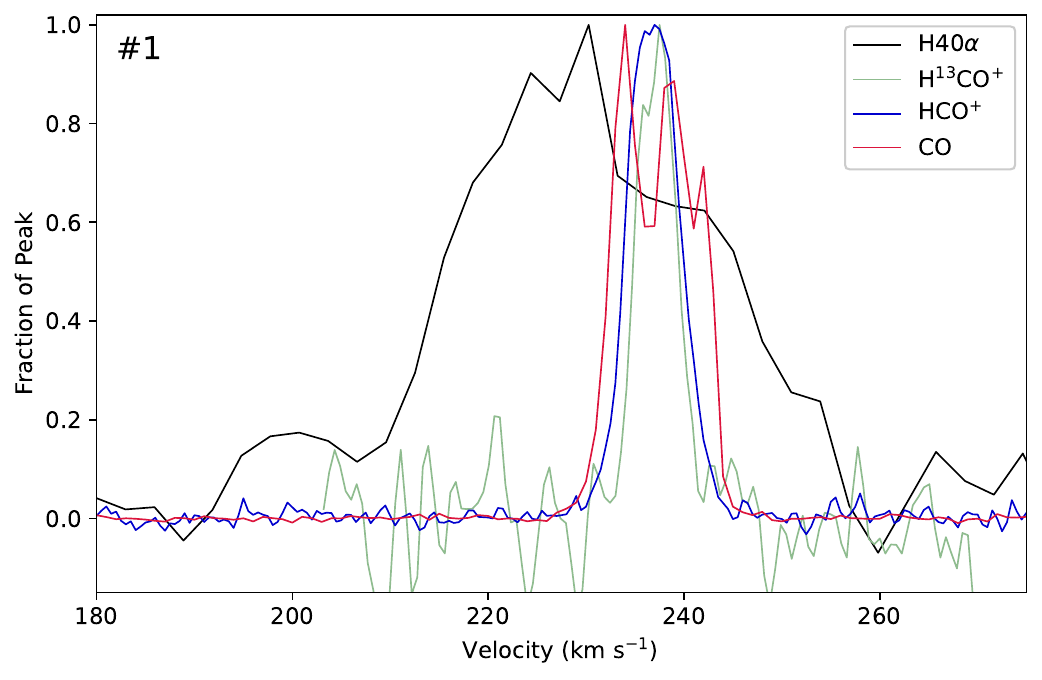}
\includegraphics[width=0.45\textwidth]{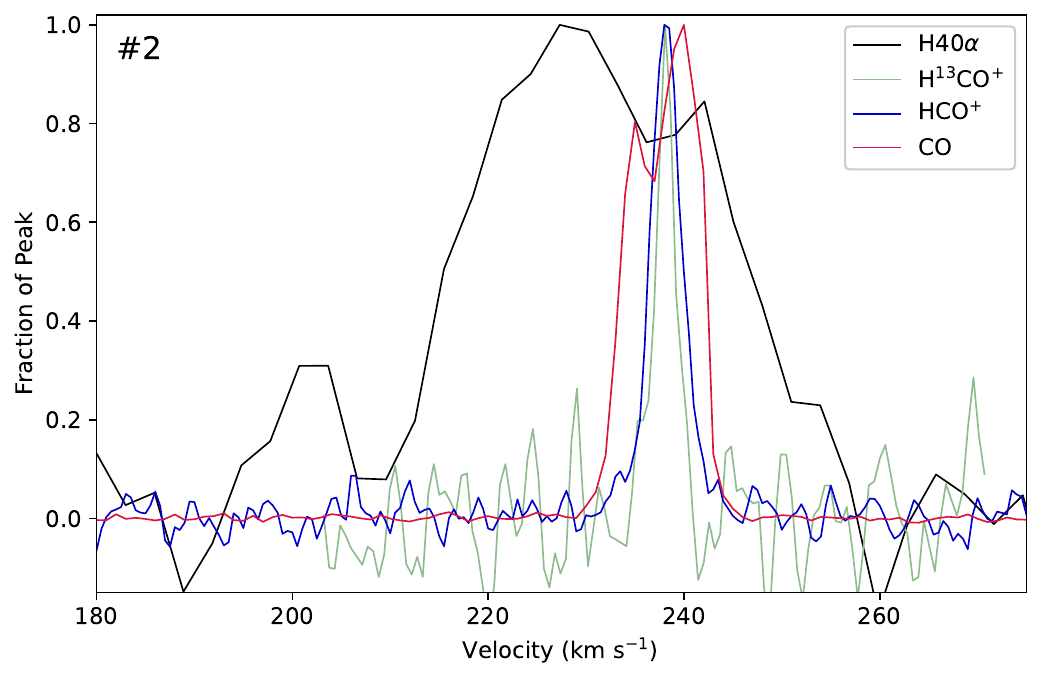}
\includegraphics[width=0.45\textwidth]{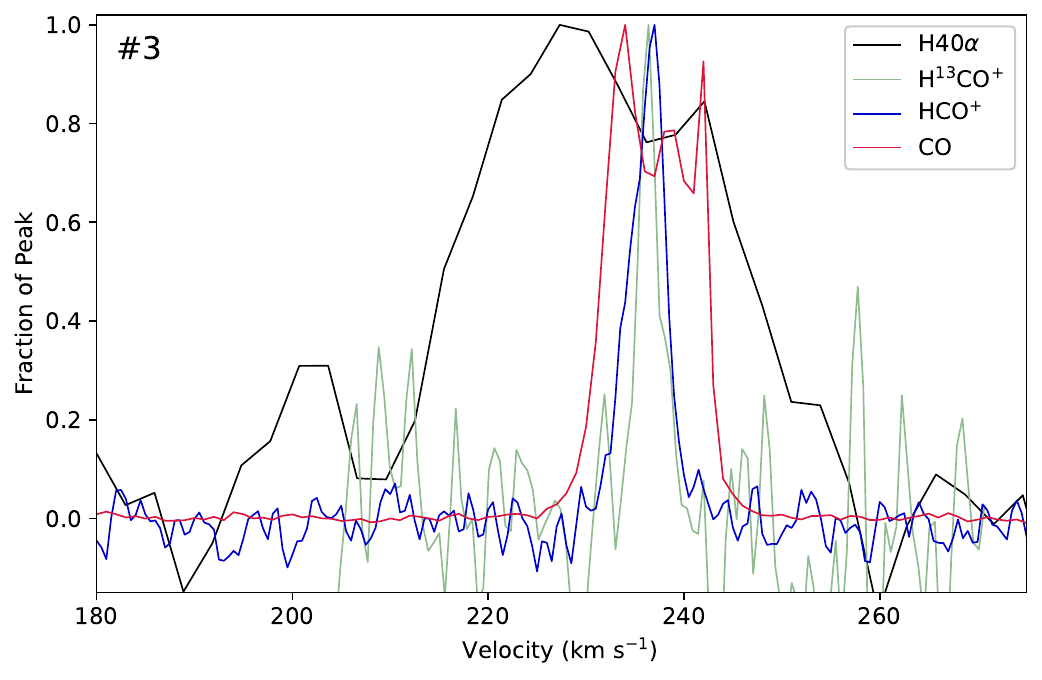}
\includegraphics[width=0.45\textwidth]{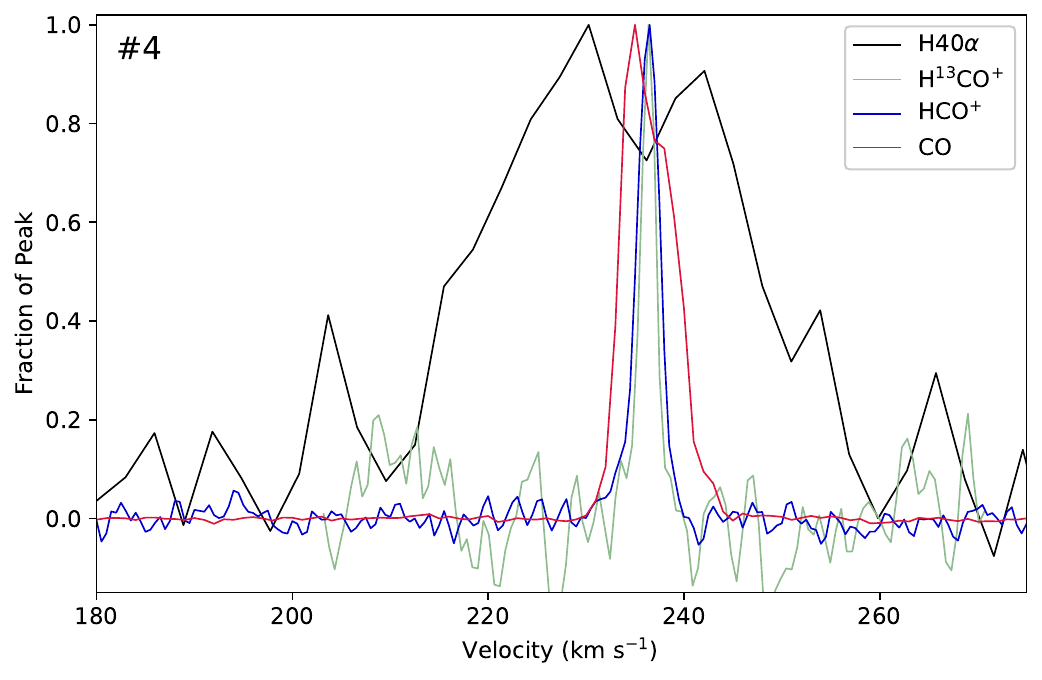}
\includegraphics[width=0.45\textwidth]{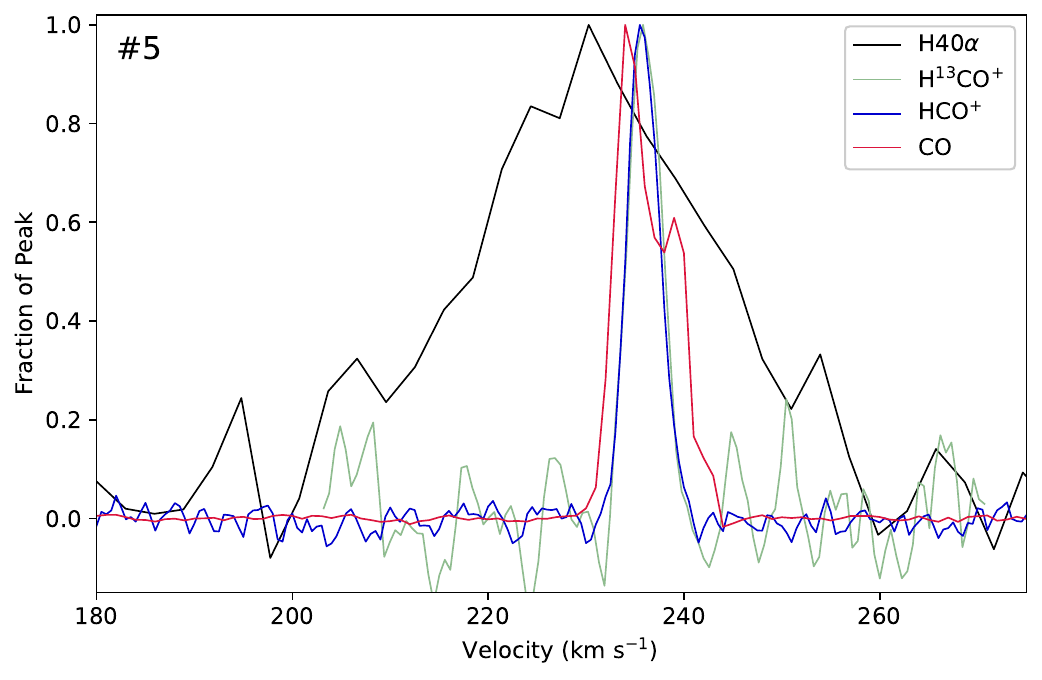}
\includegraphics[width=0.45\textwidth]{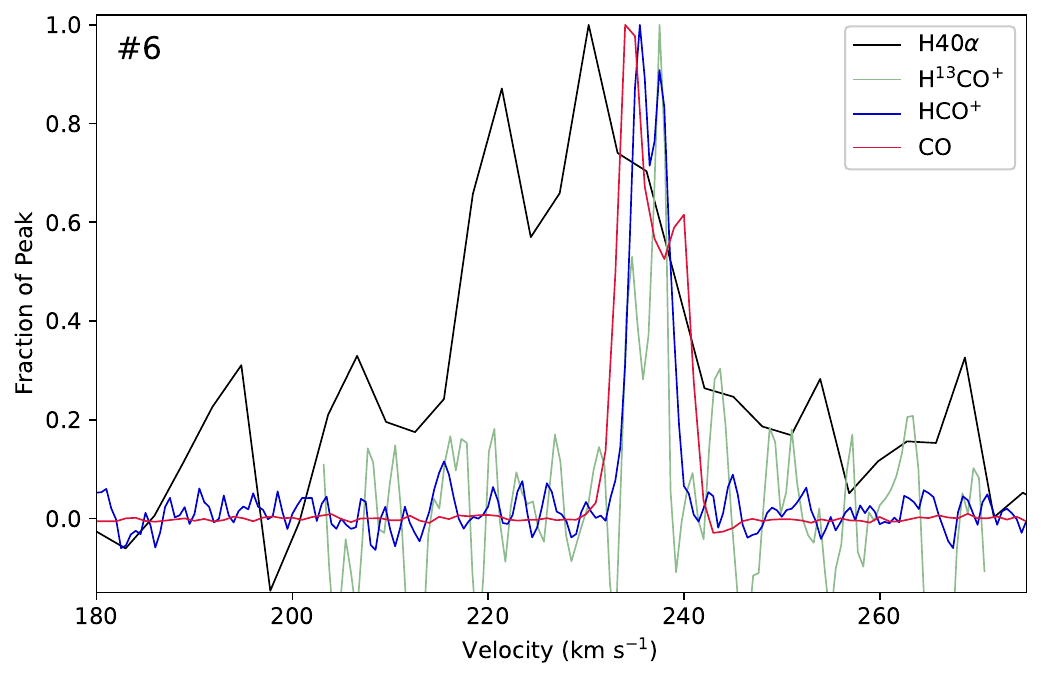}
\caption{Top panel: Three-color image combining the HCO$^{+}$ integrated intensity (red),  the 1.2 mm continuum  (green), and the H40$\alpha$ integrated intensity (blue) images with the spectral extraction regions indicated for spectra shown in the three lower panels. Lower panels: The H40$\alpha$ (unsmoothed), HCO$^{+}$ (4-3),  H$^{13}$CO$^{+}$ (3-2) \citep{sewilo2022n105}, and $^{12}$CO (3-2) line profile comparison for several locations in N\,105--1\,A.  \label{f:profiles}}
\end{figure*}

\subsubsection{Galactic H\,{\sc ii} Regions with Asymmetric Submm-RLs}
\label{s:asymmetric}

A subsample of 104 sources from the \citet{kim2017} catalog of H\,{\sc ii} regions with mm-RLs/ATLASGAL clumps was observed by  \citet{kim2018} at submm-RLs with the APEX 12\,m telescope (HPBW$\sim$16$''$--27$''$, tracing linear scales of $\sim$0.1--1 pc with the 16$''$ beam and $\sim$1.2--1.7 pc with the 27$''$ beam).  Submm-RLs H25$\alpha$, H28$\alpha$, and H35$\beta$ were observed toward the majority of the sample, and the H26$\alpha$, H27$\alpha$, and H29$\alpha$, H30$\beta$ lines toward a small subsample. Submm-RLs were detected toward 93 clumps.   \citet{kim2018} found that clumps associated with submm/mm-RLs are the most massive and luminous ($L_{\rm bol}>10^4$ $L_{\odot}$) clumps in the Galaxy.  The highest-frequency transitions observed toward N\,105--1\,A and within the frequency range of transitions observed by  \citet{kim2018} are (H36$\beta$, H41$\gamma$, H53$\eta$) at (260.03, 257.635, 257.194) GHz. The properties of 1\,A are consistent with the submm/mm-RL sample of clumps and associated H\,{\sc ii} regions from  \citet{kim2017,kim2018},

Interestingly,  \citet{kim2018} detected six sources with RL profiles that are a combination of a narrow and a broad Gaussian components. Three out of six of these sources are those with similar physical properties to 1\,A, including the integrated H$n \alpha$ line intensity above 30~K~km~s$^{-1}$:  AGAL012.804$-$00.199, AGAL034.258$+$00.154, and AGAL043.166$+$00.011.  \citet{kim2018} argue that the high-velocity components (revealed by either blue- or red-shifted wings in the RL profiles) toward four H\,{\sc ii} regions can be explained by the presence of high-velocity ionized flows. Submm-RL profiles of the other two clumps are likely the result of unresolved clusters of compact H\,{\sc ii} regions.

The inspection of the unsmoothed H40$\alpha$ lines detected toward N\,105--1\,A, shown in Fig.~\ref{f:profiles}, also reveals asymmetries in line profiles, providing an additional evidence for the presence of bulk motions in the region (see Section~\ref{s:iongaskin}).  


\subsection{Molecular Environment of N\,105--1\,A}
\label{s:molrenv}

We investigate the distribution and kinematics of the molecular gas  in the ALMA field N\,105--1 to understand the origin and nature of the source N\,105--1\,A, and identify any physical processes in addition to the photoionization by the central star that may be contributing to the measured ionizing flux (e.g., a shock ionization), producing bright enough higher-order mm-RLs to be detected outside the Galaxy.  A detailed modeling of the kinematic structure of the source is out of scope of the present paper.

For our analysis, we use the molecular gas tracers covering a wide range of gas densities. From our Band 6 observations presented in \citet{sewilo2022n105}, we utilize the CS (5--4), SO $^{3}\Sigma$ 6$_6$--5$_5$,  CH$_3$OH $5_{0,5}$--$4_{0,4}$ A, and H$^{13}$CO$^{+}$ (3--2) data. These observations trace the physical scales of $\sim$0.12 pc or $\sim$25,000 au, assuming the LMC distance of 49.59 kpc.  We also use the $^{13}$CO (1--0) data from our Band 3 observations that probe the $\sim$0.47 pc/$\sim$97,000 au scales. The archival Band 7 CO (3--2) and HCO$^{+}$ (4--3) data allow us to explore the spatial distribution and the velocity structure of the diffuse and dense gas, respectively, with the highest available spatial resolution (0.087 pc, $\sim$18,000 au). 

We present the integrated intensity (moment 0), intensity weighted velocity (moment 1), and line full-width at half-maximum (FWHM; calculated from the intensity weighted velocity dispersion, moment 2) maps in Appendix~\ref{s:appMom}: Fig. (\ref{f:mom13CO}, \ref{f:momCS}, \ref{f:momSO}, \ref{f:momCH3OH}, \ref{f:momCO}, \ref{f:momHCOp}) for ($^{13}$CO, CS, SO, CH$_3$OH, CO, HCO$^{+}$). The integrated intensity and velocity maps are available for all the species, while the FWHM maps are only available if they provide reliable results ($^{13}$CO, CS, and SO).  For CS, SO, CO, and HCO$^{+}$, we also show channel maps (Figs. \ref{f:momCSchan}, \ref{f:momSOchan}, \ref{f:COchan}, and \ref{f:HCOpchan}, respectively) that provide a more detailed view of the velocity structure in N\,105--1.

\subsubsection{Evidence for a Cloud-Cloud Collision in the Region Leading to the Formation of N\,105--1\,A}
\label{s:collision}

\begin{figure*}[th!]
\centering
\includegraphics[width=0.85\textwidth]{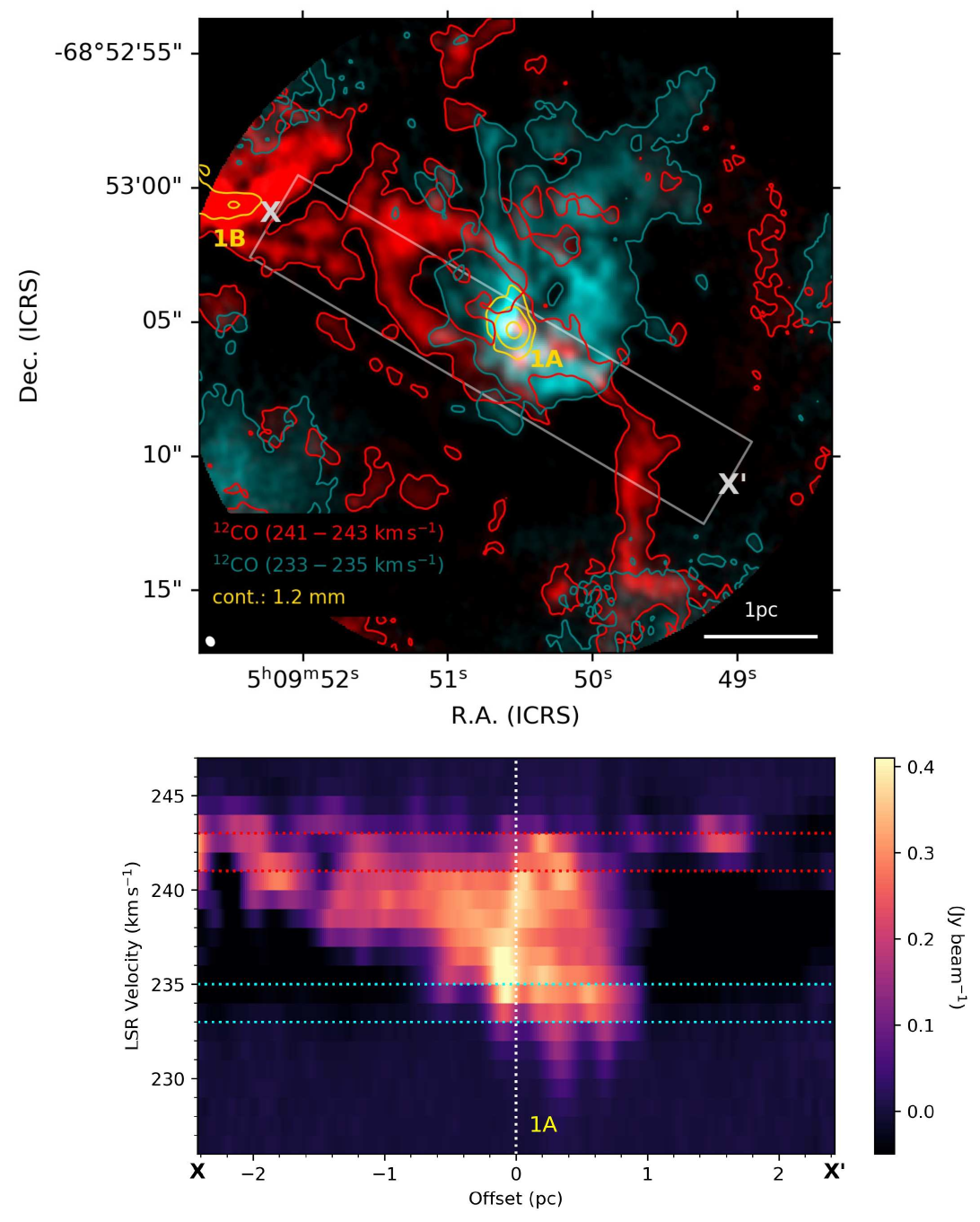}
\caption{Upper panel: Two-color image highlighting the presence of two CO (3--2) velocity components toward the 1.2 mm continuum source N\,105--1\,A. The CO (3--2) redshifted (241--243 km s$^{-1}$) and blueshifted (233--235 km s$^{-1}$) velocity components are shown in red and cyan, respectively. Both the blue and red single contours correspond to 0.17 Jy beam$^{-1}$ km s$^{-1}$ and outline the lower emission boundaries of the two velocity components.  The 1.2 mm continuum contours are shown in yellow with contour levels of (10, 30, 250)$\sigma$ (see also Fig.~\ref{f:cont}). The size of the synthesized beam ($0\rlap.{''}41\times0\rlap.{''}33$) is indicated in the lower left corner of the image. Lower panel: The CO (3--2) position-velocity diagram along the gray rectangle shown in the image in the upper panel (X--X$'$). The x-axis shows an offset from the 1.2 mm continuum peak (source 1\,A) indicated with the white dotted line; the offset directions are labeled X and X$'$ to ease a comparison to the image in the upper panel. The horizontal red and cyan dotted lines show the CO (3--2) integrated velocity ranges used to plot the redshifted and blueshifted velocity components in the image in the upper panel.  \label{f:COkin}}
\end{figure*}

\begin{figure*}[th!]
\centering
\includegraphics[width=\textwidth]{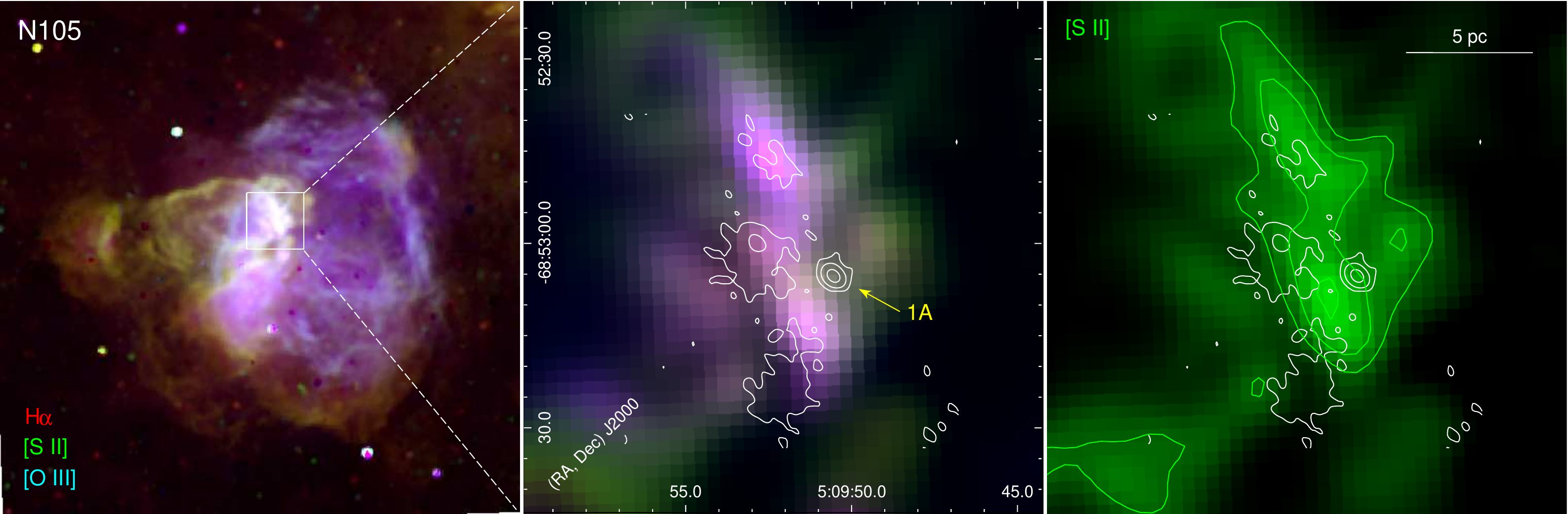}
\caption{Three-color mosaic of the star-forming region N\,105 combining the H$\alpha$ (red), [S\,{\sc ii}] (green), and [O\,{\sc iii}] (blue) images from the MCELS survey (\citealt{smith1998}; \citealt{paredes2015}; an angular resolution of $\lesssim$5$''$) is shown in the left panel.  The white box indicates the area shown in the middle panel: a zoom in on the region hosting source N\,105--1\,A with the detection of higher-order mm-RLs. The  [S\,{\sc ii}] image of the same area is presented in the right panel. White contours in the middle and right panels correspond to the 3 mm continuum emission with contour levels the same as in Fig.~\ref{f:cont}.  Green contours in the right panel correspond to (50, 70, 95)\% of the [S\,{\sc ii}] emission peak.  \label{f:mcels}}
\end{figure*}

The most striking feature in the velocity maps of all the investigated species is a velocity gradient stretching across the region with the velocity range of up to 10 km s$^{-1}$. The $^{13}$CO (1--0) observations provide the best view of the larger scale velocity structure toward N\,105--1 due to the largest field-of-view and the availability of the ALMA 7m data. The $^{13}$CO velocity map shows that the molecular gas velocity gradient extends over $\sim$28$''$ ($\sim$6.7 pc) in right ascension and $\sim$18$''$ ($\sim$4.3 pc) in declination (see Fig.~\ref{f:mom13CO}). The velocity increases with increasing RA from the area west of 1\,A to the continuum source 1\,B in the east.  At the higher-velocity end of the gradient, there is an additional U-shaped filamentary structure extending $\sim$16$''$ ($\sim$3.8 pc) north from 1\,B and then bending toward southwest in the direction of 1\,A. 

The velocity gradient across N\,105--1 is  evident in all other velocity maps, including those for species with a much more compact distribution than $^{13}$CO, i.e., CS (Fig.~\ref{f:momCS}), SO (Fig.~\ref{f:momSO}), and even CH$_3$OH (Fig.~\ref{f:momCH3OH}). 

The sharpest view of the diffuse gas kinematics toward N\,105--1 (although least extensive; see Fig.~\ref{f:mom13CO} and \ref{f:momCO}) is provided by the $^{12}$CO (3--2) observations due to the highest spatial resolution and the high dynamic range.  The inspection of the CO velocity channels reveals two distinct velocity components in N\,105--1 that seem to intersect at the position of N\,105--1\,A (Fig.~\ref{f:COchan}). These velocity components are highlighted in the two-color image shown in the upper panel of Fig.~\ref{f:COkin}, combining the redshifted (241--243 km s$^{-1}$) and blueshifted (233--235 km s$^{-1}$) CO (3--2) emission in N\,105--1. The redshifted CO emission has an S-shaped filamentary structure extending to the 1.2 mm continuum source 1\,B in the northeast. The blueshifted CO emission has a ring-like structure with some extended emission (``a tail'') pointing roughly toward northwest. Source N\,105--1\,A is located at the overlap region of the two velocity components, hinting on a possibility that the formation of the massive protostar is a consequence of the collision between the S-shaped filamentary and the ring-line clouds. 

The cloud-cloud collision has been shown to be an important mechanism in the formation of massive clusters and isolated high-mass stars (see \citealt{fukui2021} for a review). To date, evidence for cloud-cloud collision has been found in over 50 Galactic star-forming regions (e.g., W\,49\,N, e.g., \citealt{serabyn1993}; Sgr B2, e.g., \citealt{ginsburg2018}; G35.20$-$0.74, e.g., \citealt{dewangan2017}) and in several extragalactic star-forming regions (in the LMC, M33, and the Antennae Galaxies; e.g., \citealt{fukui2015,fukui2017}; \citealt{tokuda2019,tokuda2020}; see Table 1 in \citealt{fukui2021}).  

We investigate the collisional signatures in N\,105--1 to explore a possible origin of the O-star exciting the UC\,H\,{\sc ii} region N\,105--1\,A. We extracted the CO (3--2) position-velocity diagram (PV-diagram) along the rectangular region extending from northeast to southwest over 20$''$ ($\sim$4.8 pc) and centered on the proposed collision site (see Fig.~\ref{f:COkin}). At the position of the continuum source 1\,A, the PV-diagram shows the velocity width of the CO gas of $>$10 km s$^{-1}$, larger than predicted by the standard scaling relation (a correlation between the spherical radius $R$ in pc and line width $\sigma$ in km s$^{-1}$ of the form $\sigma \propto R^{\,\alpha}$ with $\alpha \approx 0.5$ for the Galaxy, e.g., \citealt{solomon1987}; see e.g., \citealt{wong2019} for the LMC clouds) for a cloud of a few pc in size under typical Galactic and LMC conditions. 

The S-shaped redshifted component is distributed over $\sim$4.5 pc in the CO PV-diagram with the central velocity of $\sim$242.5 km s$^{-1}$. The spatially less extended component with the broad velocity range suddenly appears close to the position of 1\,A (between the offset of $-$0.5 to 1 pc).  Such a feature in the PV-diagram has been observed in other LMC (e.g., N\,159\,W; \citealt{tokuda2019,tokuda2022}) and Galactic regions, and is thought to be a signature of the cloud-cloud collision. Theoretical simulations of a cloud-cloud collision predict that two colliding clouds appear as a single broad and continuous cloud in the PV-diagram (sometimes V-shaped) with the intermediate-velocity gas produced by the collisional interaction (e.g., \citealt{haworth2015a,haworth2015b}; see \citealt{fukui2021} for a review). 

A hub-filamentary structure of the molecular line emission observed toward N\,105--1\,A  (e.g., CS, CO, HCO$^{+}$, and CH$_3$OH; see Appendix~\ref{s:appMom}; see also Section~\ref{s:zoomin}) provides indirect evidence for the cloud-cloud collision event in the region.  The hub-filamentary structure of the molecular gas in the compressed layer between the colliding clouds is predicted by the cloud-cloud collision models in the presence of the magnetic field (see \citealt{inoue2018}).

The [S\,{\sc ii}] $\lambda\lambda$6716, 6731 {\AA} data from the Magellanic Cloud Emission Line Survey (MCELS, $\sim$2$''$ resolution; \citealt{smith1998}) provide additional indirect evidence for the cloud-cloud collision. The [S\,{\sc ii}] emission traces shock-excited gas and reveals ionization fronts. Figure~\ref{f:mcels} shows that in N\,105--1, the peak of the [S\,{\sc ii}] emission is located toward southeast from 1\,A, near the proposed site of the cloud-cloud collision. In addition to the [S\,{\sc ii}]  image,  Fig.~\ref{f:mcels} also shows the three-color mosaic combining the MCELS H$\alpha$, [O\,{\sc iii}] $\lambda$5007 {\AA}, [S\,{\sc ii}] $\lambda\lambda$6716, 6731 {\AA} images that allows us to further explore the stellar radiation feedback in the region. H$\alpha$ tracers all the ionized gas, while [O\,{\sc iii}] reveals high-excitation regions ionized by O stars. Since N\,105 is associated with the OB association (LH\,31), the previous generation of OB stars, it is not surprising that the [O\,{\sc iii}] emission is present throughout the entire region.

The CO (3--2) peak integrated intensity around 1\,A is $\sim$400 K km s$^{-1}$, which corresponds to the H$_2$ peak column density ($N_{\rm H_2,peak}$) of $2\times10^{23}$ cm$^{-2}$, assuming a CO-to-H$_2$ conversion factor in the LMC of $5\times10^{20}$ cm$^{-2}$ (K km s$^{-1}$)$^{-1}$ (e.g., \citealt{hughes2010}) and the CO (3--2)/CO (1--0) ratio of 1.  Observations show that molecular clouds traced by CO with $N_{\rm H_2,peak}>10^{23}$ cm$^{-2}$ and a collision velocity of $\sim$10 km s$^{-1}$ can form up to 10 OB-type stars (\citealt{enokiya2021}; \citealt{abe2022}). The empirical relation between $N_{\rm H_2,peak}$ and the number of formed OB stars predicts the formation of ~7 OB stars as a result of the collision of clouds with $N_{\rm H_2,peak}$ similar to that observed toward N\,105--1 (\citealt{enokiya2021}).  Only one O-type star is observed toward N\,105--1 which is inconsistent with that prediction. The compact nature of colliding clouds (a few pc in size) may explain the discrepancy, or the clouds may still be in an early stage of collision. There are several star-forming regions in the Galaxy with $N_{\rm H_2,peak}$ as large as $1\times10^{23}$ cm$^{-2}$ where only one OB star was formed as a result of the cloud-cloud collision (e.g., G24.85$+$0.09, G24.71$-$0.13, G24.68$-$0.16, \citealt{dewangan2018}; BD\,$+$40\,4124, \citealt{looney2006}); however, the relative velocities between the colliding CO clouds is lower than in N\,105--1. While not entirely consistent with the Galactic observations, it is certainly possible that an intense, localized cloud-cloud collision event could have been responsible for forming the massive O star in N\,105--1\,A, with sufficiently high luminosity to produce higher-order mm-RLs detectable at a distance of $\sim$50 kpc.

Other cloud-cloud collision signatures include a complementary distribution of gas from two clouds, as well as a U-shape of the bigger cloud participating in the collision at the final phase of the collision (see \citealt{fukui2021} for a review).  Identification of the cloud-cloud collision signatures is not always straightforward. Irregular shapes of the colliding clouds and/or the projection effects may affect the observed gas morphology and kinematics significantly, making them difficult to interpret. In general, not all cloud-cloud collision signatures are observed toward each region where such an event likely occurred (\citealt{fukui2021} and references therein). 

To further investigate the cloud-cloud collision event in N\,105--1, it will be necessary to obtain the ALMA Full Array observations (12m, 7m, and Total Power) with lower-$J$ CO transitions (tracing lower-density gas) to ensure that the molecular clouds are traced in their entirety. 

\subsubsection{Zooming-in on N\,105--1\,A: Dense Molecular Gas Distribution and Kinematics}
\label{s:zoomin}

\begin{figure*}
\centering
\includegraphics[width=0.493\textwidth]{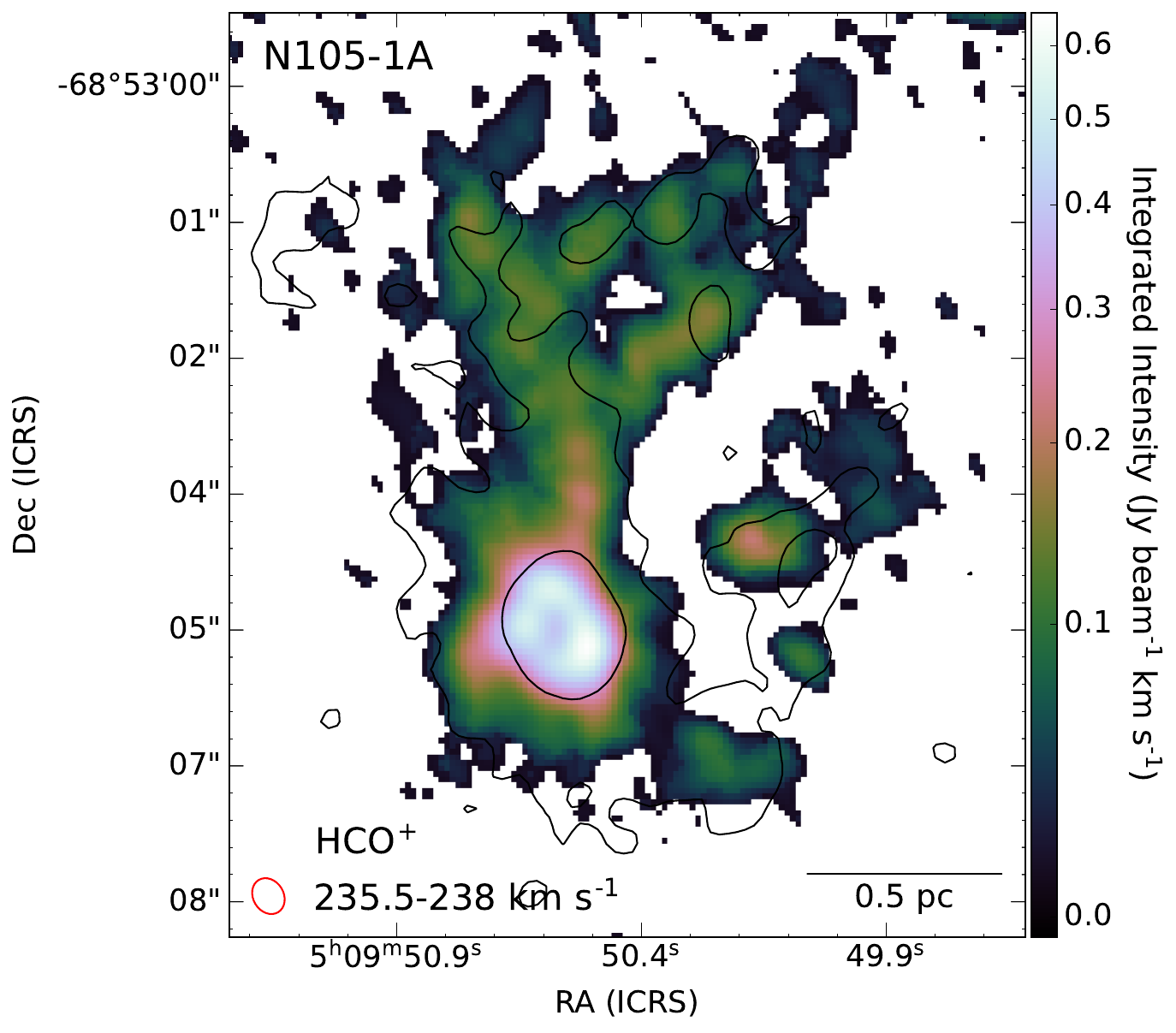}
\includegraphics[width=0.493\textwidth]{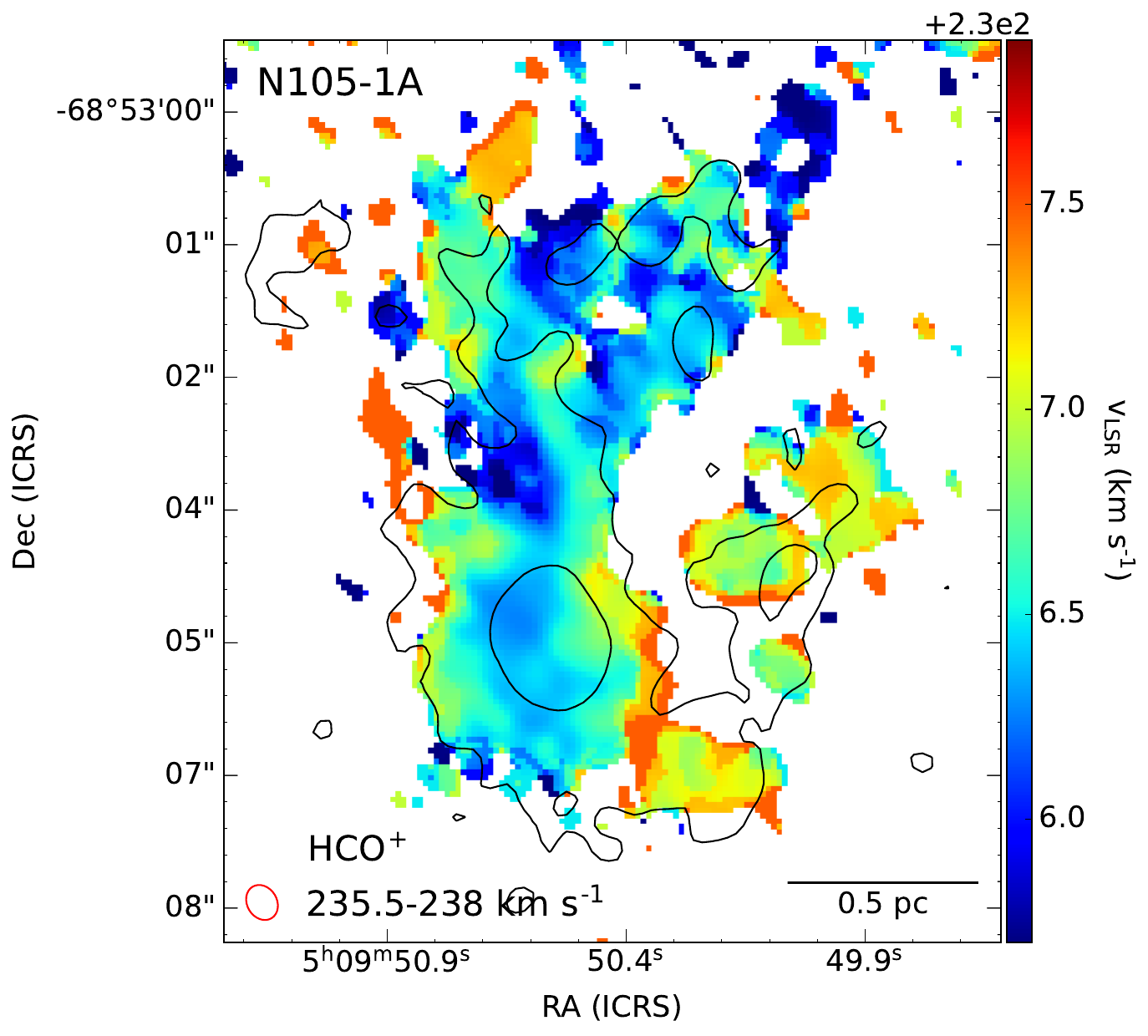}
\includegraphics[width=0.497\textwidth]{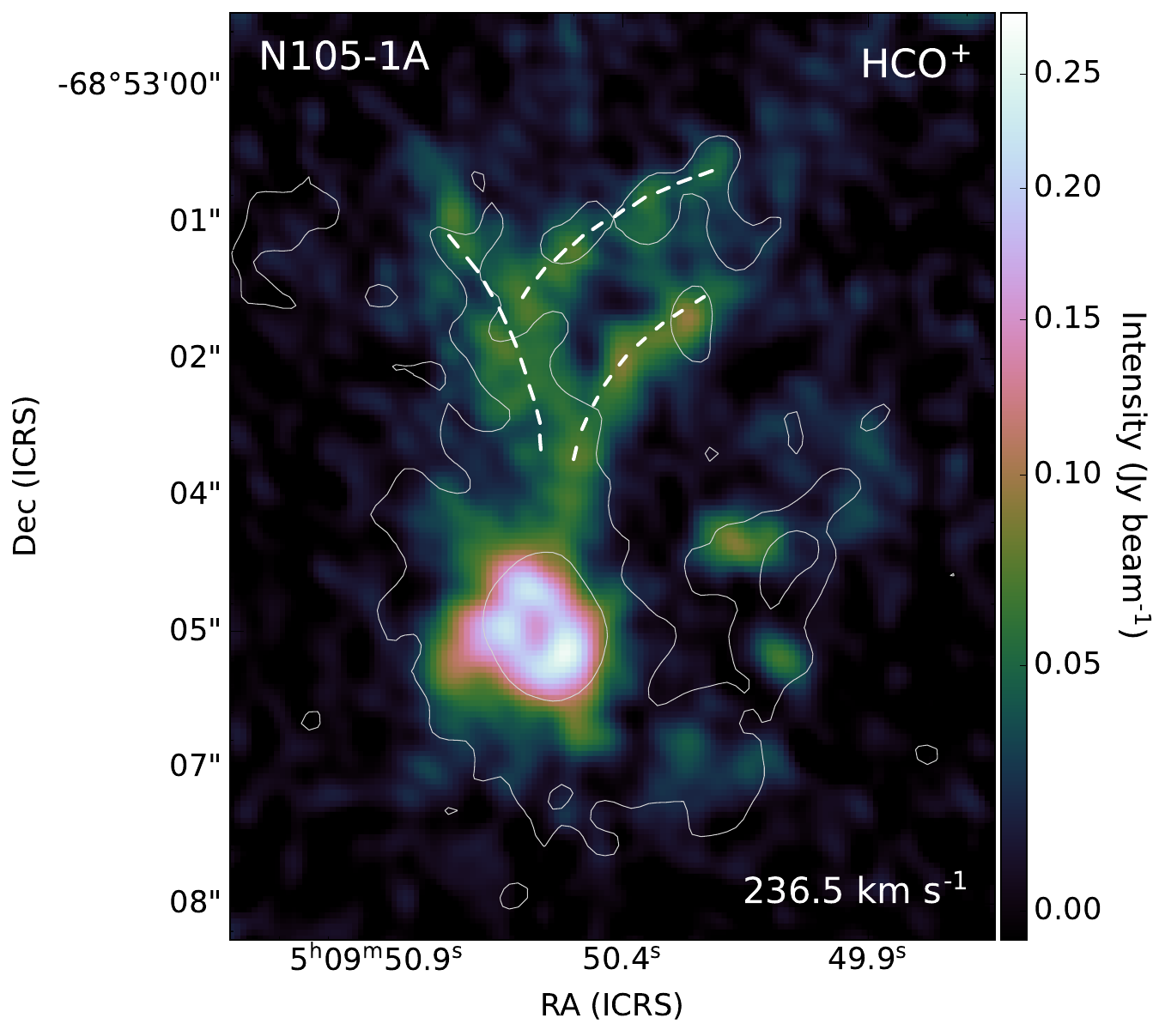}
\includegraphics[width=0.497\textwidth]{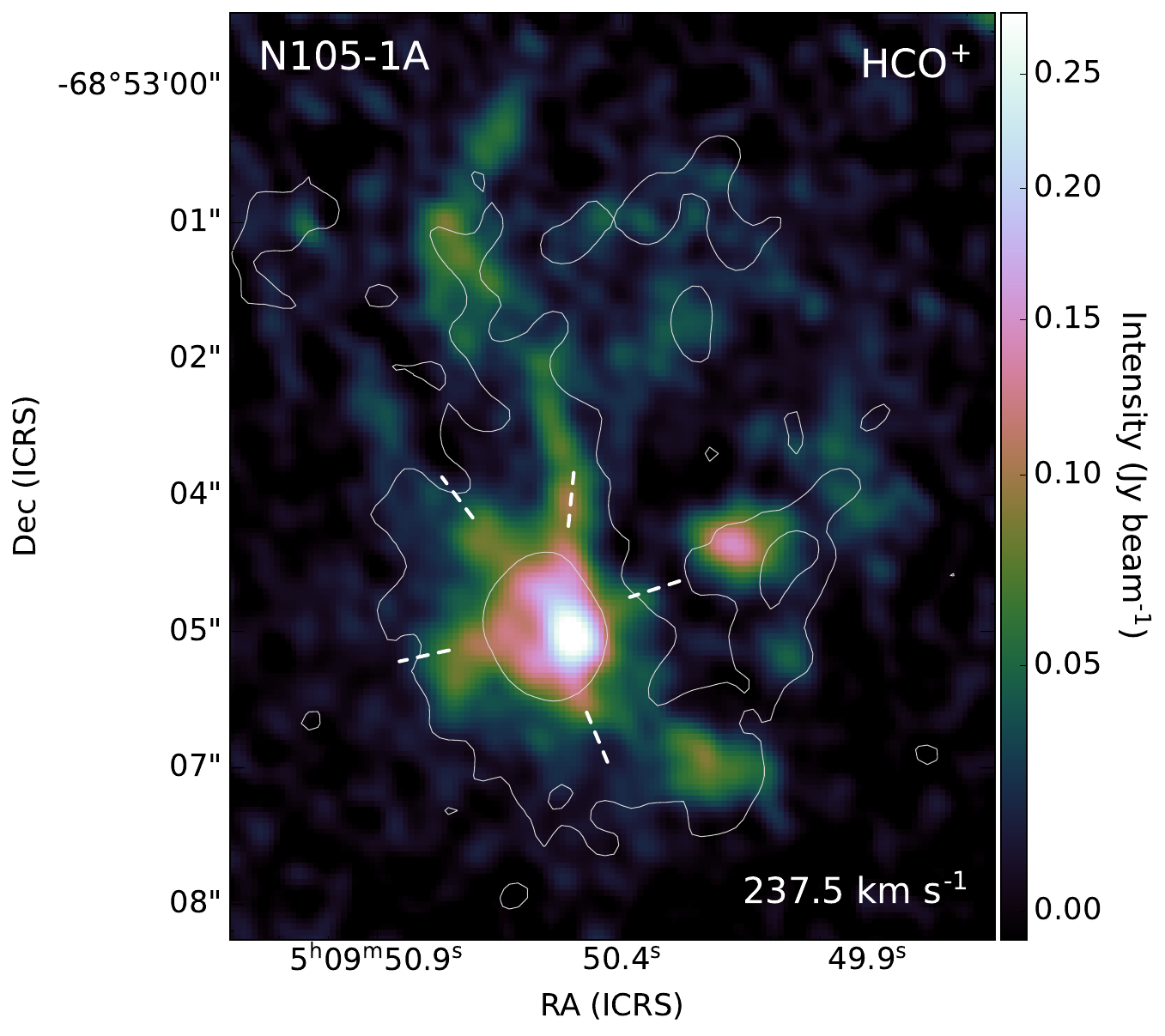}
\caption{Upper panel: The HCO$^{+}$ (4--3) integrated intensity (left) and velocity (right) images of N\,105--1\,A, incorporating the emission with velocities in the  235.5--238 km s$^{-1}$ range. Pixels corresponding to regions with the integrated intensity below 3$\sigma$ are masked in both images. Lower panel: Single channel HCO$^{+}$ (4--3) intensity images for the emission with the velocity of 236.5  km s$^{-1}$ (left) and 237.5 km s$^{-1}$ (right).  In the lower left panel, dashed lines outline the filamentary structure seen in HCO$^{+}$ and other tracers north of the mm/radio continuum peak.  In the lower right panel, dashed lines indicate possible accretion streamers. The 1.2 mm continuum contours are overlaid on all images for reference, with contour levels of (3, 30)$\sigma$.  The ALMA beam size is indicated in red in the bottom left corners of images in the upper panel.  \label{f:HCOpstreamers}}
\end{figure*}

Filaments are the prevalent feature in the images of N\,105--1\,A; their hub-like morphology provides indirect evidence that the formation of the source ionizing the UC~H\,{\sc ii} N\,105--1\,A, or a (proto)cluster it is likely a part of, was triggered by the cloud-cloud collision (Section~\ref{s:collision}). Figure~\ref{f:HCOpstreamers} shows the HCO$^{+}$ integrated intensity and velocity images, incorporating the emission in the 235.5--238 km s$^{-1}$ velocity range where the filamentary structure of the dense gas traced by HCO$^{+}$ is most evident. The filaments that are coincident with the H$\alpha$-dark region (see Section~\ref{s:1A} and Fig.~\ref{f:region}) converge in the central hub at the location of 1\,A. The emission from other molecular species also trace filaments, including CO and CS (see Appendix~\ref{s:appMom}). The CH$_3$OH emission clearly traces filaments as well, even though it is much less extended than the emission from other species (Fig.~\ref{f:momCH3OH}). Dust in the filaments is traced by the 1.2 mm continuum emission. 

In addition to large-scale filaments extending to the north from the 1\,A continuum peak, the high-resolution HCO$^{+}$ images reveal several gas streamers pointing toward the HCO$^{+}$ and the continuum emission peaks. Approximate positions of the HCO$^{+}$ filaments and streamers are outlined in the single channel images presented in Fig.~\ref{f:HCOpstreamers}. The presence of filaments and streamers indicates that the mass accretion toward the molecular clump hosting the UC H\,{\sc ii} region may still be ongoing. 

There is a strong indication in the HCO$^{+}$ data of the presence of multiple velocity components in the region that could (at least partially) explain multi-peaked and asymmetric line profiles. The HCO$^{+}$ integrated intensity (moment 0) image in Fig.~\ref{f:momHCOp} reveals three emission peaks, forming an incomplete ring roughly around the 1.2 mm continuum peak and extending over $\sim$0.37 pc (or $\sim$76,100 au), with the brightest peak to the west of the continuum peak. The velocity (moment 1) image in the same figure, shows a  sharp boundary between the higher-velocity gas in the west associated with the brightest HCO$^{+}$ emission peak and the lower-velocity gas in other directions, including the remaining two HCO$^{+}$ emission peaks. The HCO$^{+}$ channels maps in Fig.~\ref{f:HCOpchan} reveal an additional, significantly fainter lower-velocity component with the velocity peak at $\sim$236 km s$^{-1}$ located slightly to the southwest from the continuum peak. The 1.2 mm continuum peak is offset from the center of the apparent HCO$^{+}$ emission ring by $\sim$0.15 pc ($\sim$76,100 au) toward the higher-velocity component, but it is not coincident with it. 

\begin{figure*}
\centering
\includegraphics[width=0.493\textwidth]{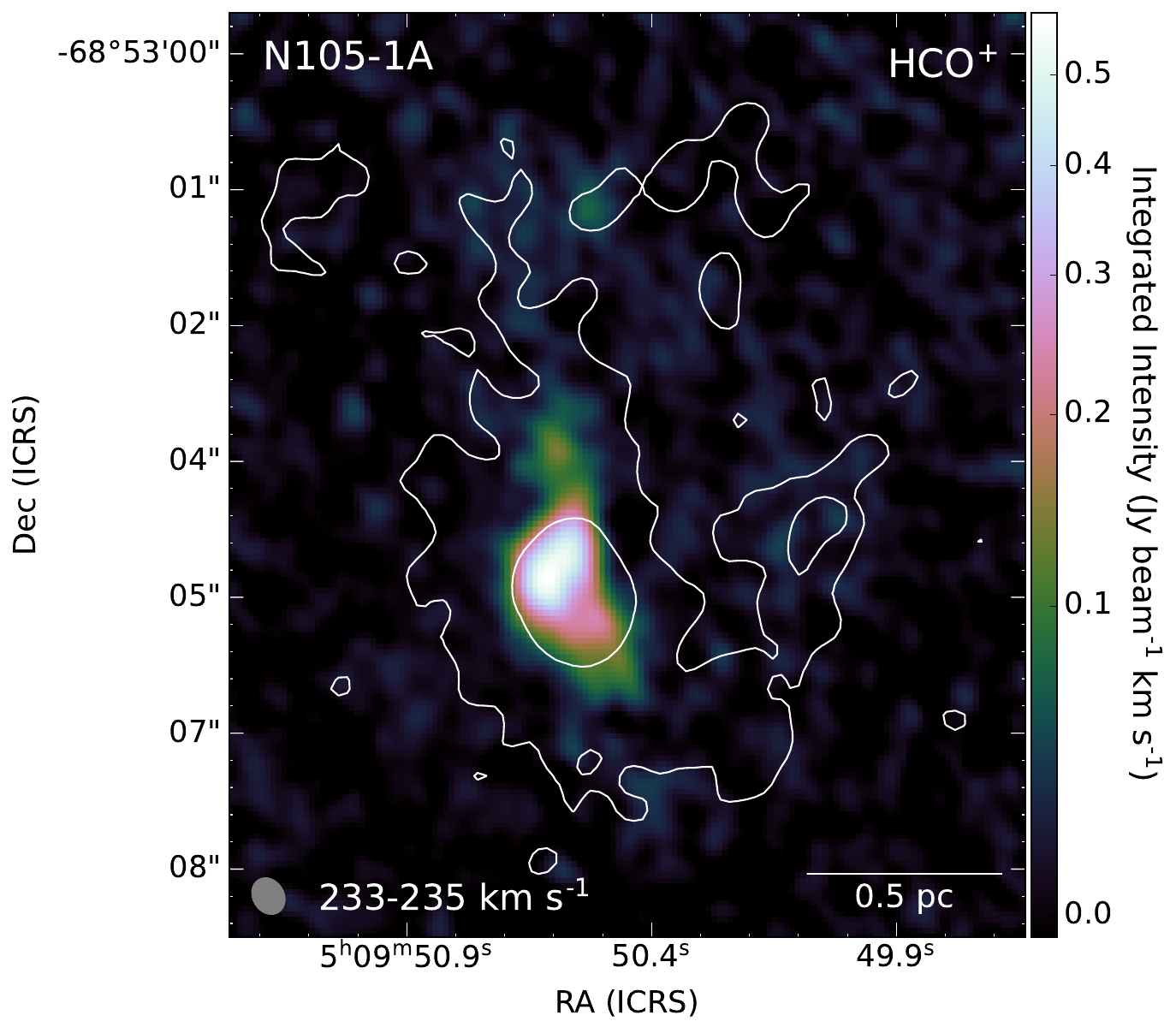}
\includegraphics[width=0.493\textwidth]{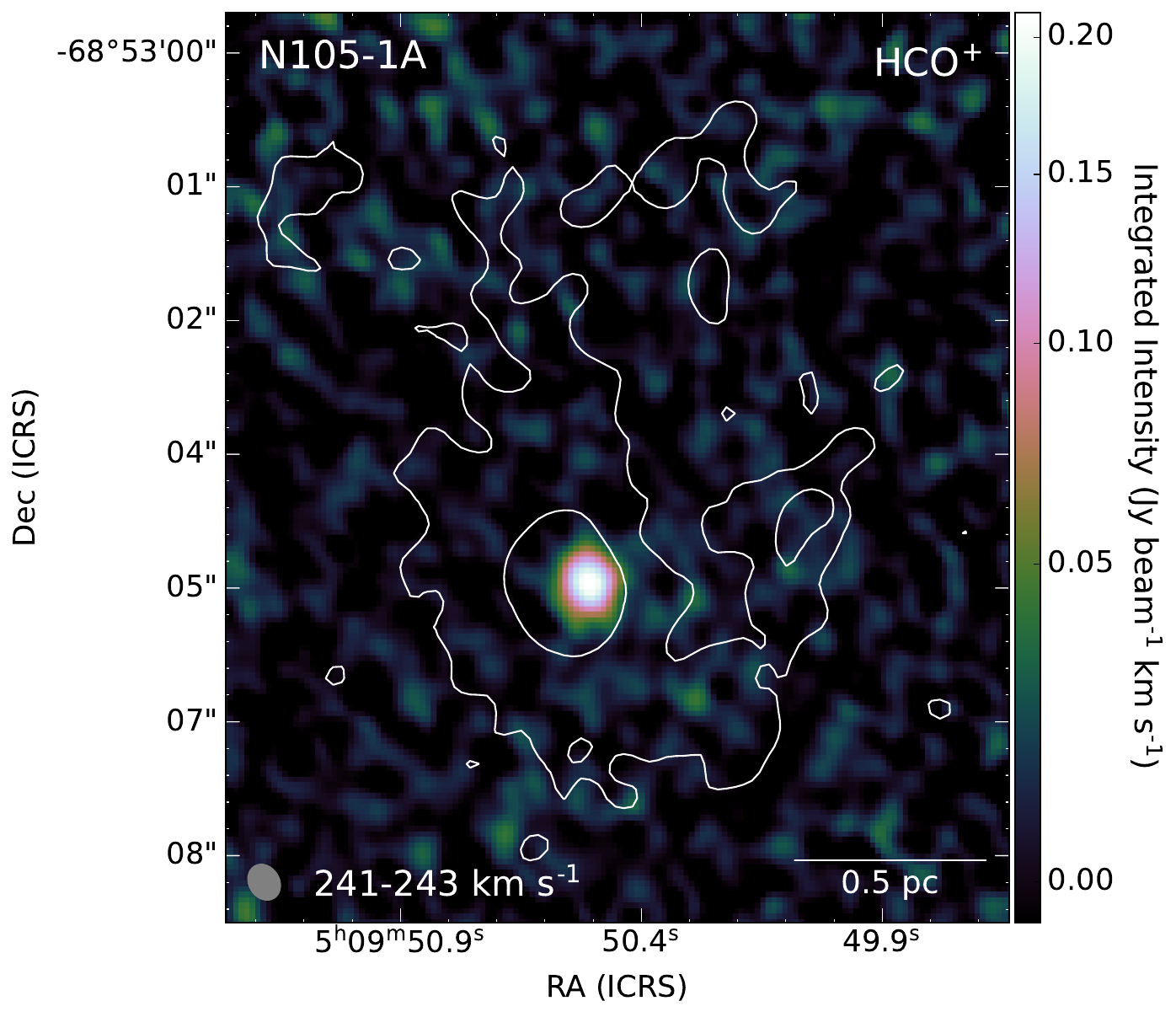}
\includegraphics[width=\textwidth]{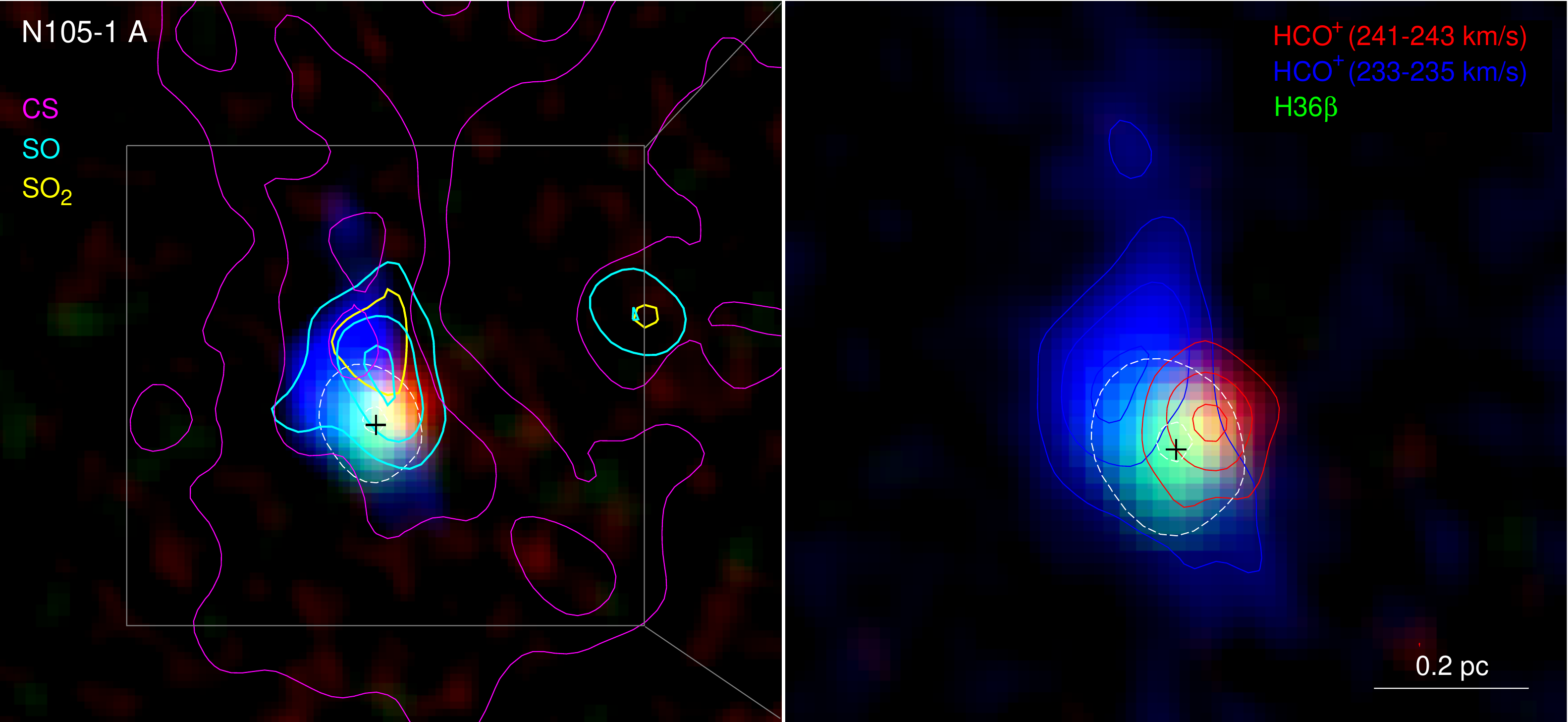}
\caption{Upper panel: The HCO$^{+}$ (4--3) integrated intensity images of N\,105--1\,A incorporating the emission with velocities in the  233--235 km s$^{-1}$ (left) and 241--243 km s$^{-1}$ (right) ranges, corresponding to the blueshifted and redshifted CO velocity components, respectively, shown in Fig.~\ref{f:COkin}. Lower left panel:  Three-color mosaic combining the HCO$^{+}$ integrated intensity images shown in the upper panel (for the velocity range 241--243 km s$^{-1}$ in red and for 233--235 km s$^{-1}$ in blue), and the H36$\beta$ recombination line (green) images. Solid magenta/cyan/yellow contours correspond to the CS/SO/SO$_2$ emission with contour levels of (20, 50, 90)\%/(20, 50, 90)\%/50\% of the CS/SO/SO$_2$ emission peak (\citealt{sewilo2022n105}).  Dashed white contours correspond to the H36$\beta$ emission with contour levels of (20, 90)\% of the H36$\beta$ emission peak.  Lower right panel: The same image as in the lower left panel, zoomed in on the central region.  Blue/red contours correspond to the HCO$^{+}$ blueshifted/redshifted emission with contour levels of (20, 90)\% times the corresponding emission peak. Dashed white contours are the same as in the lower left panel. The black cross symbol in both images in the lower panel indicate the position of the 1.2 mm continuum peak. \label{f:2velcomp}} 
\end{figure*}

We highlight the lower- and higher-velocity components observed toward 1\,A in Fig.~\ref{f:2velcomp} in two HCO$^{+}$ integrated intensity images, incorporating the emission from the velocity ranges 233--235 km s$^{-1}$ (blueshifted) and 241--243 km s$^{-1}$ (redshifted). The lower- and higher-velocity components are well-separated spatially and kinematically in these images.  On the larger scale, the velocity ranges 233--235 km s$^{-1}$ and 241--243 km s$^{-1}$ separate two CO velocity components - the ring-like and S-shaped clouds, respectively (see Fig.~\ref{f:COkin}).  The CO emission peaks corresponding to the HCO$^{+}$ peaks in Fig.~\ref{f:2velcomp} can easily be identified in Fig.~\ref{f:COkin}, indicating that the lower-velocity/higher-velocity gas traced by HCO$^{+}$ is associated with the ring-like/S-shaped cloud traced by CO.

In the lower panel in Fig.~\ref{f:2velcomp}, we compare the blueshifted and redshifted HCO$^{+}$ velocity components to the distribution of the ionized gas emission traced by H36$\beta$, and to the SO, SO$_2$, and CS emission in the region. The images clearly show the offset between the ionized gas/1.2 mm continuum and the HCO$^{+}$ peaks; the latter are shifted to the north and are associated with the SO, SO$_2$, and CS emission peaks.

\subsubsection{Accretion Shocks and Sulfur Chemistry}
\label{s:shocksS}

The spatial distribution and abundances of the S-bearing species (SO$_2$ and SO in particular) support the hypothesis that the accretion from filaments is still ongoing in N\,105--1\,A. The emission originating from warm ($T_{\rm ex}>50$ K) SO and SO$_2$ has been suggested as a good tracer of accretion shocks at the interface between the disk and the envelope in protostellar systems (e.g., \citealt{sakai2017}; \citealt{oya2019}; \citealt{artur2019}). Under comparable physical conditions, the same chemistry may originate at the interface between the filament/streamer and the clump/core where the infalling material may cause a shock, raising the temperature of gas and dust and drive endoergic chemical processes that would enhance the abundance of SO and SO$_2$. The distribution of the SO and SO$_2$ emission in N\,105--1\,A, i.e., the offset to the north from the continuum peak toward the region where the filaments seem to converge, and the high SO$_2$ temperature derived by \citet{sewilo2022n105} at this location ($\sim$100 K, see Section~\ref{s:1A}) indicate that such chemistry may be at work toward this source. 

To investigate this possibility, we compared our data to the predictions of models of nonmagnetized, irradiated $J$-type protoplanetary disk accretion shocks presented by \citet{vangelder2021}. These models were computed using the Paris-Durham shock code \citep{flower2003} and considered preshock densities, $n_0$ ($= n_{\rm H} = 2\,n({\rm H_2}) + n({\rm H}) + n({\rm H^{+}})$) of $10^{5}-10^{8}$ cm$^{-3}$, shock velocities $V_{\rm s}$ in the range 1--10 km s$^{-1}$, and irradiation by multiples of the (Mathis) UV radiation field, $G_0$. \citet{vangelder2021} discussed the viable formation routes for warm gas-phase SO and SO$_2$ (and the resulting abundances) under variation of $n_0$, $V_{\rm s}$, and $G_0$.

For the peak H$_2$ column density of $2.0\times10^{23}$ cm$^{-2}$ determined based on the CO observations (see Section~\ref{s:collision}) and the source size of 0.13 pc (Section~\ref{s:galcomp}), we estimate $n({\rm H_2})$ toward 1\,A of $\sim$$5\times10^{5}$ cm$^{-3}$, assuming that the extent of the source along the line of sight is the same as in the plane of the sky. In molecular clouds, hydrogen is predominantly in a molecular form, thus $n_{\rm H}\approx2\,n({\rm H_2})$, resulting in $n_{\rm H}$ toward 1\,A of $\sim$10$^{6}$ cm$^{-3}$, within the range of $n_0$ explored by \citet{vangelder2021}. 

To compare our results to the shock models of \citet{vangelder2021}, we calculate the ratio of the SO$_2$ and SO column densities ($N_{\rm SO_2}$ and $N_{\rm SO}$, respectively) toward 1\,A. $N_{\rm SO_2}$ and $N_{\rm SO}$ are $(1.6^{+0.3}_{-0.2})\times10^{14}$ cm$^{-2}$ and $(3.5^{+0.4}_{-0.3})\times10^{14}$ cm$^{-2}$, respectively, resulting in the $N_{\rm SO_2}$/$N_{\rm SO}$ ratio of 0.46$^{+0.4}_{-0.3}$ (\citealt{sewilo2022n105}).   For a UV radiation field strength of $G_0=1$ (in units of the Mathis interstellar radiation field, ISRF), this ratio implies low preshock densities of $n_0<10^6$ cm$^{-3}$ and shock speeds of either $V_{\rm s}<2$ km s$^{-1}$ or $V_{\rm s}>6.5$ km s$^{-1}$ (Fig. 10 of \citealt{vangelder2021}). At $G_0=100$, the $N_{\rm SO_2}$/$N_{\rm SO}$ ratio observed toward 1\,A can again be similarly reproduced with either $n_0<10^6$ cm$^{-3}$ and $V_{\rm s} <2$ km s$^{-1}$ or $n_0>10^7$ cm$^{-3}$ and $V_{\rm s}>6$ km s$^{-1}$.

We only consider the $N_{\rm SO_2}$/$N_{\rm SO}$ ratio to compare our observations to the shock model predictions. We do not compare the absolute values of $N_{\rm SO_2}$ and $N_{\rm SO}$ measured toward 1\,A to the maximum values predicted by the \citet{vangelder2021}'s shock models for different physical conditions because of the differences in elemental abundances between the Galaxy and the LMC. The abundance of atomic S and O in the LMC is lower than in the Galaxy by a factor of 2.6 and 2.2, respectively (\citealt{acharyya2015} and references therein). However, as discussed in \citet{vangelder2021}, for $n_0 \lesssim 10^6$ cm$^{-3}$, the abundances of both SO and SO$_2$ drop in a similar way in the environment with the reduced abundance of atomic S and O. The abundance of atomic C (2.5 lower in the LMC) is also relevant since it reacts with SO, forming CS, and thus is a main destruction pathway for this molecule (e.g., \citealt{hartquist1980}). Decreasing the abundance of atomic C results in higher abundances of both SO and SO$_2$ in shocks. 

Overall, our measured $N_{\rm SO_2}$/$N_{\rm SO}$ ratio in 1\,A is consistent with shocks propagating with $V_{\rm s}>6.5$ km s$^{-1}$ in gas with low preshock densities ($n_0<10^6$ cm$^{-3}$). \citet{sewilo2022n105} derived the SO$_2$ temperature toward the 1.2 mm continuum peak / SO$_2$ peak in N\,105--1\,A of $(96\pm20)$ K / $(98\pm20)$ K, indicating that the observed composition of 1\,A could be explained by simple models of postevaporation chemistry. For shocks propagating with velocities $\gtrsim$4 km s$^{-1}$, both SO and SO$_2$ are efficiently formed in the gas phase. The OH radical is crucial for the formation of SO and SO$_2$ through the route:

$\rm S \buildrel OH \over \longrightarrow SO \buildrel OH \over \longrightarrow SO_2 $. 

It is formed through photodissociation of H$_2$O that was formed before the shock cooled down to below $\sim$300 K. It can also be efficiently produced in shocks through the endothermic reaction between H$_2$ and atomic O. Atomic S, as well as SH, can be produced in hydrogen atom abstraction reactions with H$_2$S in hot gas following the thermal desorption of H$_2$S ice from grain mantles into the gas \citep{charnley1997}: 

$\rm H_2S \buildrel H \over \longrightarrow SH \buildrel H \over \longrightarrow S $.

The reactions between SH and atomic O further increase the abundance of SO and SO$_2$: 

$\rm SH \buildrel O \over \longrightarrow SO \buildrel OH \over \longrightarrow SO_2 $.

The abundance of radicals such as OH and SH depends not only on the temperature in the shock, but also on the strength of the local UV field. The UV radiation is stronger in the LMC than in the Galaxy and locally in 1\,A, there is an additional contribution from the O-type star.However,  although the source is deeply embedded and the effects of the strong UV radiation are somewhat diminished, the UV radiation may still be strong enough to promote the efficient formation of SO and SO$_2$.

The nondetection of the SiO (6--5) transition toward 1\,A by \citet{sewilo2022n105} is consistent with the \citet{vangelder2021}'s shock models since they do not include high-velocity shocks powerful enough to destroy dust grains and release the Si atoms to the gas, making them available for chemical reactions and leading to the formation of SiO (e.g., \citealt{schilke1997}; \citealt{gusdorf2008}; \citealt{sanchezmonge2013}). SiO may form (less efficiently) by grain-grain interactions in higher-density regions (e.g., \citealt{guillet2007}); however, the higher-excitation SiO transitions have been found to originate in the high velocity gas (e.g., \citealt{leurini2014}). 

\citet{vangelder2021} do not predict any significant increase in the CH$_3$OH abundance except for the highest preshock densities above $\sim$10$^8$ cm$^{-3}$ and $V_{\rm s}\gtrsim10$ km s$^{-1}$ where ice mantles containing CH$_3$OH can be thermally desorbed from warm dust grains. Our observations are consistent with this prediction since only faint CH$_3$OH lines have been detected toward the 1.2 mm continuum and SO$_2$ peaks in 1\,A. The CH$_3$OH emission is cold with the temperature of $\sim$12 K/$\sim$22 K toward the continuum/SO$_2$ peak \citep{sewilo2022n105}, thus its origin by thermal desorption from ices is unlikely. The possible origin of both methanol and SO$_2$ emission in the star-forming region is discussed in detail in \citet{sewilo2022n105}. 

\subsubsection{Shock Activity vs. Photoionization}
\label{s:shocksphot}

We explore other shock tracers covered by our observations and those available in the literature to search for strong shock activity toward N\,105--1\,A and identify potential contribution(s) from the shock ionization to the observed ionized gas emission. \citet{oliveira2006} suggested the presence of the outflow in 1\,A based on the Br$\alpha$ data (see Section~\ref{s:1A}), but we do not find any compelling evidence supporting their conclusion.  Out of all available data discussed in this paper, CO (3--2) is the best outflow tracer.  The CO line profiles (or those for any other species) do not show high-velocity ($>$10 km s$^{-1}$) wings, the outflow signature. Other classical diagnostics of shocks include the grain-destruction products (SiO and S-bearing species such as SO$_2$ and SO; e.g., \citealt{caselli1997}; \citealt{schilke1997}; \citealt{gusdorf2008}) and grain sputtering products (e.g., CH$_3$OH; e.g., \citealt{jorgensen2020} and references therein). As mentioned above, the observations of SiO (6--5) toward 1\,A resulted in a non-detection \citep{sewilo2022n105}. Faint CH$_3$OH emission is detected south of the continuum peak, but it is cold and not in the outflow direction proposed by \citet{oliveira2006}. 
 
The optical [S\,{\sc ii}] $\lambda\lambda$6716, 6731 {\AA} emission (MCELS, see Section~\ref{s:collision}) is a high-velocity shock tracer; however, a non-detection toward 1\,A can be the result of high extinction. Near-IR molecular hydrogen emission lines that are thermally excited through shocks, are also commonly used as outflow tracers. However, H$_2$ lines are also excited by fluorescence through UV photons from OB stars, therefore additional information is often needed to confirm the origin of the H$_2$ emission. \citet{sewilo2022n105} reported the detection of the H$_2$ v = 1--0\,S(1) line at 2.12 $\mu$m toward 1\,A with the Very Large Telescope/$K$-band Multi Object Spectrograph ({\it VLT}/KMOS).  The H$_2$ emission is observed over the entire $2\rlap.{''}8\times2\rlap.{''}8$ field of view with the bright peak coincident with the ionized gas emission peak (at the continuum peak), indicating that excitation from UV photons is the dominant excitation mechanism at this location. Due to a lack of other evidence for the presence of the outflow, it is unlikely that the extended H$_2$ emission detected toward 1\,A traces high-velocity shocks.  The H$_2$ 2.12 $\mu$m emission gets brighter toward the west of the KMOS field that, similarly to the region of the peak emission, may be the results of the photoexcitation from nearby OB stars. 

The spectra discussed in \citet{sewilo2022n105} extracted from the area around the 1\,A continuum emission peak (as the mean within the contour corresponding to 50\% of the 1.2 mm continuum emission peak) and around the SO$_2$ emission peak  (as the mean within the contour corresponding to 50\% of the SO$_2$ emission peak) show rather narrow molecular lines ($\sim$3--5 km s$^{-1}$; including H$^{13}$CO$^{+}$, SO$_2$, SO, CS, CH$_3$OH, see Appendix~\ref{s:appA}), indicating that no significant bulk motions are present.  However, all the molecular and hydrogen recombination lines have asymmetric profiles often with multiple peaks, depending on the position within N\,105--1\,A. Figure~\ref{f:profiles} shows examples of the CO, HCO$^{+}$, H$^{13}$CO$^{+}$, and H40$\alpha$ lines in several locations in 1\,A.  The shape of line profiles in some regions suggest the presence of an infall (two line peaks with the brighter one at lower velocity); however, the overall distribution of line profiles is inconsistent with the large-scale infall motions. It is possible that the complex kinematics at the site of the cloud-cloud collision, the projection effects, or unresolved multiple sources (or a combination of thereof) make it difficult to disentangle signatures of different kinematic motions.


\section{Summary and Conclusions}
\label{s:summary}
Using the ALMA Band 3 and Band 6 observations, we detected the H40$\alpha$, H36$\beta$, H50$\beta$, H41$\gamma$, H57$\gamma$, H49$\epsilon$, H53$\eta$, H54$\eta$, and tentatively H55$\theta$ mm-RLs toward the 1.2 mm continuum source N\,105--1\,A in the N\,105 star-forming region in the LMC. The detection of the hydrogen $\gamma$, $\epsilon$, and $\eta$ mm-RLs in the LMC constitutes the first extragalactic detection of these higher-order transitions. The $\eta$-transitions are detected for the first time at millimeter wavelengths.

The (H40$\alpha$, H50$\beta$, H57$\gamma$)/(H36$\beta$, H41$\gamma$, H49$\epsilon$, H53$\eta$, H54$\eta$, H55$\theta$) Band 3/6 observations with the spatial resolution of $\sim$2$\rlap.{''}$1/$\sim$0$\rlap.{''}$49 probe the $\sim$0.51 pc/$\sim$0.12 pc scales, similar to single-dish studies on Galactic massive star-forming regions. 

\begin{packed_enum}
\item[--]  We performed the LTE and non-LTE analysis of the mm-RLs with $\Delta n\leq8$ covered by our observations of N\,105--1\,A using the XCLASS software. We were unable to obtain reliable measurements of the physical properties of the source ($T_{\rm e}$ and EM) in any of the three sets of mm-RLs we modeled  (Band 3 only, Band 6 only, and all mm-RLs across both bands). The XCLASS analysis showed that the observed mm-RL intensities are consistent with those predicted by the LTE model, and revealed considerable beam dilution effect in the Band 3 observations. 

\item[--]  Using the H40$\alpha$ line and the 3 mm continuum data, we measured the LTE electron temperature of $\sim$10,900 K toward N\,105--1\,A.  This temperature is similar to the electron temperature observed toward Galactic H\,{\sc ii} regions at large Galactocentric distances ($\sim$18 kpc) where the oxygen abundance / metallicity is comparable to that found in the LMC. 

\item[--]   We determined the number of Lyman continuum photons ($N_{\rm Ly}$) required to ionize the source of $\sim$$1.3\times10^{49}$ s$^{-1}$ toward N\,105--1\,A, based on the derived $T_{\rm e}$ and the 3 mm free-free continuum emission.  This value of  $N_{\rm Ly}$ corresponds to the star with the O5.5 V spectral type with the luminosity of $2.6\times10^5$ $L_{\odot}$ and mass of 34.4 $M_{\odot}$ \citep{martins2005}.

\item[--]  The physical properties of N\,105--1\,A: FWHM size of $(0.13\pm0.02)$ pc, EM of $(8.9\pm1.3)\times10^7$ pc cm$^{-6}$, and $n_{\rm e}$ of $(2.6\pm0.3)\times10^4$ cm$^{-3}$, are within the size, EM, and $n_{\rm e}$ ranges observed toward a sample of 178 Galactic compact and UC H\,{\sc ii} regions with the detection of (sub)mm-RLs \citep{kim2017,kim2018}. The physical properties of 1\,A and its association with the IR and radio emission are consistent with it being an UC H\,{\sc ii} region.

\item[--]  The H40$\alpha$ line integrated intensity observed toward 1\,A is among the highest (for both H40$\alpha$ and H$n\alpha$, the stacked $\alpha$-transitions) reported for Galactic H\,{\sc ii} regions by \citet{kim2017}.  The sources with the highest values of the integrated H40$\alpha$ and/or H$n\alpha$ line intensities are among the most luminous in the \citet{kim2017} sample ($L_{\rm bol} > 10^5$ $L_{\odot}$).  They include well-known massive star-forming regions such as the central part of W\,49, G34.26$+$0.15, and G10.62$-$0.38 (unresolved in \citealt{kim2017}'s observations). 

\item[--]  The results of the SED fitting with the \citet{robitaille2017}'s 2D radiative transfer YSO SED models indicate an earlier evolutionary stage for 1\,A where both the circumstellar envelope and the disk are present (Class I YSO).  It is possible that 1\,A is in fact an unresolved protocluster, containing the UC H\,{\sc ii} region (the most massive star that is evolving the fastest), and one or more less massive protostars. We find no clear outflow signatures in the available data (including reliable outflow tracers), but our view may be obscured by geometry and overlapping velocity components. 

\item[--]  There is evidence for large scale motions of the ionized gas in N\,105--1\,A, including broadened and asymmetric RL profiles, and the H40$\alpha$ velocity gradient.  However, we are unable to unambiguously identify a single phenomenon responsible for the observed kinematic structure. In the absence of outflow signatures, the core rotation and/or higher-velocity ionized gas flows in the region are likely candidates. 

\item[--] The high-resolution CO (3--2) data reveal two distinct velocity components in N\,105--1 that appear to intersect at the position of 1\,A, hinting on a possibility that star formation in the region was triggered by the cloud-cloud collision. Such a scenario is supported by the CO gas distribution in the PV-diagram, a high relative velocity between the clouds ($>$10 km s$^{-1}$),  the filamentary structure of the molecular gas observed in multiple tracers, and the presence of the shocked gas (traced by [S\,{\sc ii}]) at a proposed collision site. 

\item[--]  The HCO$^{+}$ data reveal multiple velocity components in N\,105--1\,A that could explain multi-peaked and asymmetric line profiles observed for multiple molecular species.  We identified three low-velocity peaks and one high-velocity HCO$^{+}$ peak in 1\,A with velocities consistent with the two (blueshifted and redshifted) colliding CO clouds.  

\item[--] We identified HCO$^{+}$ gas filaments and streamers toward N\,105--1\,A indicating an ongoing accretion onto the clump harboring the UC H\,{\sc ii} region and (likely) an associated embedded (proto)cluster, formed as a result of the cloud-cloud collision.  Chemistry of S-bearing species (SO and SO$_2$) observed toward 1\,A is consistent with the accretion shock model predictions.  

\item[--] Our observations demonstrate that ALMA's high resolution and high sensitivity enable detailed Galactic-like studies of compact and UC H\,{\sc ii} regions and their molecular environments outside the Galaxy.

\end{packed_enum}

\begin{acknowledgements}
We thank the anonymous referee for the constructive report and insightful comments that helped us improve the manuscript. The material is based upon work supported by NASA under award number 80GSFC21M0002. We thank Robert Gruendl and You-Hua Chu for providing the MCELS2 H$\alpha$ image of N\,105, and Sean Points for providing the flux-calibrated, continuum-subtracted MCELS H$\alpha$, [S\,{\sc ii}], and [O\,{\sc iii}] images of the LMC. S.B.C. acknowledges support from the NASA Planetary Science Division Internal Scientist Funding Program through the Fundamental Laboratory Research work package (FLaRe).  K.T. was supported by a NAOJ ALMA Scientific Research grant No. 2022-22B, Grants-in-Aid for Scientific Research (KAKENHI) of Japan Society for the Promotion of Science (JSPS; grant Nos. JP21H00049, JP20H05645, and 21K13962). A.S.M. acknowledges support from the RyC2021-032892-I grant funded by MCIN/AEI/10.13039/501100011033 and by the European Union `Next GenerationEU'/PRTR, as well as the program Unidad de Excelencia Mar\'ia de Maeztu CEX2020-001058-M. The National Radio Astronomy Observatory is a facility of the National Science Foundation operated under cooperative agreement by Associated Universities, Inc. This paper makes use of the following ALMA data: ADS/JAO.ALMA\#2019.1.01720.S, \\ ADS/JAO.ALMA\#2017.1.00093.S, and \\ADS/JAO.ALMA\#2019.1.01770.S.  ALMA is a partnership of ESO (representing its member states), NSF (USA) and NINS (Japan), together with NRC (Canada), NSC and ASIAA (Taiwan), and KASI (Republic of Korea), in cooperation with the Republic of Chile. The Joint ALMA Observatory is operated by ESO, AUI/NRAO and NAOJ.  
\end{acknowledgements}

\clearpage 


\appendix 
\counterwithin{figure}{section}


\section{ALMA Band 3 and Band 6 Spectra of N\,105--1\,A }
\label{s:appA}

\begin{figure*}
\centering
\includegraphics[width=\textwidth]{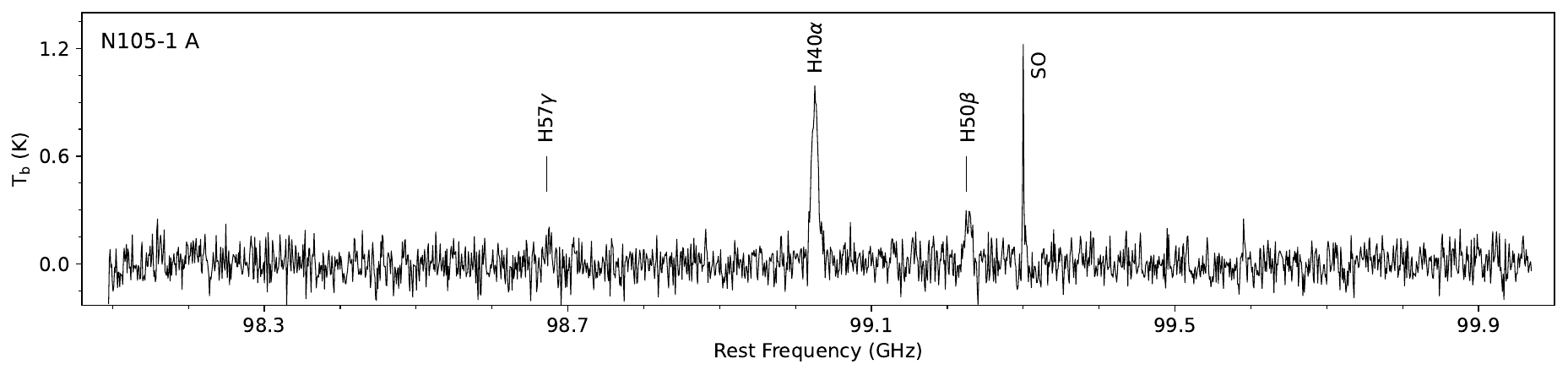}
\caption{ALMA Band 3 spectrum of N\,105--1\,A: a full spectral window covering hydrogen recombination lines. The detected spectral lines are labeled. \label{f:specB3}}
\end{figure*}

In Fig.~\ref{f:specB3}, we show the ALMA Band 3 spectrum of N\,105--1\,A covering the H57$\gamma$, H40$\alpha$, and H50$\beta$ recombination lines (see Table~\ref{t:fitRLs}), and the SO $^{3}\Sigma$ 3$_2$--2$_1$ line with the rest frequency of 99,299.87 MHz and the upper energy level ($E_{\rm U}$) of 9.2 K. 

\begin{figure*}[ht!]
\centering
\includegraphics[width=\textwidth]{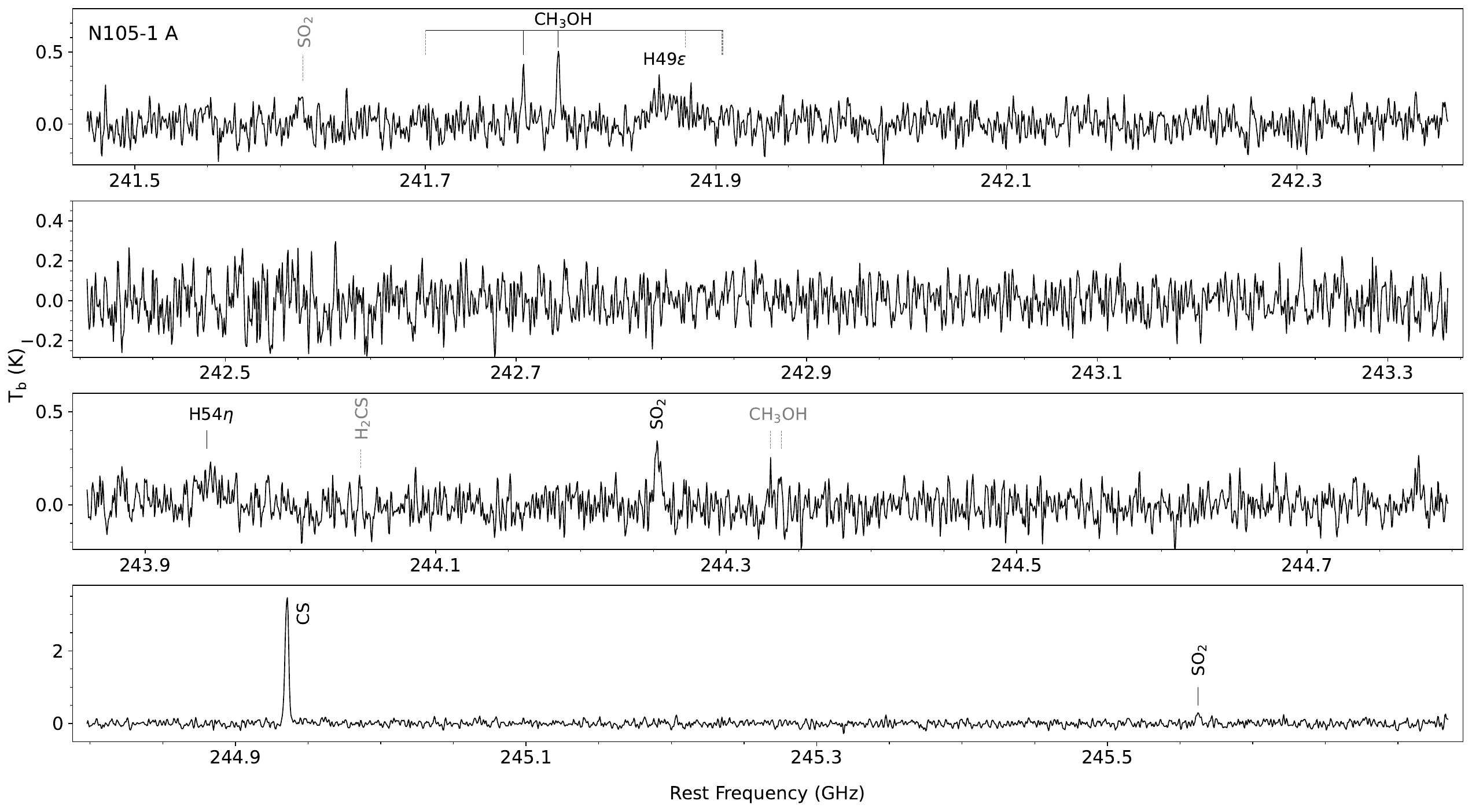}
\includegraphics[width=\textwidth]{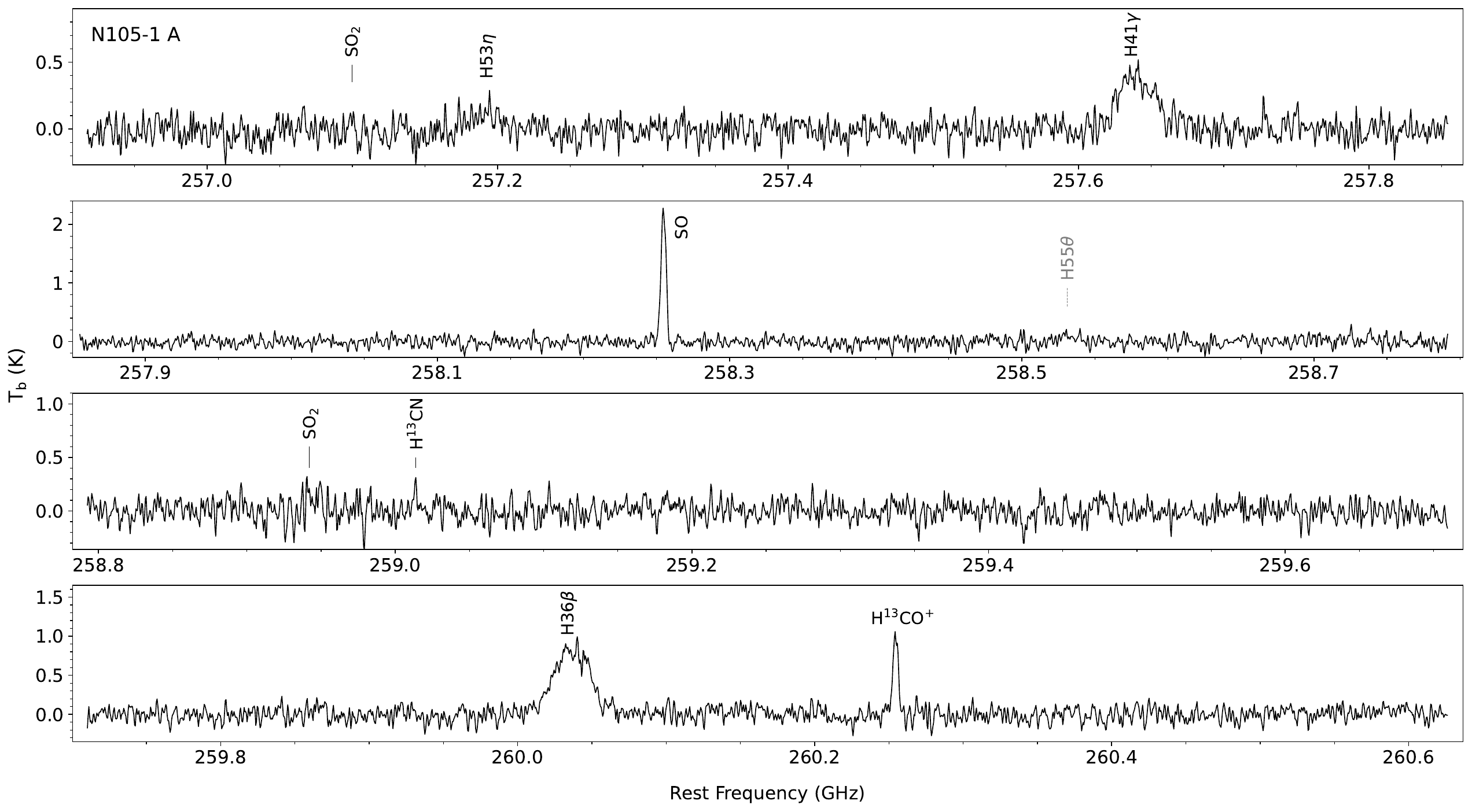}
\caption{ALMA Band 6 spectra of N\,105--1\,A. The detected ({\it black}) and tentatively detected ({\it gray}) spectral lines are labeled.  The results of the molecular line analysis are presented in \citet{sewilo2022n105}. \label{f:specB6}}
\end{figure*}

For completeness, we provide the ALMA Band 6 spectra of N\,105--1\,A in Fig.~\ref{f:specB6} that were analyzed in detail in \citet{sewilo2022n105}. The H$^{13}$CO$^{+}$ (3--2), H$^{13}$CN (3--2), CS (5--4), SO $^{3}\Sigma$ 6$_6$--5$_5$, and multiple transitions of CH$_3$OH and SO$_2$ (see Table 3 in \citealt{sewilo2022n105}) are detected toward 1\,A.   The H$_2$CS 7$_{1,6}$--6$_{1,5}$ line is tentatively detected.  Hydrogen recombination lines detected (H36$\beta$, H41$\gamma$, H49$\epsilon$, H53$\eta$, and H54$\eta$) and tentatively detected (H55$\theta$) in our Band 6 observations are indicated.

\clearpage

\section{XCLASS Modeling: Technical Details, Example Model Fit and Error Estimation}
\label{app:xclass}

As mentioned in Section~\ref{s:xclass}, a contribution of each mm-RL to a model spectrum is described in XCLASS by a certain number of components with each component characterized by a set of physical parameters. These are the source size $\theta_{\rm source}$ (in~arcsec), electron temperature $T_{\rm e}$ (K),  emission measure EM (pc~cm$^{-6}$),  line width $\Delta v$ (km~s$^{-1}$),  velocity offset $v_{\rm offset}$ (km~s$^{-1}$), and the position along the line of sight. 

XCLASS models the mm-RLs/RRLs by solving the 1D radiative transfer equation assuming LTE conditions and an isothermal source,
\begin{eqnarray}\label{myXCLASS:modelFirstDist}
  T_{\rm mb}(\nu) = \sum_{m,c \in i} &\Bigg[\eta \left(\theta_{\rm source}^{m,c}\right) \left[S^{m,c}(\nu) \left(1 - e^{-\tau_{\rm total}^{m,c}(\nu)}\right)\right. \\
  &\left. \,\, + I_{\rm bg} (\nu) \left(e^{-\tau_{\rm total}^{m,c}(\nu)} - 1\right) \right] \Bigg]\nonumber\\
  & \,\, + \left(I_{\rm bg}(\nu) - J_\mathrm{CMB} \right),\nonumber
\end{eqnarray}
where the sums go over the indices $m$ for molecule / atom, and $c$ for component, respectively. In addition, $\eta(\theta^{m,c})$ describes the beam filling (dilution) factor, $S^{m,c}(\nu)$ the source function, $I_{\rm bg}$ the background intensity, and $J_\mathrm{CMB}$ the intensity of the cosmic microwave background. The optical depth $\tau_{\rm total}^{m,c}(\nu)$ of mm-RL/RRLs (in LTE) is given by \citet{gordon2002}
\begin{eqnarray}\label{myxclass:RRLopticalDepthLine}
  \tau_{{\rm RRL}, \nu} &= \frac{\pi \, h^3 \, e^2}{(2 \pi \, m_e \, k_B)^{3/2} \, m_e \, c} \cdot {\rm EM} \cdot \frac{n_1^2 \, f_{n_1, n_2}}{T_e^{3/2}} \nonumber \\
  & \quad \times \exp \left[\frac{Z^2 \, E_{n_1}}{k_B T_e} \right] \, \left(1 - e^{-h \nu_{n_1, n_2} / k_B T_e} \right) \, \phi_\nu,
\end{eqnarray}
where the oscillator strength\footnote{For the absorption oscillator strength $f_{n_1, n_2}$, we applied Menzel's approximation \citep{menzel1968} $f_{n_1, n_2} \approx n_1 \, M_{\Delta n} \left(1 + 1.5 \frac{\Delta n}{n_1} \right)$ with $M_{\Delta n} = 0.190775$, $0.026332$, $8.105620 \cdot 10^{-3}$, $3.491680 \cdot 10^{-3}$, $1.811850 \cdot 10^{-3}$, $1.058470 \cdot 10^{-3}$, $6.712780 \cdot 10^{-4}$, and $4.521900 \cdot 10^{-4}$ for $\Delta n = n_2 - n_1$ = 1, \ldots, 8.} $f_{n_1, n_2}$, the transition frequency\footnote{The transition frequency $\nu_{n_1, n_2}$ is given as $\nu_{n, n + \Delta n} = R_M \cdot c \cdot \left[n^{-2} - (n + \Delta n)^{-2} \right]$.} $\nu_{n_1, n_2}$, and the energy of the lower state\footnote{In Eq.~\ref{myxclass:RRLopticalDepthLine}, $E_{n_1}$ is the energy of level $n_1$ below the continuum and is given by $E_{n_1} = R_M \cdot n_1^{-2}$, where $R_M$ is the Rydberg constant.} $E_{n_1}$ of a particular mm-RL/RRL with main quantum number $n_1$ are obtained from an embedded database describing mm-RL/RRL transitions up to $\Delta n = 8$ ($\theta$-transitions). This means that in our analysis, we consider all mm-RL transitions within the given frequency range up to $\theta$ transitions. In Eq.~\eqref{myxclass:RRLopticalDepthLine}, the term $\phi_\nu$ represents the line profile function.

In contrast to the LTE case, where the Saha-Boltzmann distribution was used to determine the electron population, radiative transitions dominate over collisional transitions in the non-LTE case. To describe the departures from LTE, one introduces for each electronic level $n$ a so-called departure coefficient $b_n$, relating the electron population in non-LTE ($N_n$) with that in LTE ($N_n^{*}$),
\begin{equation}
  N_n = b_n \, N_n^*.
\end{equation}
These departure coefficients have been derived by \citet{storey1995} and depend on both collisional and radiative processes. \textsc{XCLASS} uses their tabulated coefficients\footnote{For non-tabulated temperatures and densities a linear interpolation procedure is used. No extrapolation is used and the corresponding departure coefficients are set to one.} for all mm-RLs/RRLs. Due to the fact that the departure coefficients depends on the electron density, the non-LTE description of the mm-RLs/RRLs requires an additional parameter, the electron density $N_{\rm e}$ (in cm$^{-3}$) for each component.

\begin{figure*}[!h]
  \centering
  \includegraphics[width=0.85\textwidth]{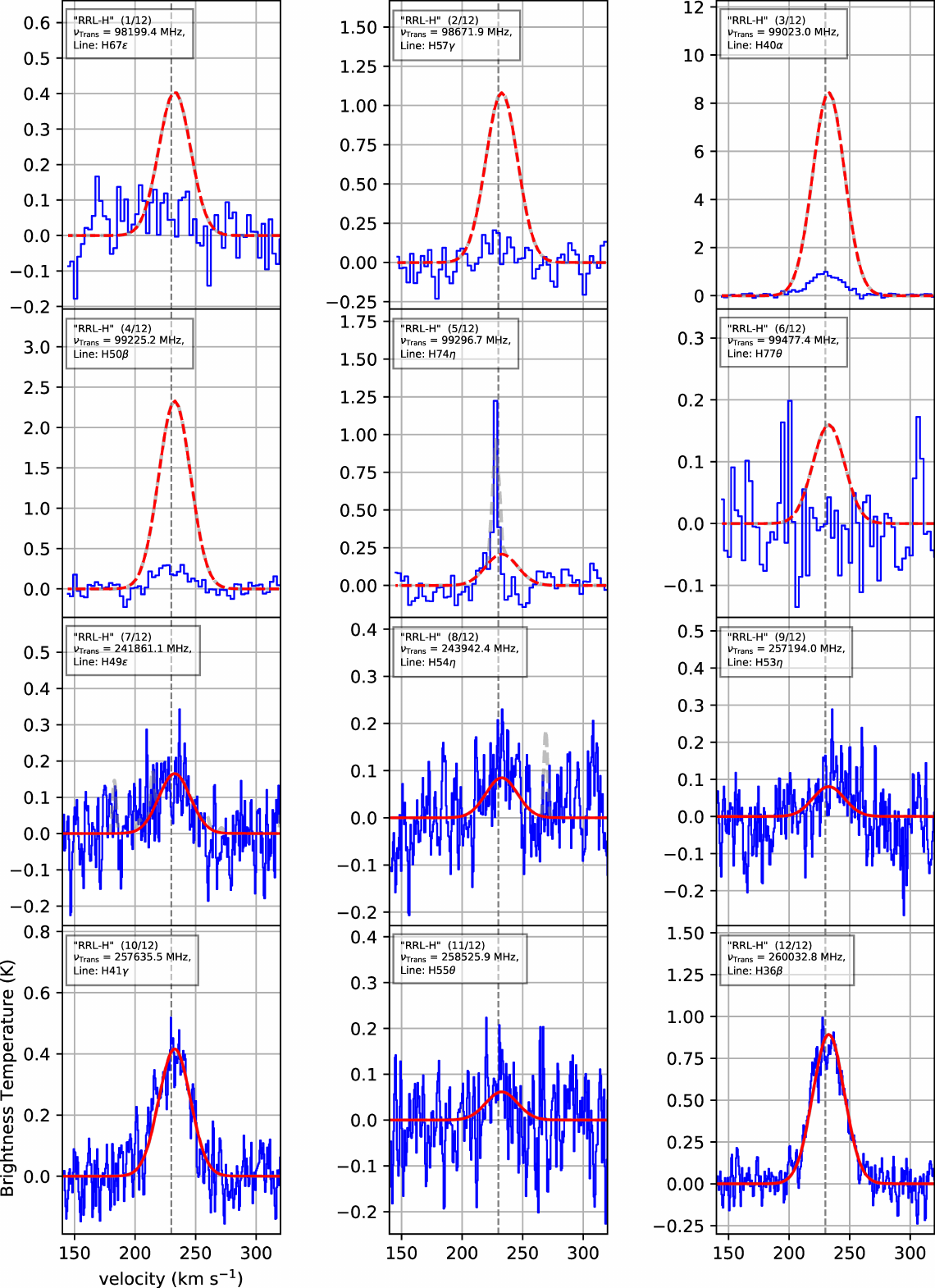}
  \caption{Spectra of N\,105--1\,A (shown in blue) centered on rest frequencies of the mm-RL transitions (up to $\theta$-transitions, $\Delta n=8$) covered by our Band 3 (top two panels) and Band 6 (bottom two panels) observations, ordered by increasing frequency (from left to right, top to bottom).  The red solid line shows the XCLASS synthetic spectrum fitted to Band 6 mm-RL transitions; it represents the pure mm-RL contribution assuming LTE with a single component. The XCLASS spectrum is extended to lower frequencies (shown as the red dashed line) to compare line intensities predicted by the LTE model to our observed Band 3 spectra, revealing the beam dilution effects in  Band 3.  The XCLASS spectrum taking into account both mm-RLs and all molecular species detected toward 1\,A (discussed in \citealt{sewilo2022n105}; see also Appendix~\ref{s:appA}) is shown as the gray solid line.  The gray vertical dashed line indicates the source velocity.  \label{f:xclassLTE}}
\end{figure*}

\begin{figure*}[!htp]
  \centering
  \includegraphics[width=0.8\textwidth]{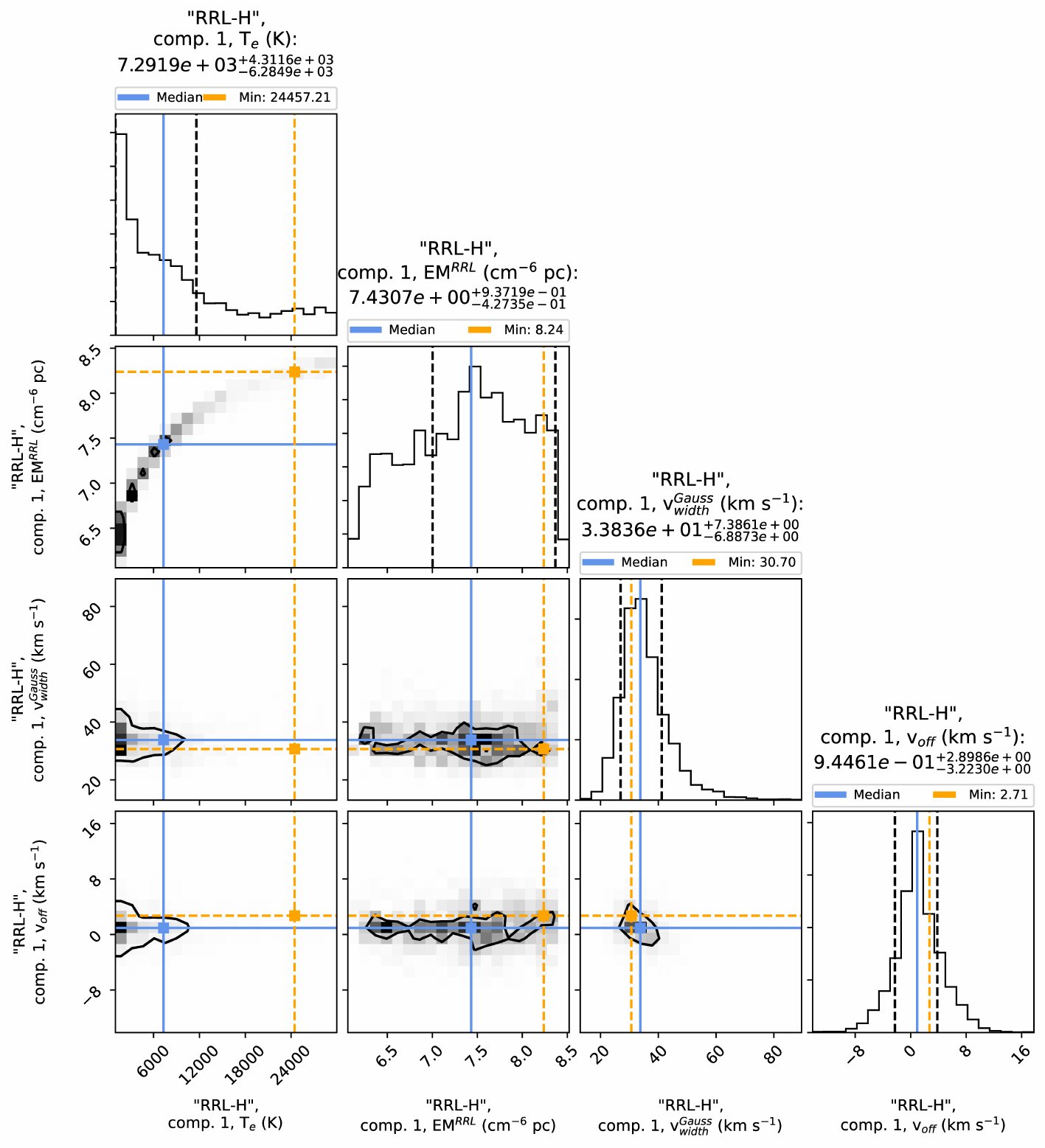}
  \caption{Corner plot of the MCMC error estimate for the XCLASS fit of the LTE model with one component shown in Fig.~\ref{f:xclassLTE}. The 16\% and 84\% quantiles are plotted as black dashed lines, the 50\% quantile (median) is shown in blue, while the lowest $\chi^2$ value is shown in orange. The plot demonstrates that we cannot reliably determine either the electron temperature ($T_{\rm e}$) or  emission measure (EM); it is likely due to the signal-to-noise ratio being too low for most of the Band 6 transitions.  \label{f:xclasscornerLTE}}
\end{figure*}

The general results of the XCLASS analysis of our Band 3 and Band 6 data for N\,105--1\,A are described in Section~\ref{s:xclass}. Here, we present the example model fit and error estimation for the XCLASS fit to the Band 6 transitions. Figure~\ref{f:xclassLTE} shows spectra centered on rest frequencies of the mm-RL transitions (up to $\theta$-transitions, $\Delta n=8$) covered by our Band 6 (H36$\beta$, H41$\gamma$, H49$\epsilon$, H53$\eta$, H54$\eta$, and H55$\theta$)  and Band 3 (H40$\alpha$, H50$\beta$, H57$\gamma$, H67$\epsilon$, H74$\eta$, and H77$\theta$) observations, with the XCLASS ``LTE with one component'' model fit overlaid (`Scenario~1' in Section\ref{s:xclass}).  The corner plot illustrating the error estimation for this model is shown in Fig.~\ref{f:xclasscornerLTE}. 

\clearpage

\section{Multiwavelength Images and Photometry of N\,105--1\,A}
\label{s:appPhot}

\begin{figure*}[h!]
\centering
\includegraphics[width=\textwidth]{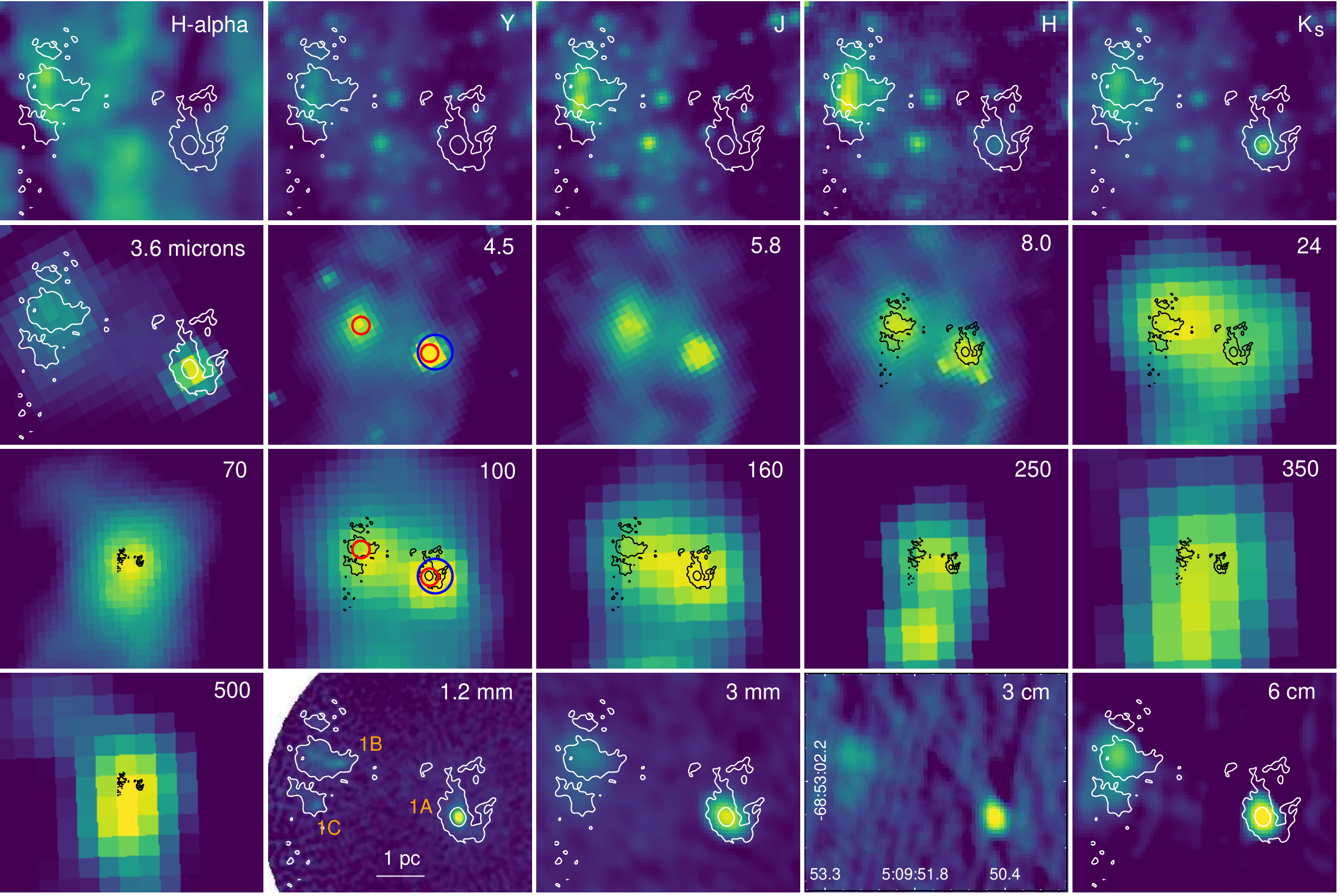}
\caption{Multiwavelength images of N\,105--1 (from left to right, top to bottom):  the H$\alpha$ image from the MCELS2 survey (PI: You-Hua Chu),  VMC $Y$- and $J$-band \citep{cioni2011}, IRSF $H$-band \citep{kato2007}, VMC $K_{s}$-band, {\it Spitzer}/SAGE IRAC 3.6--8.0 $\mu$m and MIPS 24 and 70 $\mu$m (\citealt{meixner2006}; \citealt{sage}), {\it Herschel}/HERITAGE PACS 100 and 160  $\mu$m and SPIRE 250--500 $\mu$m (\citealt{meixner2013}; \citealt{heritage}), ALMA 1.2 mm \citep{sewilo2022n105}, ALMA 3 mm (this paper), and ATCA 3 and 6 cm \citep{indebetouw2004}. The 1.2 mm continuum contours with contour levels of (3, 30) $\times$ the image rms of $6.9\times10^{-5}$ mJy beam$^{-1}$ are overlaid on selected images for reference.  Red circles in the 4.5 and 100 $\mu$m images indicate the positions of {\it Spitzer} YSOs associated with 1\,A and 1\,B (\citealt{whitney2008}; \citealt{gruendl2009}; \citealt{seale2009}), while the blue circle shows the position of the {\it Herschel} YSO candidate (\citealt{seale2014}). \label{f:grid}}
\end{figure*}

We provide the multiwavelength photometry (from $\sim$1 $\mu$m to 6 cm) for N\,105--1\,A in Table~\ref{t:photometry}.  The image cutouts are shown in Fig.~\ref{f:grid}. The references and technical details are provided in the Table~\ref{t:photometry} footnotes, Fig.~\ref{f:grid} caption, and the text below. The spectral energy distribution (SED) of N\,105--1\,A for a wavelength range from $\sim$1.3 $\mu$m to 3 mm is presented in Fig.~\ref{f:sed}. 

In Table~\ref{t:photometry}, we provide the following photometric data: 

\begin{packed_enum}
\item[--] the $JHK_S$ data from the Infrared Survey Facility (IRSF) Magellanic Clouds Point Source Survey (\citealt{kato2007}); we treat the $J$-band flux density as an upper limit for the SED fitting since no source is visible in the image at the position of 1\,A.  The Visible and Infrared Survey Telescope for Astronomy (VISTA) $YJK_S$ survey of the Magellanic Clouds system (VMC; \citealt{cioni2011}) only provides the $K_s$-band photometry and it is consistent with the IRSF measurement;
 
\item[--]  {\it Spitzer} Space Telescope Infrared Array Camera (IRAC) 3.6--8.0 $\mu$m data from the LMC-wide {\it Spitzer} ``Surveying the Agents of Galaxy Evolution'' (SAGE; \citealt{meixner2006}; \citealt{sage}) survey (e.g., \citealt{whitney2008}; \citealt{gruendl2009}; \citealt{carlson2012}).  Both the point spread function (PSF; \citealt{sage}) and aperture photometry (\citealt{gruendl2009}) are available (see Table~\ref{t:photometry}).  The PSF and aperture photometry agree within the uncertainties. We have adopted the aperture photometry fluxes for the fitting since it better accounts for the extended emission surrounding the source at longer IRAC bands.  No Multiband Imaging Photometer for Spitzer (MIPS) 24 or 70 $\mu$m photometry is available in the existing catalogs. Source 1\,A is unresolved from the neighboring YSO in these bands (see Fig.~\ref{f:grid});

\item[--]  the five-band {\it Herschel} Space Observatory photometric measurements from \citet{seale2014} covering 100--500 $\mu$m; the catalog is based on the Photoconductor Array Camera and Spectrometer (PACS; 100 and 160 $\mu$m) and Spectral and Photometric Imaging Receiver (SPIRE; 250--500 $\mu$m) data from the ``{\it Herschel} Inventory of the Agents of Galaxy Evolution'' (HERITAGE; \citealt{meixner2013}; \citealt{heritage}) survey. Source 1\,A is unresolved from the neighboring source between 160 and 500 $\mu$m, thus we use the {\it Herschel} flux densities in these bands as upper limits; 

\item[--] the ALMA 870 $\mu$m continuum flux density based on the reprocessed archival data;

\item[--] ALMA 1.2 mm and 3 mm flux densities reported in \citet{sewilo2022n105} and measured in the present paper, respectively, after removing the contribution from the free-free emission. 

\end{packed_enum}

In addition to using the {\it Spitzer} catalog data, we have used the {\it Spitzer} InfraRed Spectrograph (IRS) spectrum from \citet{seale2009} to better constrain the SED between 5.2 and 37.9 $\mu$m.  We extracted 11 data points from the IRS spectrum that were selected at wavelengths free of fine-structure emission lines to delineate silicate features and the underlying continuum: 6.89, 8.96, 9.44, 9.95, 10.75, 11.64, 13.63, 15.95, 19.68, 25.58, 29.95 $\mu$m. 

In the analysis of the IRS spectra, the scaling factors were applied to match smoothly the spectrum segments taken under different modules across the full wavelength range (\citealt{seale2009}).  For the fitting, we have reverted these IRS fluxes to their original values for three affected spectrum segments by removing the corresponding scaling factors, i.e., dividing the fluxes within SL1 (short wavelength, low resolution; 7.6--14.6 $\mu$m), SH (short wavelength, high resolution; 9.9--19.3 $\mu$m), and LH (long wavelength, high resolution; 18.9--36.9 $\mu$m) modules by 0.9704, 0.8937, and 1.4173, respectively. 

The SED fitting procedure and the fitting results are discussed in Section~\ref{s:yso}.

\begin{deluxetable}{lcclc}
\centering
\rotate
\tablecaption{Multiwavelength Photometry for N\,105--1\,A  \label{t:photometry}}
\tablewidth{0pt}
\tablehead{
\multicolumn{1}{c}{Instruments and Bands} &
\multicolumn{1}{c}{Flux density (uncertainty)} &
\colhead{FWHM\tablenotemark{\footnotesize a} } &
\multicolumn{1}{c}{Source ID} &
\colhead{Ref.}  \\
\colhead{} &
\colhead{(mJy)} &
\colhead{($''$)} &
\colhead{}  &
\colhead{}
}
\startdata
VMC $Y$ (1.02), $J$ (1.25), $K_s$ (2.15 $\mu$m) & -999.9, -999.9, 1.78\,(0.10) & $\lesssim$1 & 558354728325 & 1\\
IRSF $J$ (1.25), $H$ (1.63) , $K_s$ (2.14 $\mu$m) & 0.04 (0.01), 0.20 (0.02), 1.85 (0.07) & $\sim$1.3 & 05095050-6853052 & 2\\
2MASS $J$ (1.25), $H$ (1.65) , $K_s$ (2.16 $\mu$m) & -999.9, -999.9, 2.26 (0.10) & $\sim$2.5 & J05095054$-$6853052 & 3 \\
{\it Spitzer}/IRAC 3.6, 4.5, 5.8,  8.0 $\mu$m\tablenotemark{\footnotesize b}  & 24.45\,(0.94), 64.85\,(2.54), 156.50\,(3.25), 504.70\,(14.69) & 1.7, 1.7, 1.9, 2.0 & SSTISAGE1C\,J050950.53$-$685305.4 & 4\\
 & 28.61\,(1.58), 65.25\,(3.00),  177.30\,(9.80), 514.12\,(23.68)  & \ditto &050950.53$-$685305.5\tablenotemark{\footnotesize c} & 5 \\
{\it Herschel}/PACS 100, 160 $\mu$m & 41390\,(3013), 23080 (1612) & 8.6, 12.6 & HSOBMHERICC\,J77.459799$-$68.884809 & 7\\
{\it Herschel}/SPIRE 250, 350, 500 $\mu$m & 10060\,(652), 4967\,(309), 7710\,(552) & 18.3, 26.7, 40.5 & \ditto & 7\\
ALMA 870 $\mu$m ( GHz) & 161.7\,(0.1)\tablenotemark{\footnotesize d} & $0.47\times0.39$, 37$\rlap.^{\circ}$0 & N\,105--1\,A & 9\\
ALMA 1.2 mm (242.4 GHz) & 71.8\,(2.2)\tablenotemark{\footnotesize d} & $0.51\times0.47$, 37$\rlap.^{\circ}$2 &  \ditto & 8, 9\\
ALMA 3 mm (99 GHz) & 48.6\,(0.3)\tablenotemark{\footnotesize d} & $2.49\times1.83$, 46$\rlap.^{\circ}$1 & \ditto & 9\\
ATCA 3 cm (8.6 GHz) & 39\,(1) & $1.82\times1.24$, 14$\rlap.^{\circ}$8 & B0510$-$6857\,W &  10 \\
ATCA 6 cm (4.8 GHz) & 26\,(1) & $2.19\times1.70$, 16$\rlap.^{\circ}$7 & \ditto &  10 \\
\hline
\enddata
\tablerefs{(1) \citet{cioni2011}; (2) \citet{kato2007}; (3) \citet{skrutskie2006}; (4) \citet{whitney2008}, {\it Spitzer} PSF photometry from the SAGE Epoch 1 Catalog;  (5) \citet{gruendl2009}, {\it Spitzer} aperture photometry; (6) \citet{wright2010}; (7) \citet{seale2014}; (8) \citet{sewilo2022n105}; (9) this paper; (10) \citet{indebetouw2004}}
\tablenotetext{\footnotesize a}{The image angular resolution at full-width half maximum (FWHM). For the near-IR ground-based surveys VMC, IRSF, and 2MASS, the angular resolution depends on the atmospheric seeing. For the ALMA and ATCA observations, we provide the synthesized beam sizes and position angles: $\theta_{\rm min}\times\theta_{\rm maj}, PA$. }
\tablenotetext{\footnotesize b}{No MIPS photometry is available for 1\,A in  \citet{whitney2008} and  \citet{gruendl2009}.}
\tablenotetext{\footnotesize c}{The magnitudes provided in \citet{gruendl2009} were converted to fluxes using the zero-magnitude fluxes of (280.9, 179.7, 115.0, 64.13) Jy for the IRAC (3.6, 4.5, 5.8, 8.0) $\mu$m bands (the same as those used in the SAGE survey).}
\tablenotetext{\footnotesize d}{Measured within a 3$\sigma$ contour using the CASA task {\sc imstat}. Both the dust and free-free emission contribute to the measured flux densities at 1.2 and 3 mm. The contribution of the free-free emission was estimated to be $\sim$50\% at 242.4 GHz \citep{sewilo2022n105} and $\sim$3.7\% at 99 GHz (see Section~\ref{s:temp}), respectively.}
\end{deluxetable}

\clearpage


\section{ALMA Integrated Intensity, Velocity, and Line Width Images of N\,105--1}
\label{s:appMom}

\begin{figure*}
\centering
\includegraphics[width=0.5\textwidth]{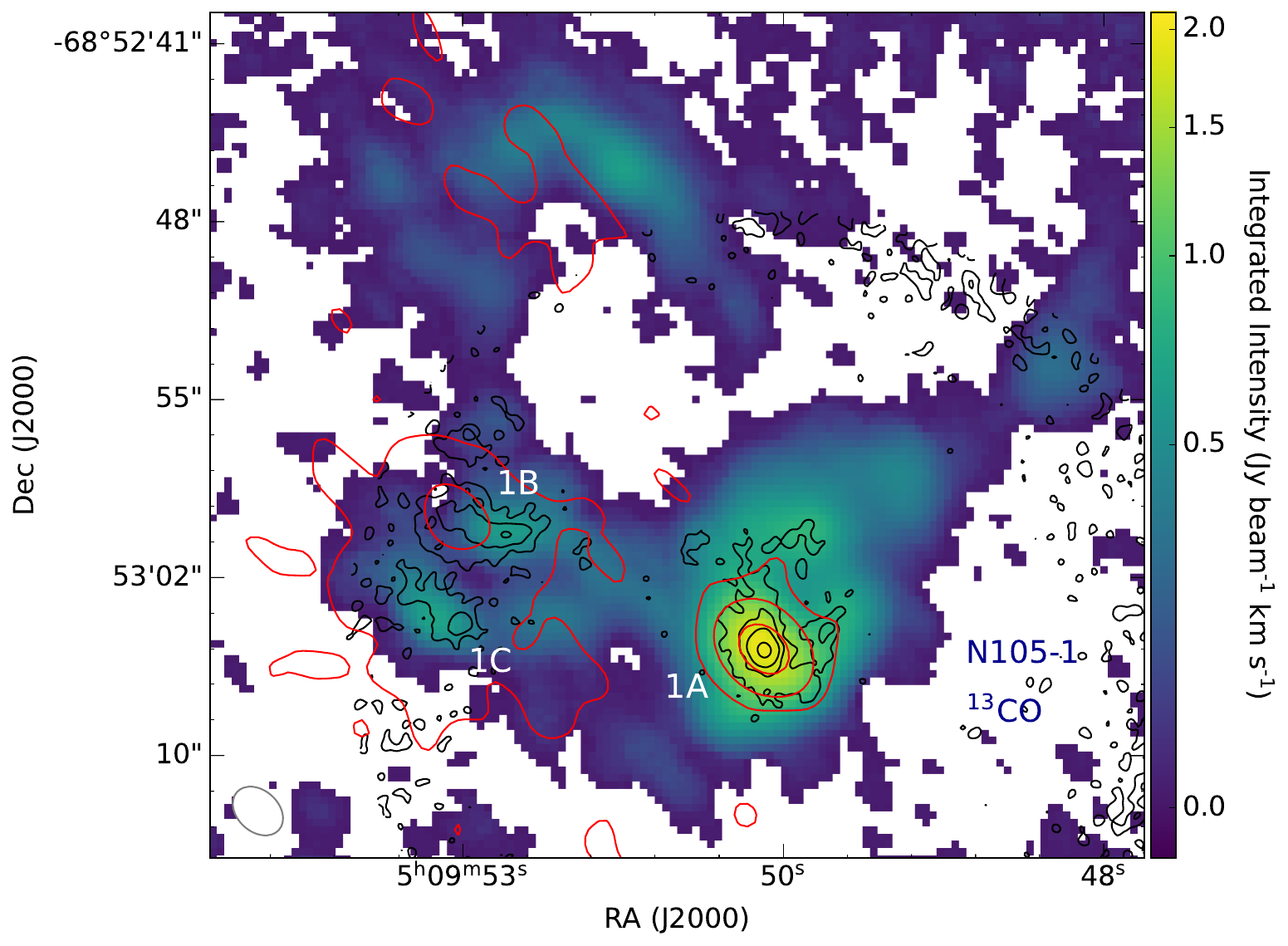} \\
\hfill
\includegraphics[width=0.485\textwidth]{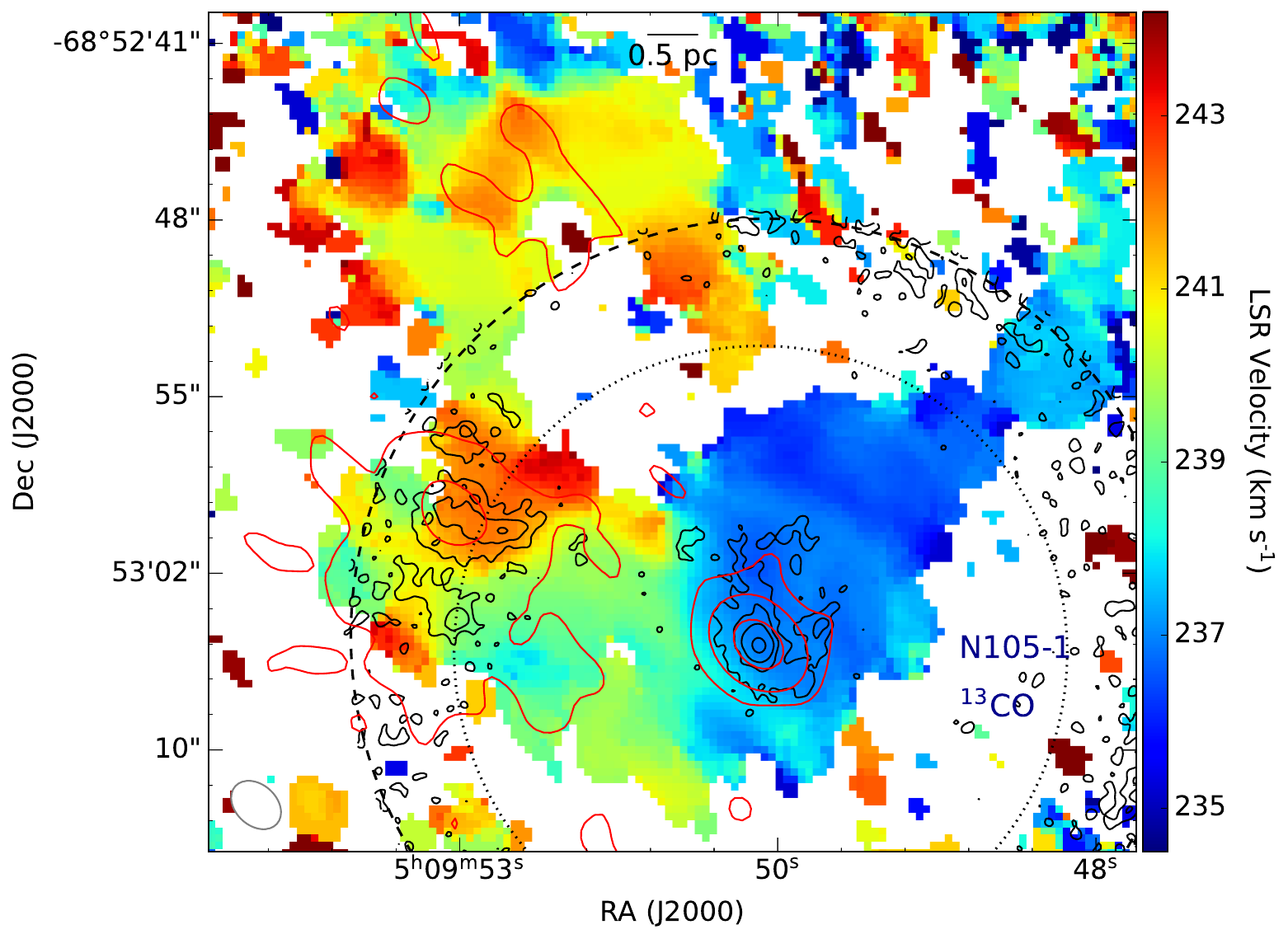}
\hfill
\includegraphics[width=0.485\textwidth]{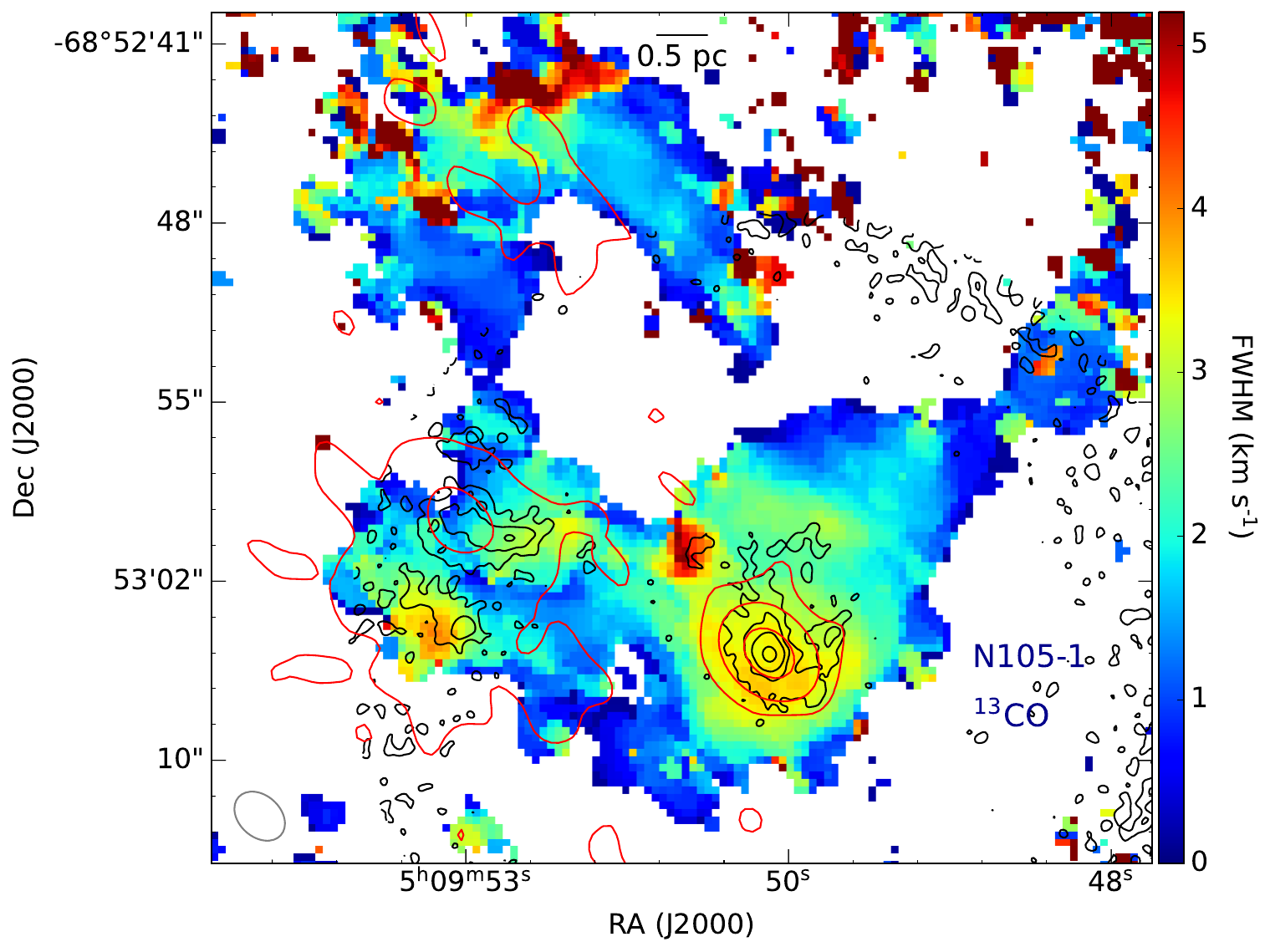}
\caption{The $^{13}$CO (1--0) integrated intensity (top panel), velocity (lower left), and line width (lower right)  images of N\,105--1.  The emission was integrated between 233.4 and 246.4 km s$^{-1}$. The dashed and dotted circles in the velocity map indicate the Band 6 and Band 7 field of view, respectively. The 1.2 mm and 3 mm continuum contours are overlaid on all the images in black and red, respectively, with contour levels the same as in Fig.~\ref{f:cont}.   \label{f:mom13CO}}
\end{figure*}

\begin{figure*}[h!]
\centering
\includegraphics[width=0.7\textwidth,trim=0 50 0 60, clip]{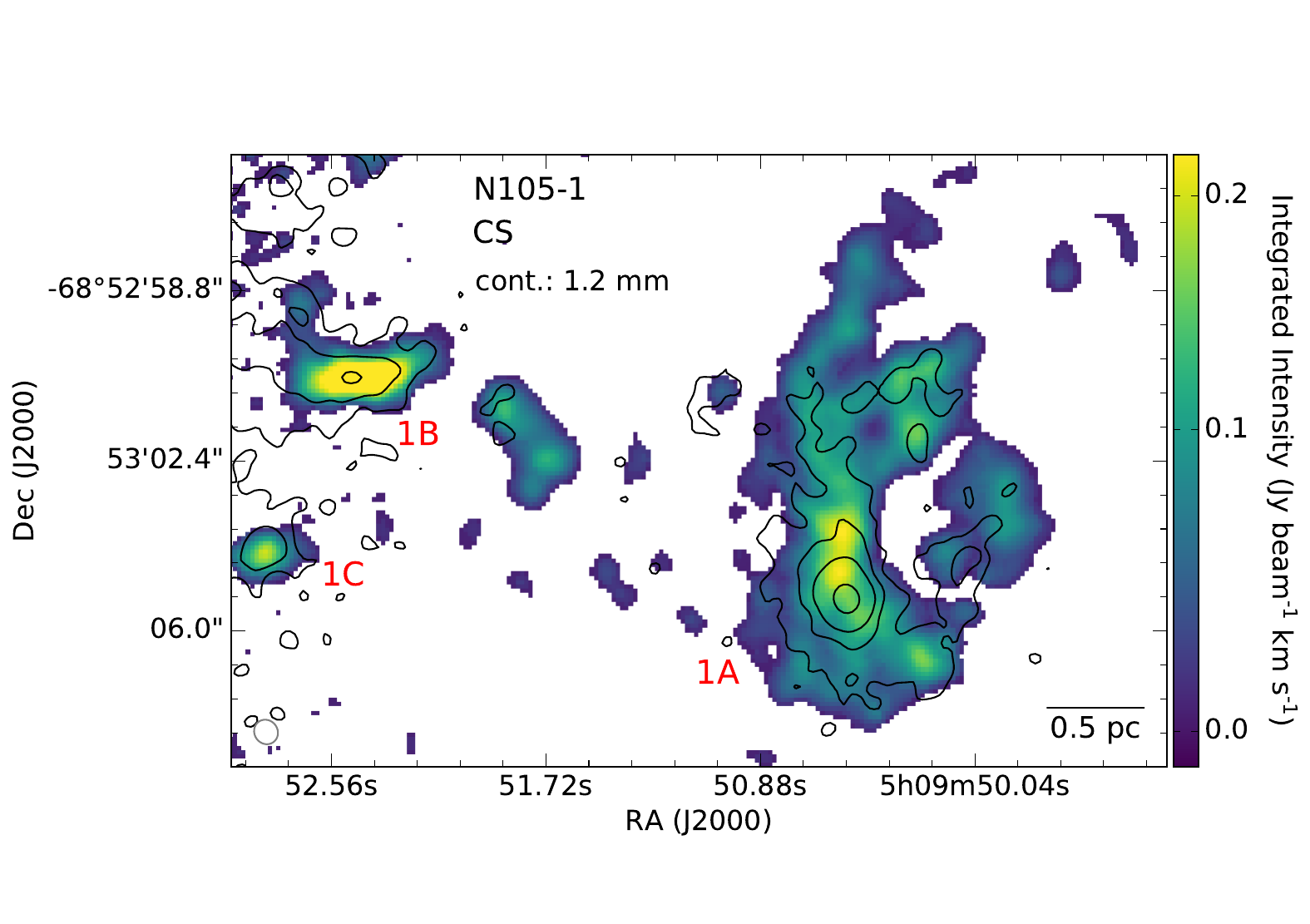}
\includegraphics[width=0.7\textwidth, trim=0 50 0 60, clip]{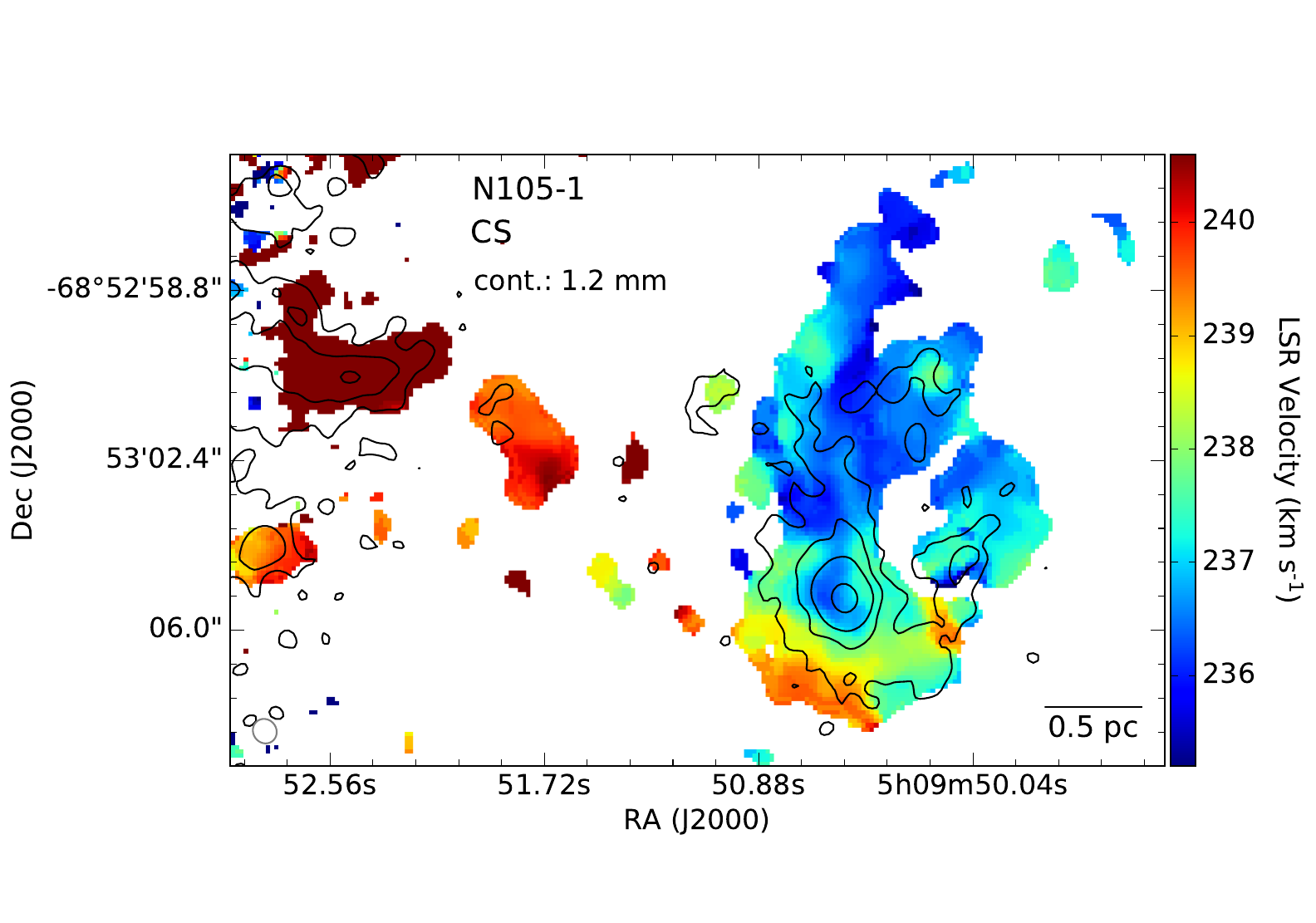}
\includegraphics[width=0.7\textwidth, trim=0 50 0 60, clip]{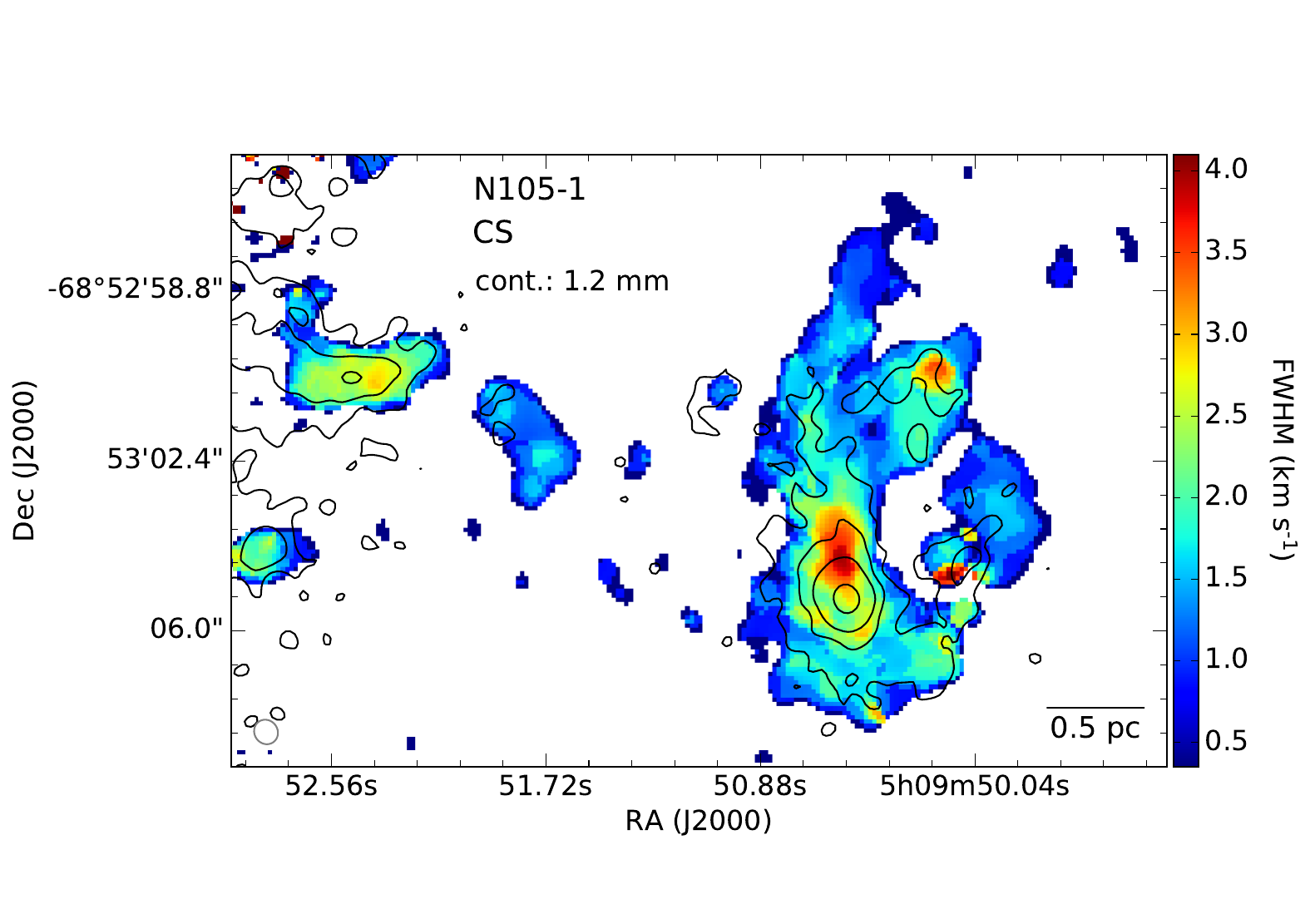}
\caption{The CS (5--4) integrated intensity (top panel), velocity (center panel), and line width (bottom panel) images of N\,105--1.  The CS emission was integrated over the velocity range 232.7--244.7 km s$^{-1}$. The 1.2 mm continuum contours are overlaid with contour levels the same as in Fig.~\ref{f:cont}.  \label{f:momCS}}
\end{figure*}

\begin{figure*}
\centering
\includegraphics[width=\textwidth,trim=40 0 0 0, clip]{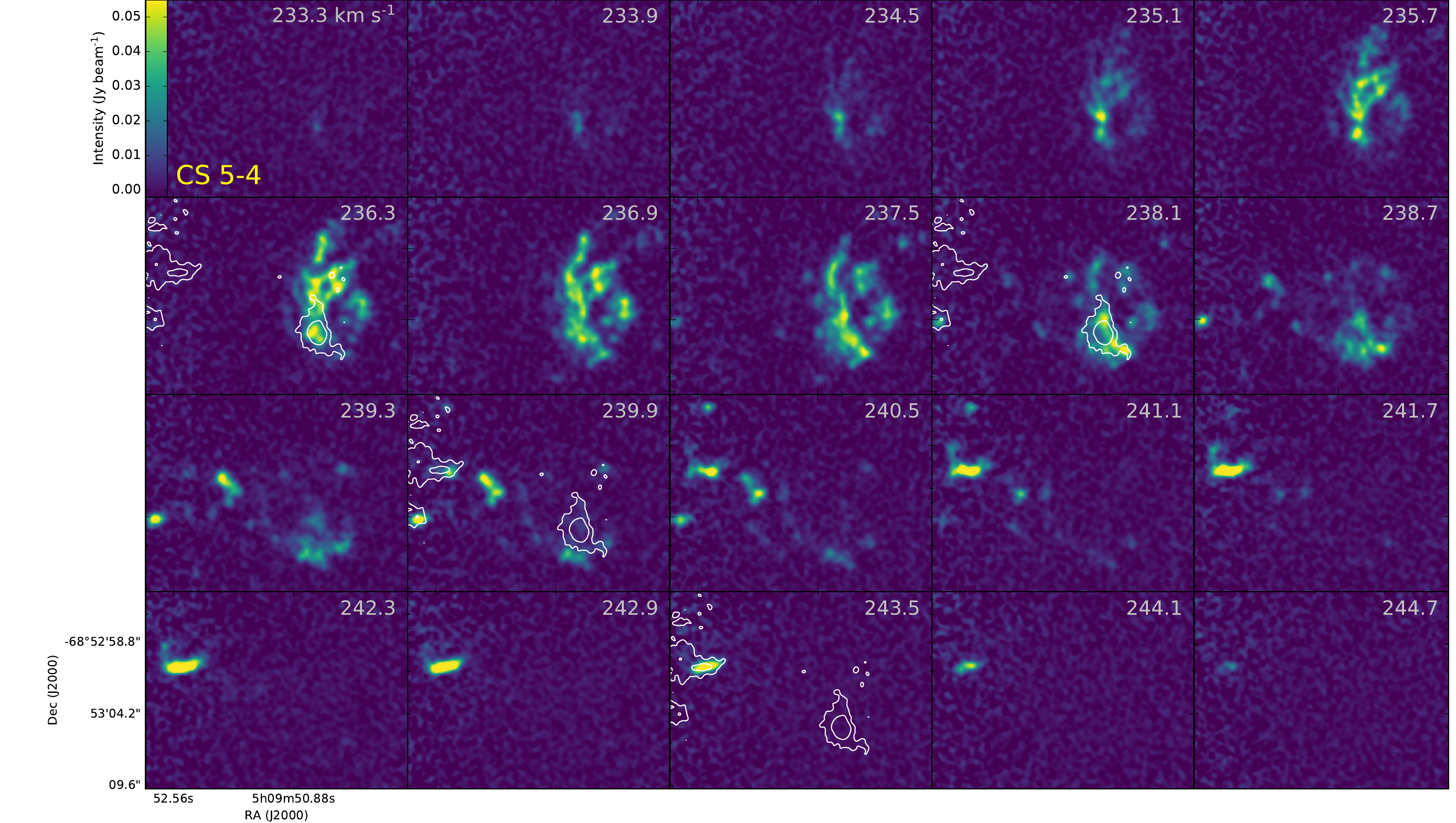}
\caption{The CS (5--4) channel maps. The 1.2 mm continuum contours are overlaid on selected channel maps with contour levels of (5, 20)$\sigma$. The brightest CS emission is associated with the continuum source 1\,B with the peak CS intensity of 0.12 Jy beam$^{-1}$. \label{f:momCSchan}}
\end{figure*}

\begin{figure*}
\centering
\includegraphics[width=0.7\textwidth,trim=0 50 0 60, clip]{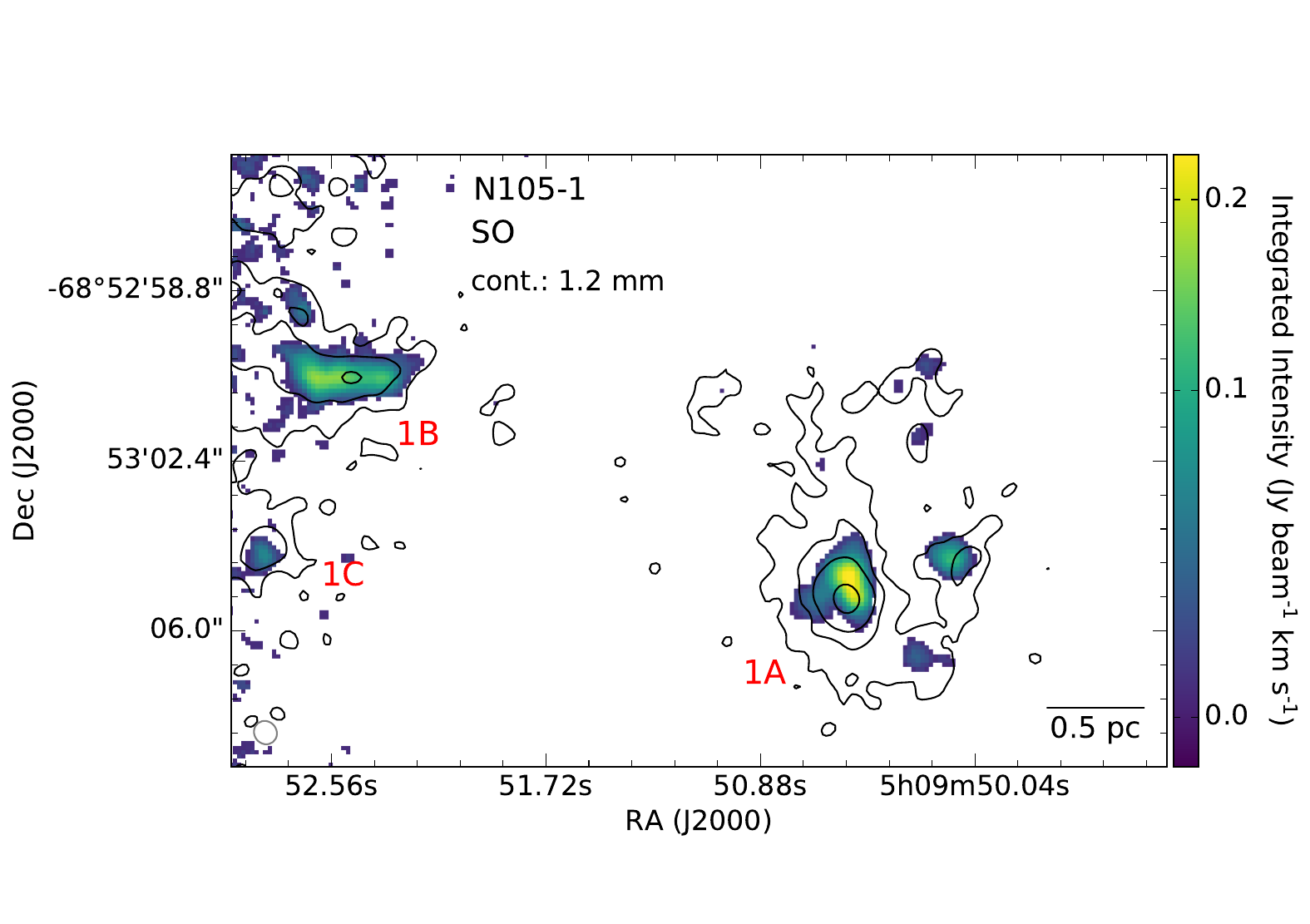}
\includegraphics[width=0.7\textwidth, trim=0 50 0 60, clip]{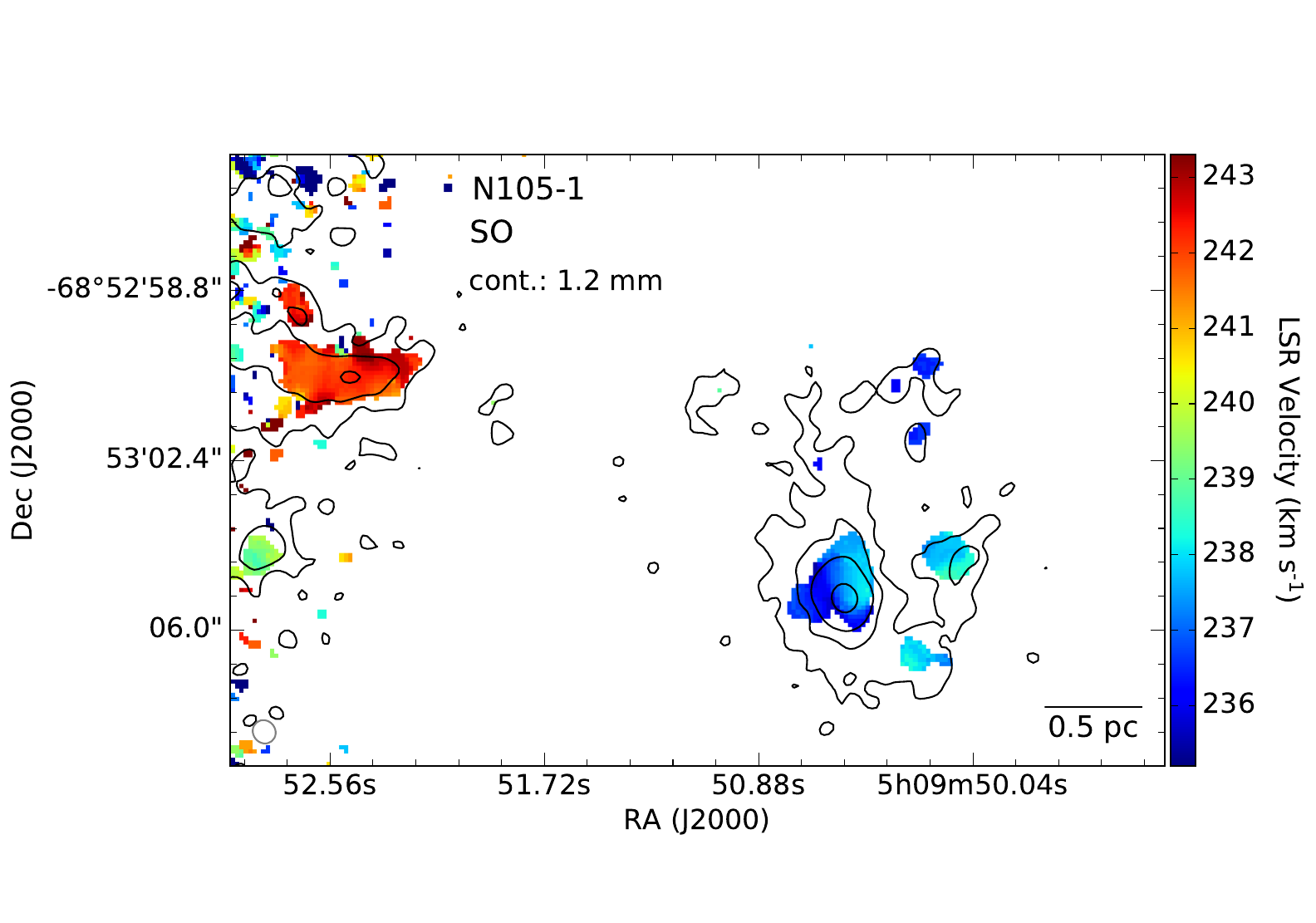}
\includegraphics[width=0.7\textwidth, trim=0 50 0 60, clip]{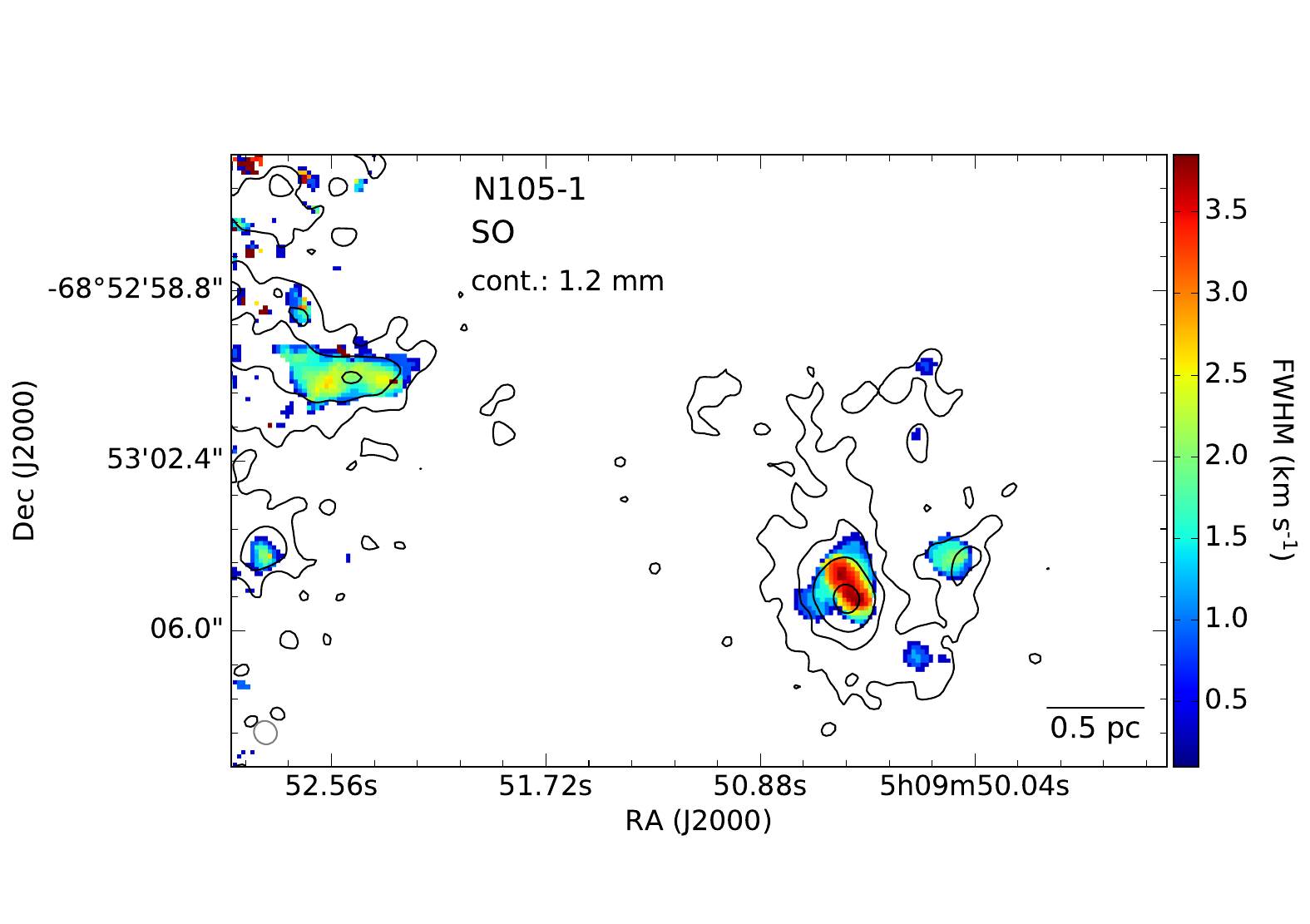}
\caption{The SO $^{3}\Sigma$ 6$_6$--5$_5$ integrated intensity (top panel),  velocity (center panel), and line width (bottom panel) images of N\,105--1.  The SO emission was integrated over the velocity range 233.2--244.6 km s$^{-1}$. The 1.2 mm continuum contours are overlaid with contour levels the same as in Fig.~\ref{f:cont}. \label{f:momSO}}
\end{figure*}

\begin{figure*}
\centering
\includegraphics[width=\textwidth,trim=40 0 0 0, clip]{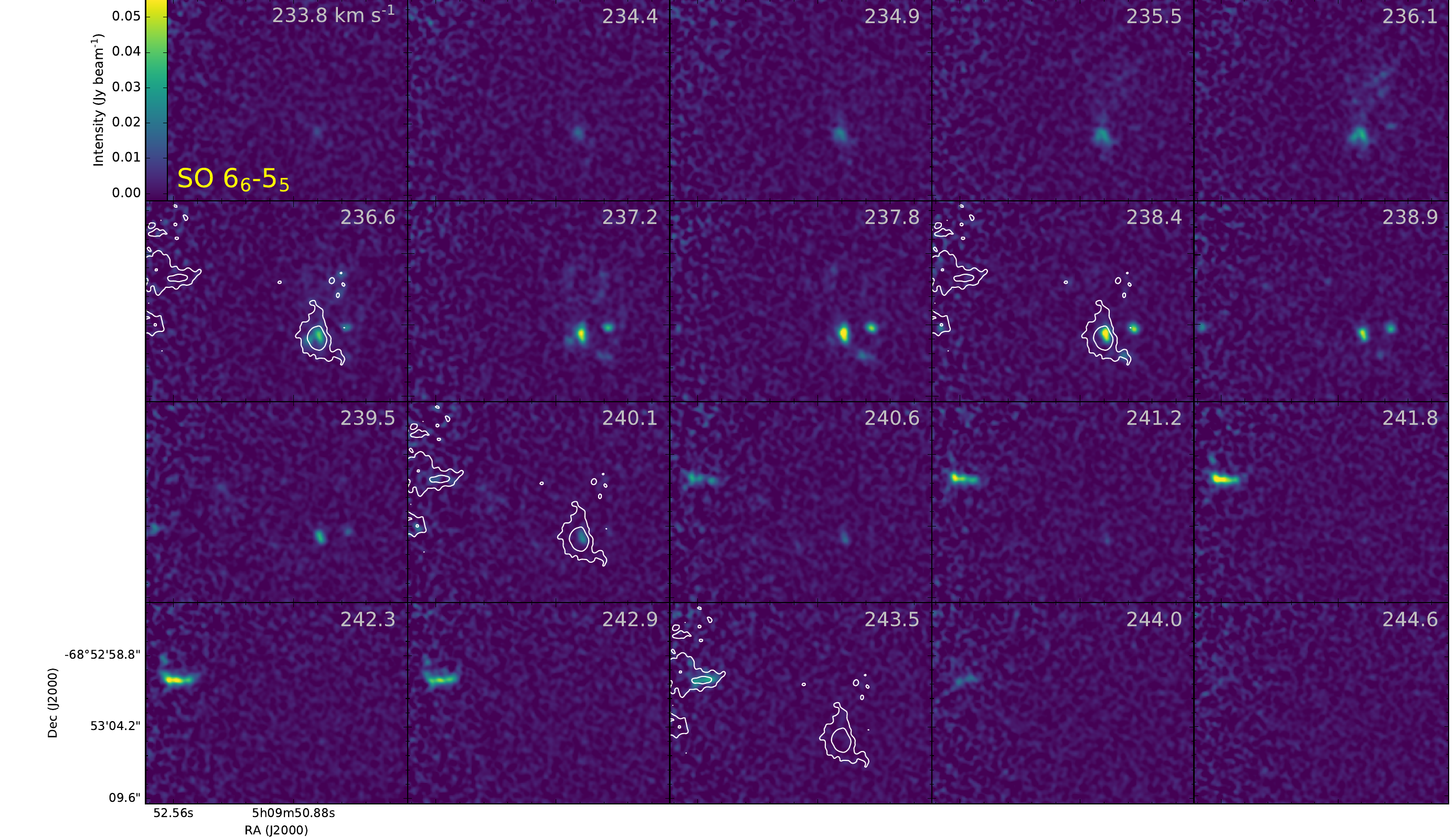}
\caption{The SO $^{3}\Sigma$ 6$_6$--5$_5$ channel maps. The 1.2 mm continuum contours are overlaid on selected channel maps with contour levels of (5, 20)$\sigma$.  \label{f:momSOchan}}
\end{figure*}

\begin{figure*}
\centering
\includegraphics[width=0.7\textwidth,trim=0 50 0 60, clip]{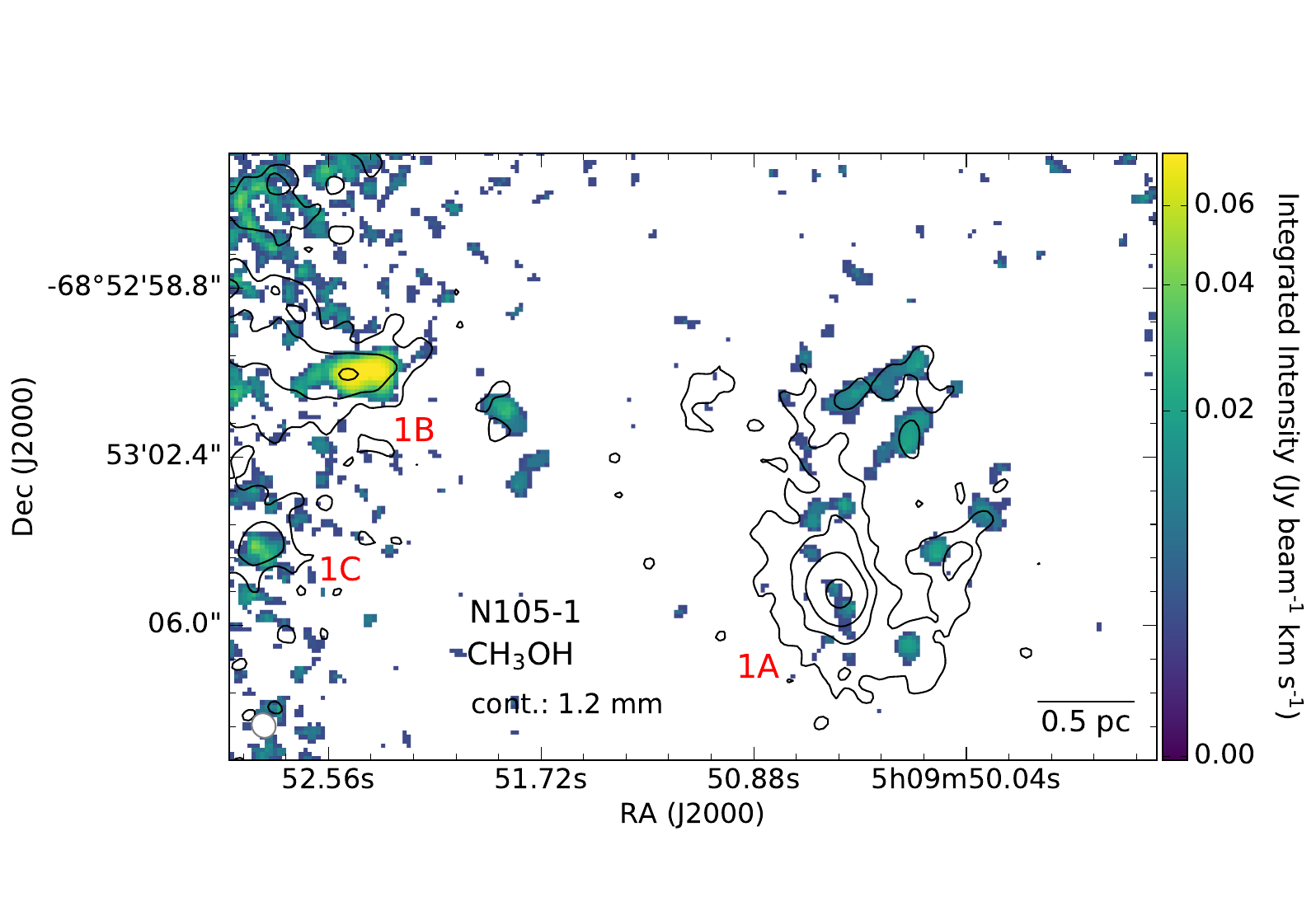}
\includegraphics[width=0.7\textwidth,trim=0 50 0 60, clip]{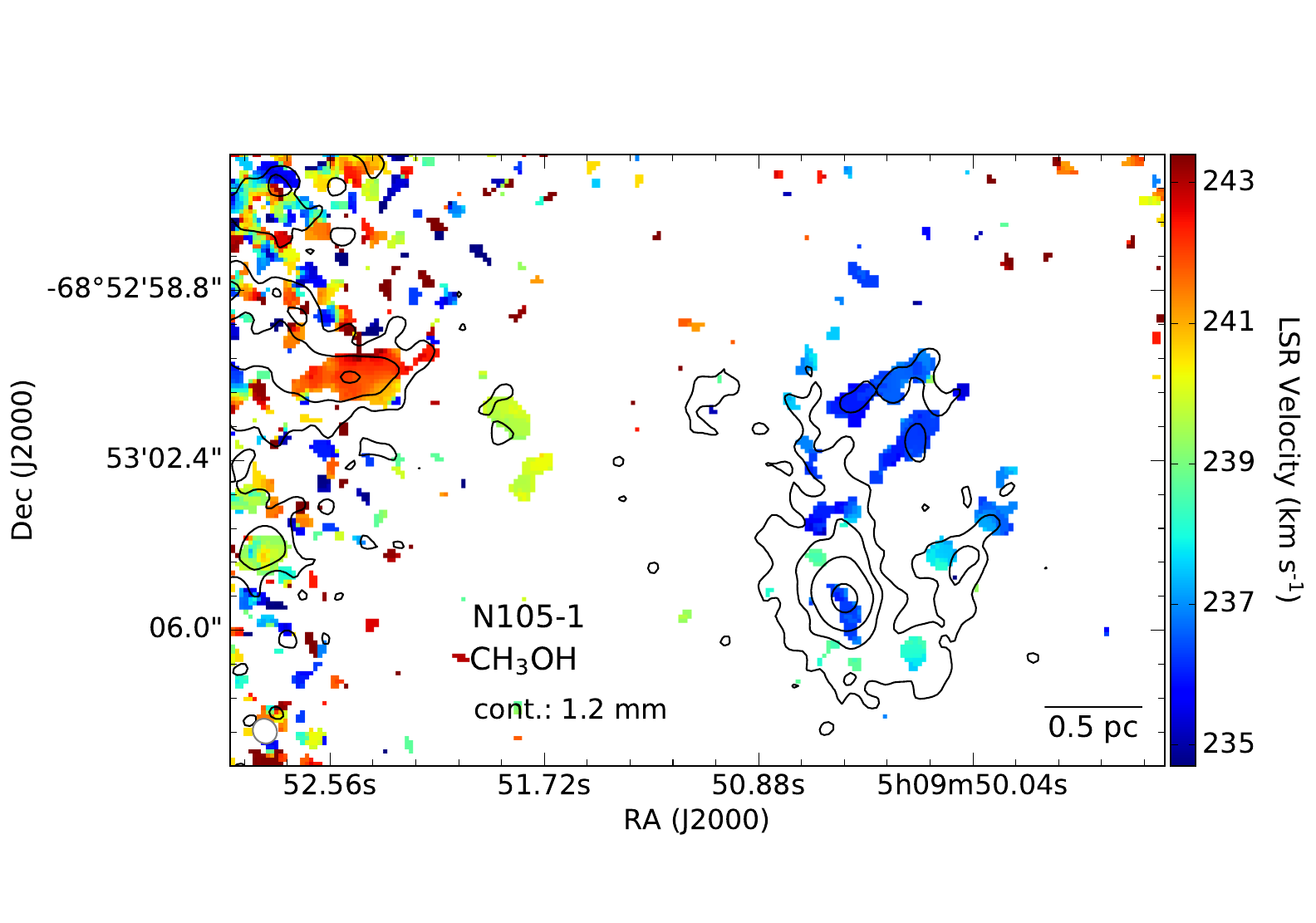}
\caption{The CH$_3$OH $5_{0,5}$--$4_{0,4}$ A integrated intensity (upper panel) and velocity (lower panel) images of N\,105--1. The CH$_3$OH emission was integrated over the velocity range 233.9--244.2 km s$^{-1}$.  The 1.2 mm continuum contours are overlaid with contour levels the same as in Fig.~\ref{f:cont}.  \label{f:momCH3OH}}
\end{figure*}

We present the integrated intensity (moment 0), intensity weighted velocity (moment 1), and line width (the full width at half maximum, FWHM, calculated from the intensity weighted velocity dispersion, or moment 2, image and deconvolved from the instrumental broadening arising from a finite channel width) images of the ALMA field N\,105--1 centered on the continuum source 1\,A for three molecular species discussed in \citet[CS, SO, and CH$_3$OH]{sewilo2022n105}, and for $^{13}$CO, the data reported in this paper.  All three images are shown for the CS (5--4) transition in Fig.~\ref{f:momCS}, for SO $^{3}\Sigma$ 6$_6$--5$_5$ in Fig.~\ref{f:momSO}, and for $^{13}$CO (1--0)  in Fig.~\ref{f:mom13CO}. The CH$_3$OH $5_{0,5}$--$4_{0,4}$ emission is the faintest and spatially most limited out of the four species and no reliable line width image was obtained; the CH$_3$OH integrated intensity and velocity images are shown in Fig.~\ref{f:momCH3OH}.  In addition, in Fig.~\ref{f:momCSchan} and \ref{f:momSOchan}, we show the channel maps for CS and SO, respectively. The kinematics of the molecular gas in N\,105--1 is discussed in Section~\ref{s:molrenv}.

\begin{figure*}
\centering
\includegraphics[width=0.491\textwidth]{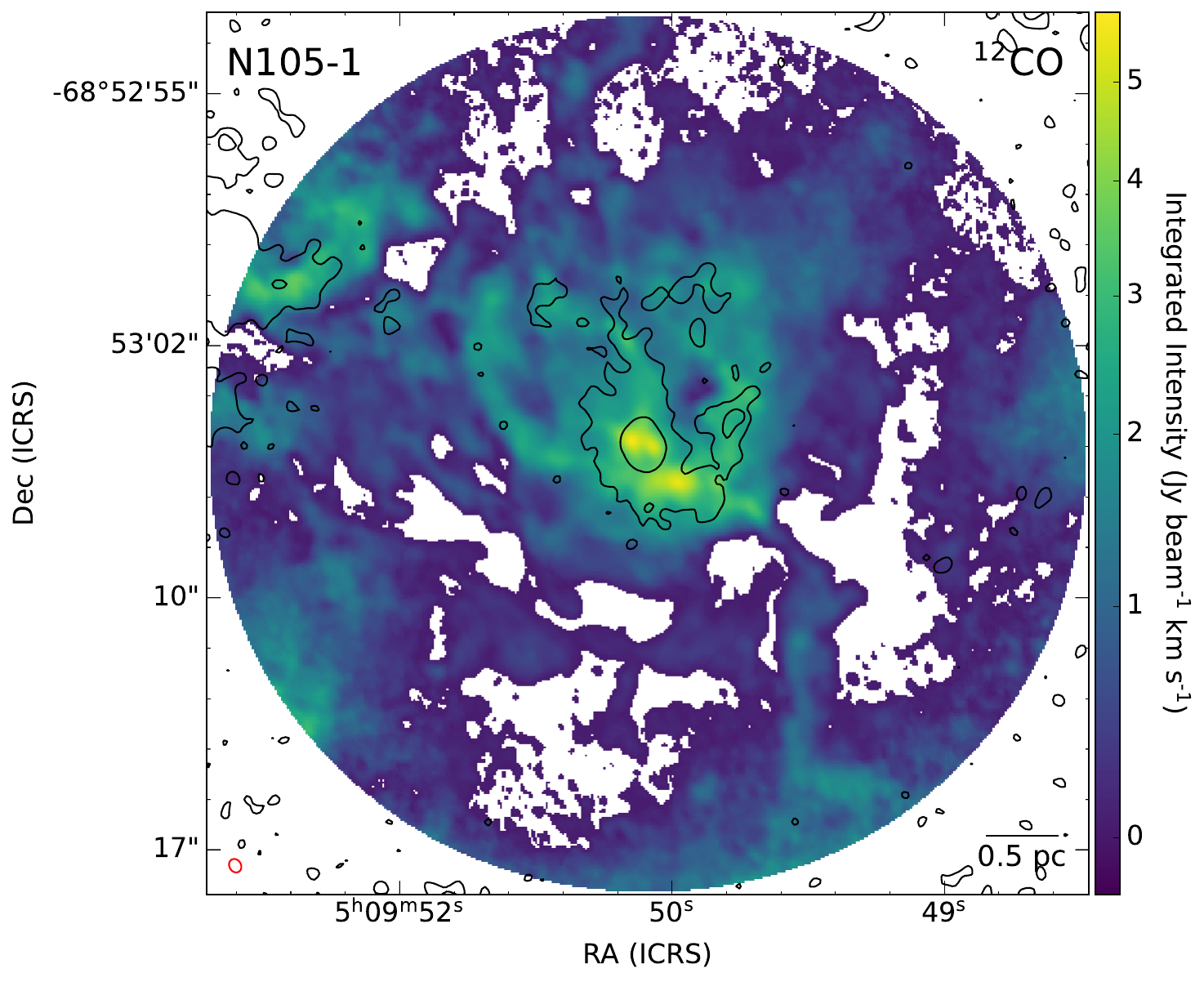}
\includegraphics[width=0.497\textwidth]{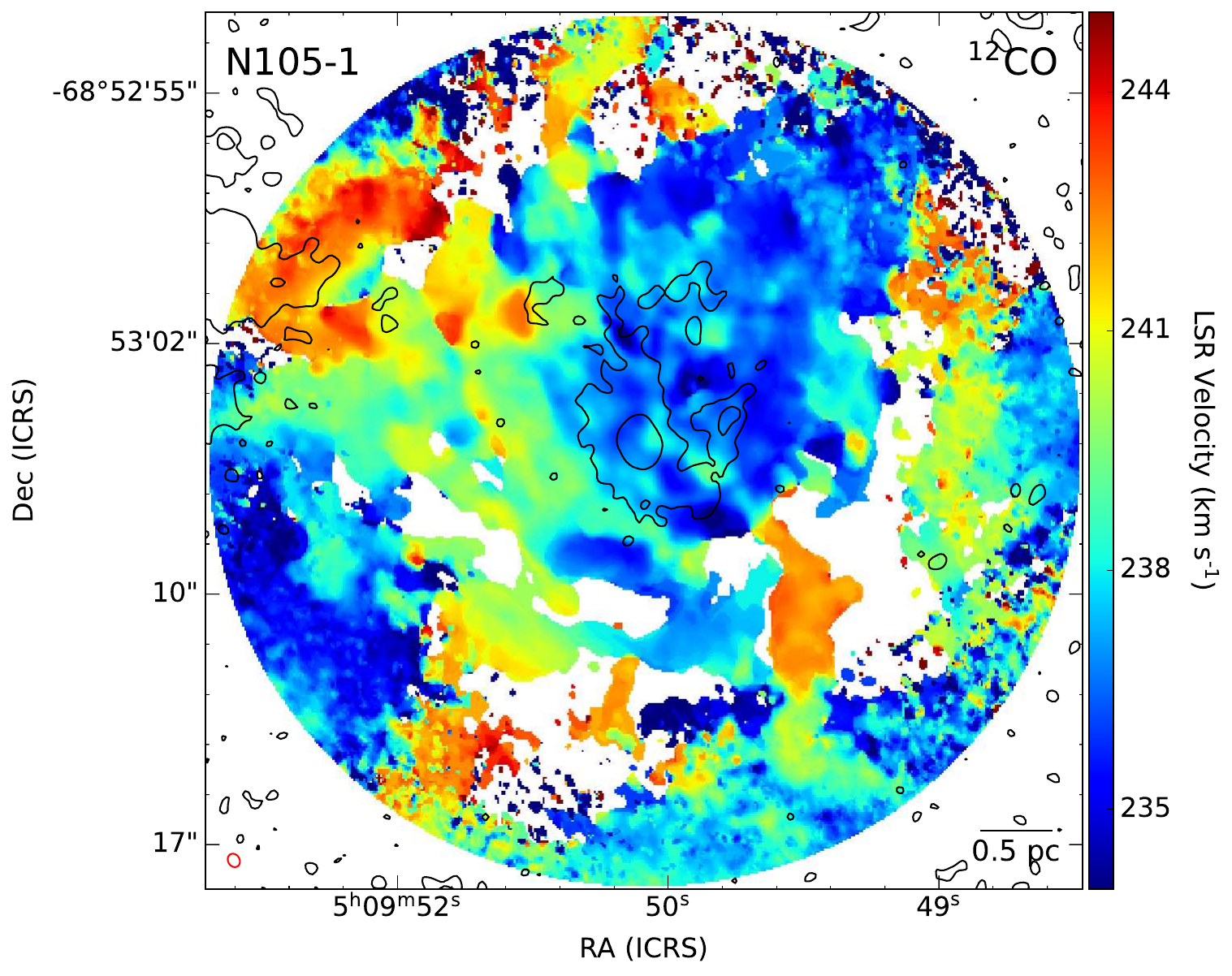}
\caption{The $^{12}$CO (3--2) integrated intensity (left) and velocity (right) images of N\,105--1.  The emission was integrated between 219 and 251 km s$^{-1}$.  Pixels with the signal-to-noise value less than five in the $^{12}$CO (3--2) integrated intensity image are masked in both images. The ALMA beam size is indicated in red in the lower left corner in each image. The 1.2 mm continuum contours are overlaid on both images with contour levels of (3, 30)$\sigma$. \label{f:momCO}}
\end{figure*}

\begin{figure*}
\centering
\includegraphics[width=\textwidth]{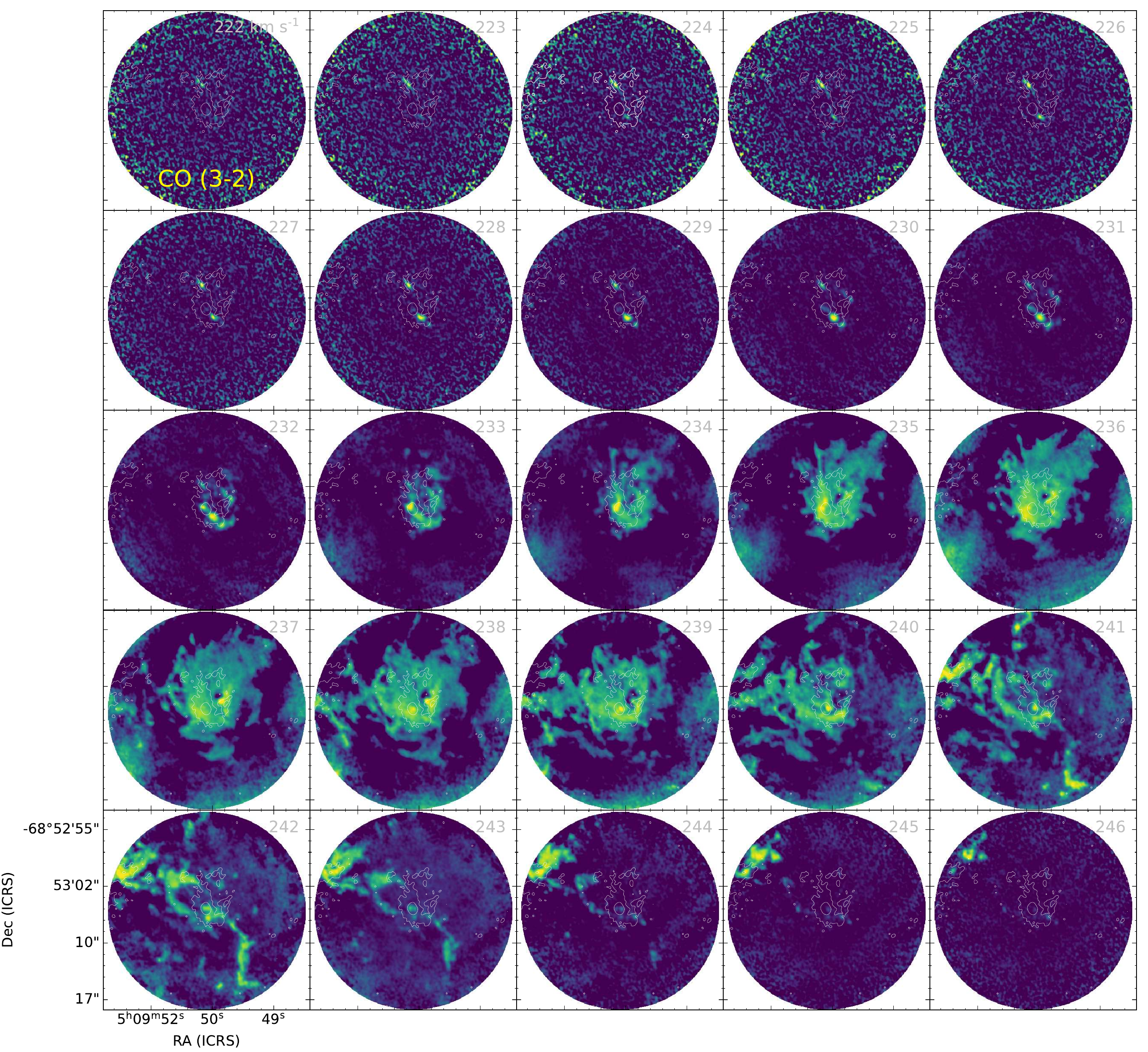}
\caption{The $^{12}$CO (3--2) channel maps. The 1.2 mm continuum contours are overlaid on all channel maps for reference; the contour levels are (3, 30)$\sigma$.  \label{f:COchan}}
\end{figure*}

\begin{figure*}
\centering
\includegraphics[width=0.8\textwidth]{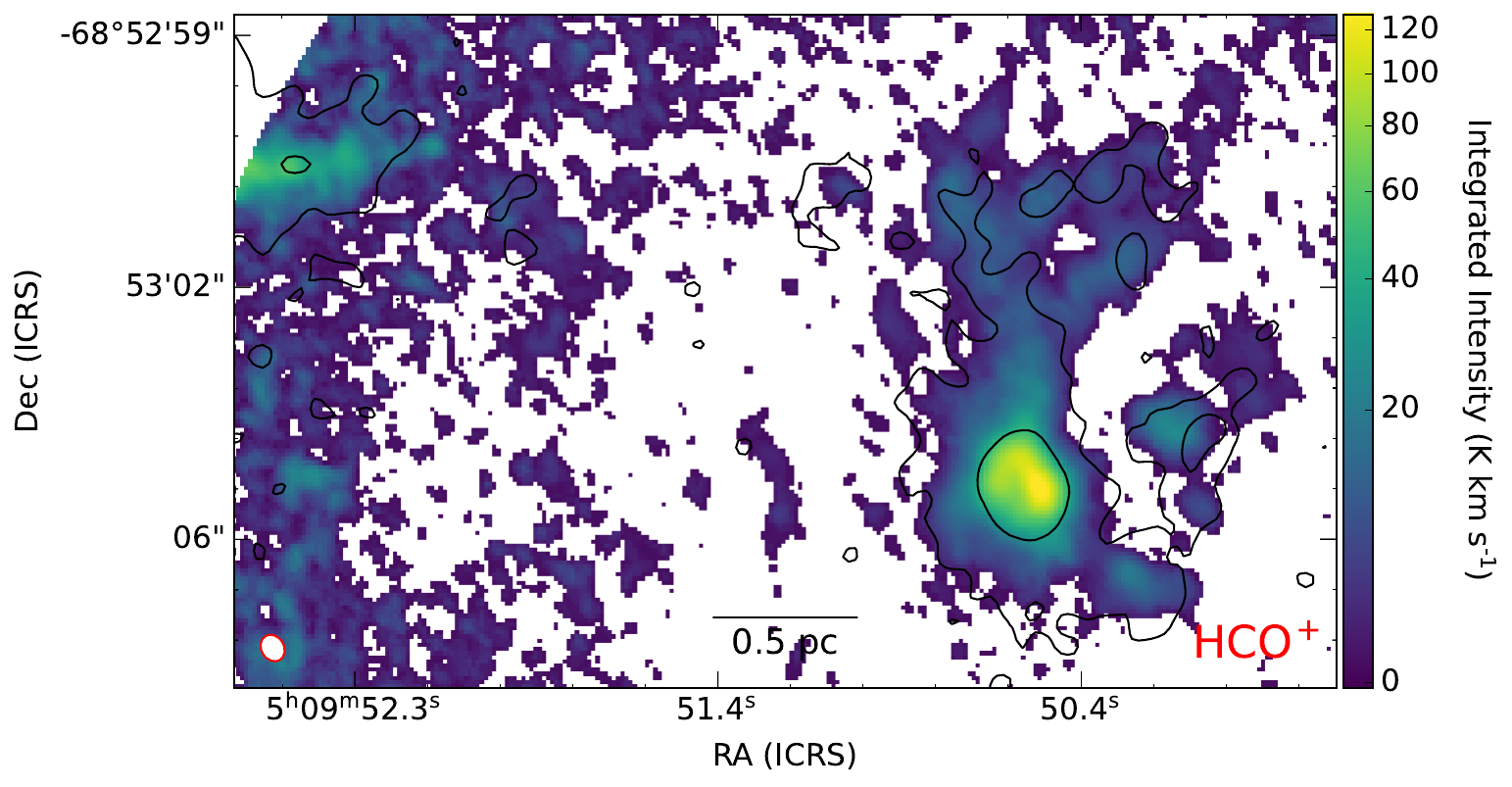}
\includegraphics[width=0.8\textwidth]{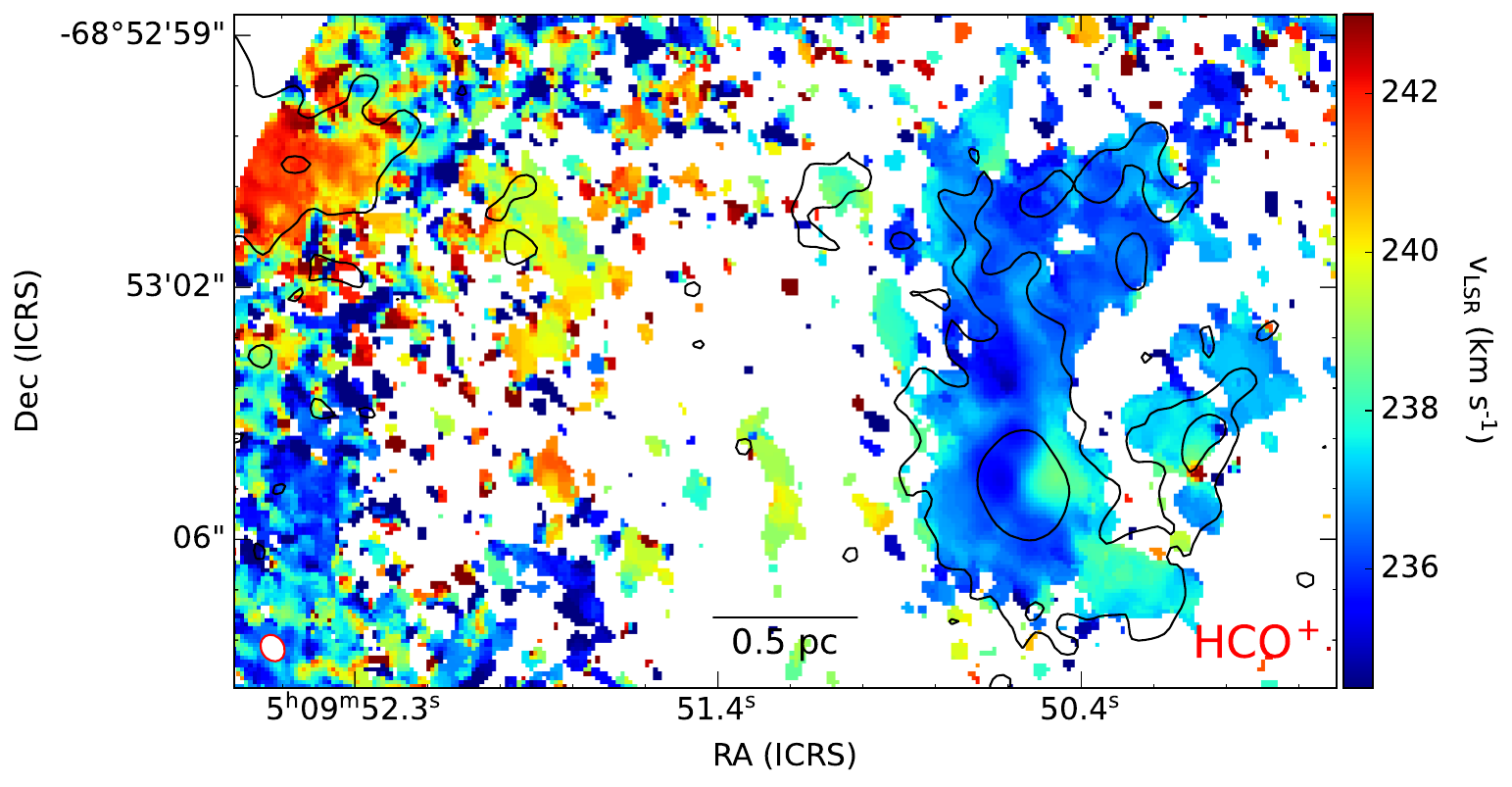}
\caption{The HCO$^{+}$ (4--3) integrated intensity (top) and velocity (bottom) images of N\,105--1.  The emission was integrated between 232 and 244 km s$^{-1}$. Pixels with the signal-to-noise value less than three in the HCO$^{+}$ (4--3) integrated intensity image are masked in both images. The ALMA beam size is shown in the lower left corner in each image.  The 1.2 mm continuum contours are overlaid on both images with contour levels of (3, 30)$\sigma$.  \label{f:momHCOp}}
\end{figure*}

\begin{figure*}
\centering
\includegraphics[width=\textwidth]{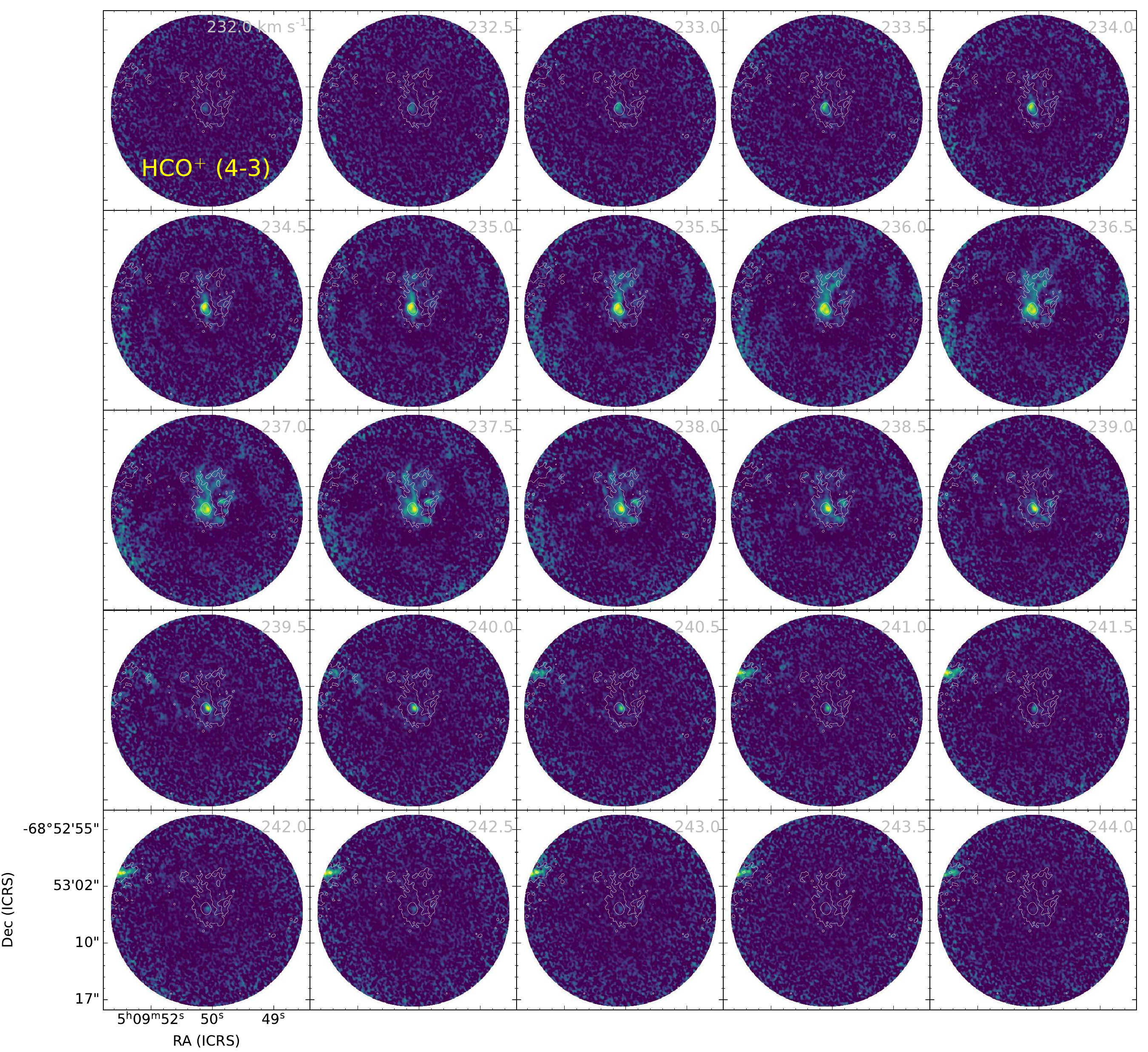}
\caption{The HCO$^{+}$ (4--3) channel maps. The 1.2 mm continuum contours are overlaid on all channel maps for reference; the contour levels are (3, 30)$\sigma$.  \label{f:HCOpchan}}
\end{figure*}

\clearpage


\bibliographystyle{aasjournal}
\bibliography{refs.bib}

\end{document}